\documentclass[prd,amsmath,amssymb,showpacs,nofootinbib]{revtex4}

\usepackage{graphicx}
\usepackage{dcolumn}
\usepackage{bm}
\usepackage{mathrsfs} 
\usepackage{amsthm}
\usepackage{xspace}
\usepackage{enumitem}
\usepackage{expdlist}

\mathchardef\mhyphen="2D

\newcommand{\dzero}{D0\xspace}

\newcommand{\W}{\ensuremath{W}\xspace}
\newcommand{\Z}{\ensuremath{Z}\xspace}

\newcommand{\uquark}{\ensuremath{u}\xspace}

\newcommand{\dquark}{\ensuremath{d}\xspace}

\newcommand{\squark}{\ensuremath{s}\xspace}

\newcommand{\cquark}{\ensuremath{c}\xspace}

\newcommand{\bquark}{\ensuremath{b}\xspace}

\newcommand{\tauptaum}{\ensuremath{\tau^+ \tau^-}\xspace}
\newcommand{\emu}{\ensuremath{e\mu}\xspace}

\newcommand{\sigmaP}{\ensuremath{\sigma_{\!P}}\xspace}
\newcommand{\dsigmaP}{\ensuremath{{\rm d}\sigmaP}\xspace}

\newcommand{\sigmaPobs}{\ensuremath{\sigmaP^{\rm obs}}\xspace}
\newcommand{\sigmaPobsprime}{\ensuremath{{\sigmaP^{\rm obs}}'}\xspace}

\newcommand{\sigmaobsprime}{\ensuremath{{\sigma^{\rm obs}}'}\xspace}

\newcommand{\sigmattbarobsprime}{\ensuremath{{\sigma_{\!\ttbar}^{\rm obs}}'}\xspace}
\newcommand{\lsample}{\ensuremath{L_{\rm sample}}\xspace}
\newcommand{\pevt}{\ensuremath{L_{\rm evt}}\xspace}
\newcommand{\pevtP}{\ensuremath{L_P}\xspace}
\newcommand{\pevtttbar}{\ensuremath{L_{\ttbar}}\xspace}

\newcommand{\pevtwjjjj}{\ensuremath{L_{\wjjjj}}\xspace}
\newcommand{\pevtztautaujj}{\ensuremath{L_{\ztautaujjshort}}\xspace}
\newcommand{\Pzero}{\ensuremath{P_0}\xspace}
\newcommand{\fPzero}{\ensuremath{f_{\Pzero}}\xspace}
\newcommand{\pevtPzero}{\ensuremath{L_{\Pzero}}\xspace}
\newcommand{\nfP}{\ensuremath{n_{f_P}}\xspace}
\newcommand{\lsampleoned}{\ensuremath{L_{\rm sample}^{\rm 1d}}\xspace}

\newcommand{\mtop}{\ensuremath{m_{t}}\xspace}
\newcommand{\jes}{\ensuremath{S_l}\xspace}
\newcommand{\bjes}{\ensuremath{S_b}\xspace}
\newcommand{\phijes}{\ensuremath{S_\phi}\xspace}

\newcommand{\sjes}{\jes}
\newcommand{\sbjes}{\bjes}
\newcommand{\mt}{\ensuremath{m_{t}}\xspace}
\newcommand{\ftop}{\ensuremath{f_{\ttbar}}\xspace}

\newcommand{\fttj}{\ensuremath{f_{\ttbar j}}\xspace}
\newcommand{\fwjjjj}{\ensuremath{f_{\wjjjj}}\xspace}
\newcommand{\fwbbjj}{\ensuremath{f_{\wbbjj}}\xspace}
\newcommand{\fztautaujj}{\ensuremath{f_{\ztautaujj}}\xspace}
\newcommand{\fwwjj}{\ensuremath{f_{WWjj}}\xspace}

\newcommand{\mW}{\ensuremath{m_{W}}\xspace}

\newcommand{\qoverptmu}{\ensuremath{\left(q/\pt\right)_\mu}\xspace}
\newcommand{\qoverptmuprime}{\ensuremath{\left(q/\pt\right)_{\mu'}}\xspace}
\newcommand{\gen}{\ensuremath{\rm mat}\xspace}
\newcommand{\qoverptmurec}{\ensuremath{\qoverptmu^{\rm rec}}\xspace}
\newcommand{\qoverptmugen}{\ensuremath{\qoverptmu^{\gen}}\xspace}
\newcommand{\qoverptmuprimegen}{\ensuremath{\qoverptmuprime^{\gen}}\xspace}

\newcommand{\Et}{\ensuremath{E_T}\xspace}

\newcommand{\ptttbar}{\ensuremath{p_{T,\ttbar}}\xspace}

\newcommand{\et}{\ensuremath{E_{T}}\xspace}
\newcommand{\etmiss}{\ensuremath{E \kern-0.6em\slash_{T}}\xspace}
\newcommand{\etmissx}{\ensuremath{E \kern-0.6em\slash_{x}}\xspace}
\newcommand{\etmissy}{\ensuremath{E \kern-0.6em\slash_{y}}\xspace}
\newcommand{\ptmiss}{\ensuremath{p \kern-0.45em\slash_{T}}\xspace}
\newcommand{\ptmissx}{\ensuremath{p \kern-0.45em\slash_{x}}\xspace}
\newcommand{\ptmissy}{\ensuremath{p \kern-0.45em\slash_{y}}\xspace}
\newcommand{\ptmissvec}{\ensuremath{\vec{p} \kern-0.45em\slash_{T}}\xspace}

\newcommand{\pt}{\ensuremath{p_{T}}\xspace}

\newcommand{\glueglue}{\ensuremath{gg}\xspace}
\newcommand{\qqbar}{\ensuremath{q\bar{q}}\xspace}

\newcommand{\ppbar}{\ensuremath{p\bar{p}}\xspace}

\newcommand{\bbbar}{\ensuremath{b\bar{b}}\xspace}
\newcommand{\ttbar}{\ensuremath{t\bar{t}}\xspace}
\newcommand{\ttj}{\ensuremath{t\bar{t}j}\xspace}

\newcommand{\ztautaujj}{\ensuremath{Z/\gamma^*(\to\tau^+\tau^-)jj}\xspace}
\newcommand{\ztautaujjshort}{\ensuremath{Zjj}\xspace}
\newcommand{\ztautaubb}{\ensuremath{Z/\gamma^*(\to\tau^+\tau^-)\bbbar}\xspace}

\newcommand{\wjets}{\ensuremath{W}+{\rm jets}\xspace}
\newcommand{\zjets}{\ensuremath{Z}+{\rm jets}\xspace}
\newcommand{\wwjj}{\ensuremath{WWjj}\xspace}
\newcommand{\Wjets}{\wjets}

\newcommand{\wjjjj}{\ensuremath{\W jjjj}\xspace}

\newcommand{\wbbjj}{\ensuremath{\W\!\bbbar jj}\xspace}

\newcommand{\ejets}{\ensuremath{e}+\rm jets\xspace}
\newcommand{\mujets}{\ensuremath{\mu}+\rm jets\xspace}
\newcommand{\ljets}{lepton+jets\xspace}
\newcommand{\Ljets}{Lepton+jets\xspace}
\newcommand{\ztautau}{\ensuremath{\Z\to\tauptaum}\xspace}

\newcommand{\mthadsquare}{\ensuremath{m_{t_1}^2}\xspace}
\newcommand{\mtlepsquare}{\ensuremath{m_{t_2}^2}\xspace}
\newcommand{\mwhadsquare}{\ensuremath{m_{W_{\rm had}}^2}\xspace}
\newcommand{\mtonesquare}{\ensuremath{m_{t_1}^2}\xspace}
\newcommand{\mttwosquare}{\ensuremath{m_{t_1}^2}\xspace}
\newcommand{\pqonemag}{\ensuremath{\left|\vec{p}_u\right|}\xspace}
\newcommand{\pbonemag}{\ensuremath{\left|\vec{p}_{b_1}\right|}\xspace}
\newcommand{\pbtwomag}{\ensuremath{\left|\vec{p}_{b_2}\right|}\xspace}
\newcommand{\pzblepnu}{\ensuremath{\left(\vec{p}_{b\nu}\right)_z}\xspace}
\newcommand{\deltapxnuonetwo}{\ensuremath{\left(\vec{p}_{\nu_1}-\vec{p}_{\nu_2}\right)_x}\xspace}
\newcommand{\deltapynuonetwo}{\ensuremath{\left(\vec{p}_{\nu_1}-\vec{p}_{\nu_2}\right)_y}\xspace}
\newcommand{\qptmu}{\ensuremath{\left(q/p_T\right)_\mu}\xspace}

\newcommand{\detJ}{\ensuremath{{\rm det}\left(J\right)}\xspace}

\newcommand{\fpdf}{\ensuremath{f_{\rm PDF}}\xspace}
\newcommand{\qone}{\ensuremath{\xi_1}\xspace}
\newcommand{\qtwo}{\ensuremath{\xi_2}\xspace}
\newcommand{\qgeneral}{\ensuremath{\xi}\xspace}

\newcommand{\madgraph}{{\sc madgraph}\xspace}
\newcommand{\alpgen}{{\sc alpgen}\xspace}

\newcommand{\vecbos}{{\sc vecbos}\xspace}
\newcommand{\geant}{{\sc geant}\xspace}
\newcommand{\vegas}{{\sc vegas}\xspace}

\newcommand{\MeV}{\ensuremath{\mathrm{Me\kern-0.1em V}}\xspace}
\newcommand{\GeV}{\ensuremath{\mathrm{Ge\kern-0.1em V}}\xspace}
\newcommand{\GeVc}{\ensuremath{\mathrm{Ge\kern-0.1em V}}\xspace}
\newcommand{\GeVcc}{\ensuremath{\mathrm{Ge\kern-0.1em V}}\xspace}
\newcommand{\TeV}{\ensuremath{\mathrm{Te\kern-0.1em V}}\xspace}

\newcommand{\DeltaR}{\ensuremath{\Delta {\cal R}}\xspace}

\newcommand{\Eref}[1]{(\ref{#1})}
\newcommand{\Fref}[1]{Figure~\ref{#1}}

\newcommand{\thirdorder}{$3^{rd}${-order}\xspace}

\newcommand{\captionfont}{\small}
\newcommand{\MC}{Monte~Carlo\xspace}

\newcommand{\bID}{\ensuremath{b}~identification\xspace}

\newcommand{\bpartons}{\ensuremath{b}~partons\xspace}

\newcommand{\gsim}{\lower.25ex\hbox{\ensuremath{\stackrel{\normalsize >}{\scriptstyle \sim}}}}
\newcommand{\lsim}{\lower.25ex\hbox{\ensuremath{\stackrel{\normalsize <}{\scriptstyle \sim}}}}

\newcommand{\newlineonlyintwocol}{}
\newcommand{\nonumberonlyintwocol}{}

\begin{document}

\title{The Matrix Element Method\\ and its Application to
Measurements of the Top Quark Mass}

\author{Frank Fiedler}
\affiliation{Johannes Gutenberg University Mainz}
\author{Alexander Grohsjean}
\affiliation{CEA, Irfu, SPP, Saclay}
\author{Petra Haefner}
\affiliation{Max Planck Institute Munich}
\author{Philipp Schieferdecker}
\affiliation{Karlsruhe Institute of Technology (KIT)}
           
\begin{abstract}
The most precise measurements of the top quark mass are
based on the Matrix Element method.
We present a detailed description of this analysis method, 
taking the measurements of the top quark 
mass in final states with one and two charged leptons
as concrete examples.
In addition, we show how the Matrix Element method
is suitable to reduce the dominant systematic uncertainties related
to detector effects, by treating the absolute energy 
scales for \bquark-quark and light-quark jets independently as free parameters
in a simultaneous fit together with the top quark mass.
While the determination of the light-quark jet energy scale has
already been applied in several recent measurements, the separate
determination of the absolute \bquark-quark jet energy scale is 
a novel technique
with the prospect of reducing the overall uncertainty on the top quark
mass in the final measurements at the Tevatron and in analyses at the
LHC experiments.
The procedure is tested on Monte Carlo generated events with a realistic
detector resolution.
\end{abstract}

\maketitle

\newpage

\section{Introduction}
\label{sec:intro}
The Matrix Element method is unique among the analysis methods used
in experimental particle physics because of the direct link it
establishes between theory and event reconstruction.
Originally developed to minimize the 
statistical uncertainty in measurements of \ttbar events at the 
Tevatron experiments \dzero and CDF~\cite{bib-originalmem}, it
has since been applied with 
great success to measurements of the top quark
mass \mtop~\cite{bib-mtopmem} and also
in the discovery of electroweak production of single top 
quarks~\cite{bib-singletop}.
The method can in principle be used for any
measurement, with the largest gain 
compared to cut-based analysis techniques expected for processes involving
intermediate resonances and leading to many-particle final states.
In general, the Matrix Element method can be 
used to determine several unknown parameters (theoretical parameters
describing the physics processes measured as well as
experimental parameters describing the detector response)
at the same time in one measurement,
thus also allowing for a reduction of systematic uncertainties.
This paper presents the analysis method in general and also
gives an example of how the determination of such 
additional parameters can be implemented.

Recent measurements in \ttbar events containing one leptonic and one hadronic 
\W decay (``\ljets events'') already exploit the known \W mass 
to constrain the energy scale for light-quark jets and significantly
reduce the main systematic uncertainty of early measurements of the top quark
mass.
Among the largest remaining systematic uncertainties is the 
uncertainty on potential
differences between the energy scales \bjes and \jes
for \bquark- and light-quark 
jets\footnote{The quantities \bjes and \jes denote scale
  factors relative to the nominal detector 
  calibration.
  In the definition of these factors, 
  it is assumed that all experimental corrections to 
  measured jet energies have been applied to the events that enter the 
  analysis, such that \bjes and \jes do not
  depend on quantities like the jet direction or energy.
  Uncertainties on such dependencies give rise to additional systematic
  uncertainties of the measurement e.\ g.\ of the top quark mass.}.
Without further improvements, it will soon 
become a limiting uncertainty for those
measurements that dominate the world-average value~\cite{bib-habil}.
In this paper we show how 
together with the top quark mass, 
a simultaneous additional measurement of the 
\bquark-quark jet energy scale,
which was first proposed in~\cite{bib-mtopbjes} for the \ljets channel,
can be incorporated naturally in the Matrix Element technique -- not
only for measurements in the \ljets channel,
but also in \ttbar events with two leptonic \W decays (``dilepton events'')
where the quantities to be measured cannot be
reconstructed based on the kinematical information of a single event
alone.
This study considers the case of the Tevatron (proton-antiproton collisions at a
center-of-mass energy of $1.96\,\TeV$) as a concrete
example but is applicable to the two LHC experiments ATLAS and CMS
as well.

The paper is structured as follows: 
In Section~\ref{method.sec}, 
an overview of the Matrix Element method is given.
Section~\ref{samples.sec} discusses the generation and selection of
\ttbar events used for the studies described in the further sections.
This is followed by a 
discussion of the implementation of the likelihood calculation 
for signal and background processes 
for \ttbar measurements in the \ljets and dilepton
channels in Section~\ref{application.sec}.
We then describe studies of the performance of the new
measurement technique, separately for \ljets
(Section~\ref{enstestljets.sec}) and dilepton 
(Section~\ref{enstestdilepton.sec})
\ttbar events.
In Section~\ref{systuncs.sec}, systematic uncertainties on \mtop
are addressed, with an emphasis on the effect of events with
significant initial- and final-state radiation.
Section~\ref{conclusions.sec} summarizes the findings and gives an outlook.

\section{The Matrix Element Method}
\label{method.sec}
The Matrix Element method is based on the likelihood \lsample to observe 
a sample of selected events in the detector.
The likelihood is obtained directly from the 
theory prediction for the differential cross-sections of the relevant
processes and the detector resolution
and is calculated as a function of 
the assumed values for each of the parameters to be measured.
The minimization of $-\ln\lsample$ yields the measurement of the
parameters, where the likelihood \lsample for the entire event sample
is computed as the product of likelihoods to observe each individual event.
This is in contrast to most analysis methods 
used in experimental particle physics, where
distributions from observed events in the detector are compared with 
corresponding distributions obtained from simulated events that have
been generated
according to theory and then passed through a detector simulation and 
the same event reconstruction software.

This paper concentrates on 
the case of the measurement of the top quark mass in \ljets and
dilepton events at a hadron collider, where the parameters to be
measured are
the top quark mass \mtop, factors \bjes and \jes describing the energy
scales for \bquark- and light-quark jets relative to the default
energy scale\footnote{In dilepton \ttbar events, only \bquark-quark jets
  occur and thus only \bjes is determined.
  Only one overall scale factor \bjes is determined
  for all \bquark jets; thus for example no 
  distinction is made between \bquark jets with and without a reconstructed
  muon inside the jet.}, and the fraction \ftop
of signal events in the channel under consideration.
A comparison of the Matrix Element method with 
other methods to measure the top quark mass can be found in~\cite{bib-habil}.

\subsection{The Event Likelihood}
\label{method.pevt.sec}
The sample likelihood \lsample for $N$ measured events 
to have measured properties $x_1, ..., x_N$
can be written as
\begin{equation}
\label{lsample.eqn}
    \lsample(x_1,..,x_N;\,\vec{\alpha},\vec{\beta},\vec{f})
  = 
    \prod_{i=1}^{N}\pevt(x_i;\,\vec{\alpha},\vec{\beta},\vec{f})
  \, ,
\end{equation}
where the symbol $\vec{\alpha}$ denotes assumed values of the physics
parameters to be measured, $\vec{\beta}$ stands for 
parameters describing the detector response that are to be determined,
and $\vec{f}$ is defined below.
The likelihood $\pevt(x_i;\,\vec{\alpha},\vec{\beta},\vec{f})$ 
to observe event $x_i$ under the assumption of parameter values $\vec{\alpha}$, 
$\vec{\beta}$, and $\vec{f}$ is given as the linear combination 
\begin{equation}
\label{pevt.eqn}
    \pevt(x_i;\,\vec{\alpha},\vec{\beta},\vec{f}) 
  = 
    \sum_{{\rm processes}\, P} f_P \pevtP(x_i;\,\vec{\alpha},\vec{\beta})
  \ ,
\end{equation}
where the sum is over all individual processes $P$ that could have led to 
the observed event $x_i$, 
$\pevtP(x_i;\,\vec{\alpha},\vec{\beta})$ is the likelihood to observe this
event under the assumption that it was produced via process $P$, and
$f_P$ denotes the fraction of events from process $P$
in the entire event sample, with $\sum_{P} f_P = 1$.
In total, the physics parameters $\vec{\alpha}$, the detector
response described by $\vec{\beta}$, and the event fractions $\vec{f}$ 
are to be determined simultaneously from the minimization of
$-\ln\lsample$.

The likelihood $\pevtP(x;\,\vec{\alpha},\vec{\beta})$ 
in turn is given by the theoretical description of the process 
and the resolution of the concrete experiment, and is
computed as the convolution of the differential 
partonic cross section with the parton distribution functions of the colliding
hadrons and with the detector response.
To make the likelihood calculation manageable, simplifying assumptions
are introduced.
This concerns the description of both the detector 
response and the physics processes $P$, where only the 
dominant ones are considered explicitly, and where
the effects of parton shower and 
hadronization are accounted for with a simple parametrization.
Because of these simplifications, the technique 
has to be calibrated using fully simulated events
before applying it in an actual measurement on data.
In this paper, a conceptual study is presented.
We show how the method can be validated with events that have
been generated 
under the same assumptions as made in the likelihood calculation, which 
allows to demonstrate that 
the measurement method as such is unbiased.

In Equation~(\ref{pevt.eqn}), not
all likelihoods $\pevtP$ necessarily depend on all parameter values;
e.g.\ for the measurement of the top quark mass, the
likelihood $\pevtttbar$ depends
on the assumed top quark mass, while the likelihoods for an event to be 
produced via a background process (which by definition does not involve
top quark production or decay) do not.
Even if a likelihood does depend on a certain parameter, this dependency does 
not necessarily have to be taken into account explicitly; for example,
for the top quark mass measurement
it will be shown in Sections~\ref{enstestljets.sec} 
and~\ref{enstestdilepton.sec} that the dependency of the likelihoods for
background processes on the jet energy scales can be neglected
without introducing large biases on the top quark mass and energy scale
measurements.

\subsection{The Likelihood for one Process}
\label{method.pprc.sec}
\begin{figure}
\begin{center}
\includegraphics[width=\textwidth]{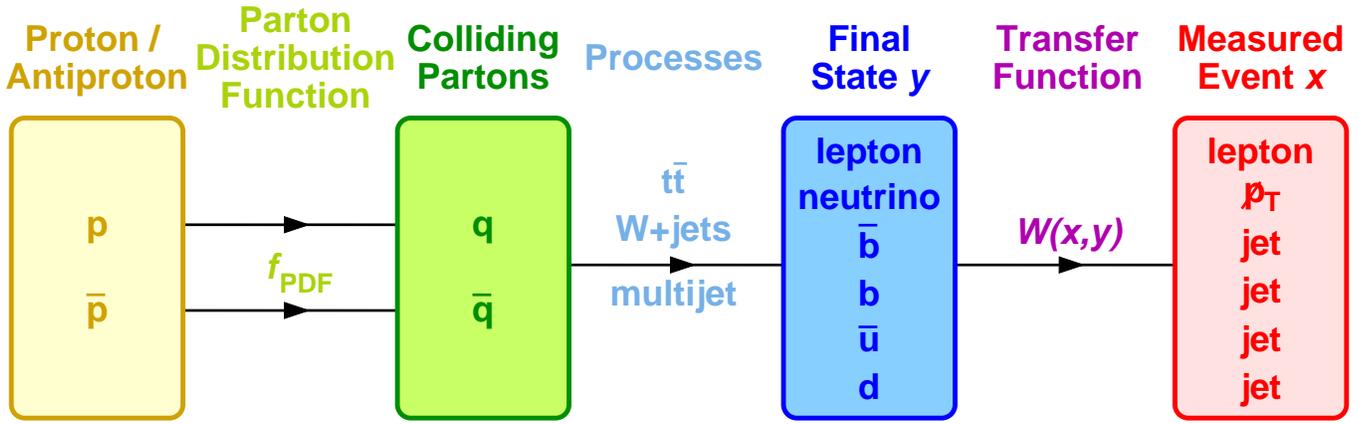}
\caption{\captionfont\label{psgn_schematic.fig}Schematic representation 
  of the calculation of the likelihood 
  to obtain a given observed
  \ljets event at a proton-antiproton collider
  (similar figures apply to dilepton events and to other 
  processes).
}
\end{center}
\end{figure}

The individual contributions to the likelihood 
for an observed event $x$ to be produced via
a given process $P$ are described in this section.
They are visualized schematically
in Figure~\ref{psgn_schematic.fig} for the example of a \ljets \ttbar
event at the Tevatron.
The observed event $x$, shown at the right,
is fixed while integrating over all possible momentum configurations 
$y$ of final-state particles.  
The differential cross section for the process 
is convoluted with the probability for the final-state
partons to yield the observed event (transfer function), and with
the probability to find initial-state partons of given flavor
and momenta inside the colliding proton and antiproton (parton
distribution function).  
All possible
assignments of final-state particles to
measured objects in the detector are
considered by the transfer function.
For each
partonic final state under consideration, the initial-state 
parton momenta are determined by energy and momentum conservation.

The likelihood for a final state with $n_f$ partons and 
given four-momenta $y$ to be produced in the hard-scattering process is 
proportional to the differential cross section ${\rm d}\sigma_P$ of the 
corresponding process, given by
\begin{eqnarray}
  \label{dsigmahs.eqn}
    \dsigmaP(a_1 a_2 \to y;\,\vec{\alpha})
   = 
    \frac{(2 \pi)^{4}\! 
    \left| \mathscr{M}_P\left(a_1 a_2 \to y;\,\vec{\alpha}\right) \right|^{2}}
         {\qone \qtwo s}
    {\rm d}\Phi_{n_f}
   \ , \nonumberonlyintwocol \newlineonlyintwocol
\end{eqnarray}
where 
$a_1 a_2$ and $y$ stand for the kinematic variables of the partonic
initial and final states, respectively.
The symbol $\mathscr{M}_P$ denotes the matrix element for this process,
$s$ is the center-of-mass energy squared of the collider, $\qone$ and $\qtwo$ are the
momentum fractions of the colliding partons $a_1$ and $a_2$
(which are assumed to be 
massless) within the colliding proton and antiproton\footnote{
This discussion is based on the situation at the Tevatron \ppbar collider as a
concrete example but is equally valid for the LHC when the antiproton
is replaced with a proton and the appropriate PDF is used.}, and
${\rm d}\Phi_{n_f}$ is an element of $n_f$-body phase space.

To obtain the differential cross section 
$\dsigmaP(\ppbar\to y;\,\vec{\alpha})$
in \ppbar collisions, the differential cross section
from Equation~\Eref{dsigmahs.eqn} is convoluted 
with the parton density functions (PDF) and summed over all possible flavor
compositions of the colliding partons,
\begin{equation}
  \label{dsigmaPpp.eqn}
    \dsigmaP(\ppbar\to y;\,\vec{\alpha})
  = 
    \int\limits_{\qone, \qtwo} \sum_{a_1, a_2}
    {\rm d}\qone {\rm d}\qtwo\ 
    f_{\rm PDF}^{a_1} (\qone)\ 
    \bar{f}_{\rm PDF}^{a_2} (\qtwo)\ 
    \dsigmaP(a_1 a_2 \to y;\,\vec{\alpha})
  \ ,
\end{equation}
where $f_{\rm PDF}^{a_{1}} (\qgeneral_{1})$ 
and $\bar{f}_{\rm PDF}^{a_{2}} (\qgeneral_{2})$ denote
the probability densities to find a parton of
given flavor $a_{1}$ and momentum fraction
$\qgeneral_{1}$ in the proton and one of flavor
$a_{2}$ and momentum fraction
$\qgeneral_{2}$ in the antiproton, respectively.
This equation reflects QCD factorization~\cite{bib-pdg}.

The finite detector resolution is taken into account via a
convolution with
a transfer function $W(x,y;\,\vec{\beta})$ that describes the probability to
reconstruct a partonic final state $y$ as $x$ in the detector, given
the values $\vec{\beta}$ of the
parameters describing the detector response.
The differential cross section to observe a given reconstructed 
event $x$ then becomes
\begin{eqnarray}
  \label{dsigmapp.eqn}
    \dsigmaP(\ppbar\to x;\,\vec{\alpha},\vec{\beta})
  & = &
    \int\limits_{y} \dsigmaP(\ppbar\to y;\,\vec{\alpha})\
    W(x,y;\,\vec{\beta})
  \ .
\end{eqnarray}

Only events that are inside the 
detector acceptance and that pass the trigger conditions and offline
event selection are used in the measurement.
To obtain a properly normalized likelihood, the
overall cross section of events observable in the detector,
\begin{equation}
  \label{sigmaobs.eqn}
    \sigmaPobs(\vec{\alpha},\vec{\beta})
  =
    \int\limits_{x,y} {\rm d}\sigmaP(\ppbar\to y;\,\vec{\alpha})\
    W(x,y;\,\vec{\beta})\
    f_{\rm acc}(x)\
    {\rm d}x
  \ ,
\end{equation}
is used, where 
$f_{\rm acc}=1$ for selected events 
and $f_{\rm acc}=0$ otherwise.
One then obtains
\begin{eqnarray}
\label{normpevtP.eqn}
    \pevtP(x;\,\vec{\alpha},\vec{\beta})\ {\rm d}x
  & = &
    \frac{\dsigmaP(x;\,\vec{\alpha},\vec{\beta})}
         {\sigmaPobs(\vec{\alpha},\vec{\beta})}
\end{eqnarray}
as the (differential) likelihood that an event produced via process $P$ has
measured properties $x$ (and not other properties that would still
lead to an event passing the event selection criteria).

\subsection{Description of the Detector Response}
\label{method.tf.sec}
This section describes a
parametrization of the transfer function $W(x,y;\,\vec{\beta})$
appropriate for measurements of high-\pt objects at a hadron collider.
The matrix element method is based on a fast parametrization 
that reproduces the basic properties of the detector.
Any biases introduced can be determined (and then corrected for)
when ensemble tests as described in Section~\ref{method.enstest.sec}
are performed with events generated with a full detector simulation,
typically based on the \geant~\cite{bib-geant} package.
This section is written with measurements in \ttbar events in 
mind but is applicable to other processes as well.

The transfer function $W(x,y;\,\vec{\beta})$ describes 
the probability density ${\rm d}P$
to reconstruct an assumed partonic final state $y$ as a measurement $x$ in the 
detector:
\begin{equation}
  \label{tfdef.eqn}
    {\rm d}P
  = 
    W(x,y;\,\vec{\beta}) {\rm d}x
  \ .
\end{equation}
Because the final-state partons are assumed to result in some
measured event $x$, the normalization condition
\begin{equation}
  \label{tfnorm_general.eqn}
    \int_x W(x,y;\,\vec{\beta}) {\rm d}x
  =
    1
\end{equation}
holds, where the integral is over all possible events $x$.
Effects due to selection cuts or finite detector acceptance are discussed
in Section~\ref{method.normL.sec}.

The transfer function is assumed to factorize into contributions from 
each measured final-state particle.
Aspects to be considered in the transfer function are in principle the
measurement of the momentum of a particle (both of its energy and of its
direction) as well as its identification.
Thus \bquark-tagging information
for the jets 
can be included, which can help to 
distinguish signal from background events.

In many applications like the description of \ttbar events, a number of 
assumptions can be made~\cite{bib-me} about how final-state particles are 
measured in the detector, such that the dimensionality
of the integration over the final-state particle phase space described in 
Section~\ref{method.pprc.sec} is reduced.
Individual particles can be described in the transfer function as follows:
\begin{list}{$\bullet$}{\setlength{\itemsep}{0.5ex}
                        \setlength{\parsep}{0ex}
                        \setlength{\topsep}{0ex}}
\item{\bf Isolated energetic electrons:}
  Electrons are assumed to be
  unambiguously identified (i.e.\ an electron is not reconstructed as
  a muon or a jet).
  The electron direction and energy are both assumed to be
  well-measured, i.e.\ during integration, the final-state
  electron is assumed to be identical to the measured particle.
  This is justified for \ttbar events
  since the resolution for electrons is far better than
  that for jets, and the jet energy resolution will dominate all
  effects due to the finite detector resolution.
\item{\bf Isolated energetic muons:}
  As for electrons, muons are assumed to be unambiguously
  identified, and their directions to be precisely measured.
  However, instead of the energy
  the detector typically yields a measurement of $\qoverptmu$, the muon
  charge divided by the transverse momentum.
  Consequently, the muon energy resolution can be poor for high-\pt
  muons, and thus a
  transfer function $W_{\mu}$ allowing for a finite resolution is introduced.

  In the studies presented in this paper, the function
  \begin{equation}
    \label{tfmu.eqn}
      W_{\mu}\left( \qoverptmurec,\qoverptmugen \right)
    =
      \frac{1}{\sqrt{2\pi}\sigma} 
      \exp\left( -\frac{1}{2} \left( \frac{ \qoverptmurec - \qoverptmugen } 
                                          { \sigma }
                              \right)^2
          \right)
  \end{equation}
  is used to describe the likelihood that a muon with 
  charge and momentum \qoverptmugen (described by the {\em mat}rix element)
  is {\em rec}onstructed with \qoverptmurec.
  The resolution $\sigma$ 
  depends on the pseudorapidity $\eta$ to account for muon tracks
  at large $|\eta|$ that do not reach the full radius of the tracking detector.
  The parameter values are taken from~\cite{bib-me}.
\item{\bf Energetic {\boldmath$\tau$} leptons:}
  Events with energetic $\tau$ lepton decays are typically selected if
  the visible decay products pass a minimum energy cut.
  In this case, the directions of the visible decay
  products are close to that of the original $\tau$ lepton, but only 
  a fraction of the $\tau$ energy can be measured in the detector.

  In this paper, only leptonic decays $\tau\to\ell\overline{\nu}_\ell\nu_\tau$
  are considered, where the symbol $\ell$ denotes an electron or muon.
  Consequently, a transfer function
  $W_{\tau}\left( E_\ell^{\rm rec}/E_\tau^{\rm mat} \right)$
  is introduced to describe
  the likelihood to obtain a charged lepton with a given 
  energy fraction $E_\ell/E_\tau$ of the decaying $\tau$ lepton.
  For the study presented here, it is parametrized as a 
  \thirdorder polynomial as shown in \Fref{fig:tautransferfunction}.
  The $\tau$ direction is taken to be 
  well approximated by the direction of the reconstructed charged lepton.

\begin{figure}[ht]
\begin{center}
\includegraphics[width=0.55\textwidth]{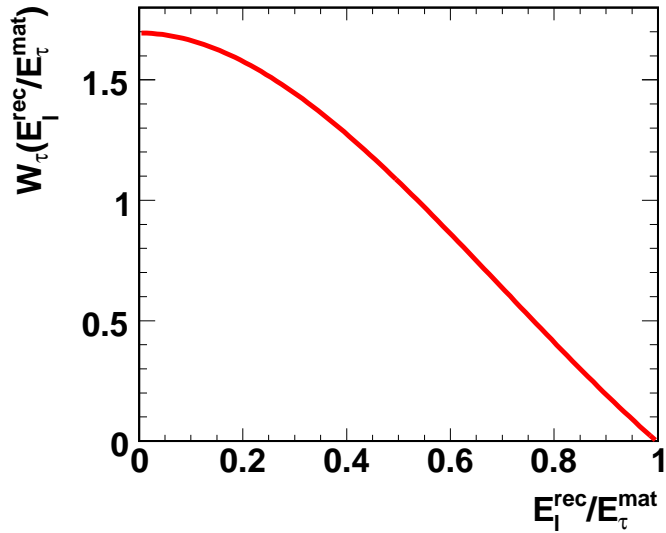}
\caption{\label{fig:tautransferfunction} 
  Likelihood for an electron or muon to carry a given energy fraction of the
  initial $\tau$ lepton energy.}
\end{center}
\end{figure}

  For muonic $\tau$ decays, the muon transfer function introduced above 
  in principle has to be taken into account as well to describe the 
  transition from the assumed to the reconstructed muon transverse momentum.
  However, muons from $\tau$ decays typically have low enough transverse
  momenta 
  so that the muon \pt 
  can be assumed to be well-measured in most applications.
  In the following, the muon transfer function is omitted 
  for muonic $\tau$ decays.
  Also, in this study
  we consider the reconstruction efficiency to be independent of
  the lepton energy.
\item{\bf Energetic quarks and gluons:}
  The directions of final-state quarks
  and gluons are assumed to be well-measured by the jet directions, and
  transfer functions are introduced for the jet energy measurement.
  The probability density for a jet energy measurement 
  $E_{j}^{\,\rm rec}$ in the
  detector if the true quark energy is $E_{j}^{\,\gen}$ 
  (depending on the overall jet energy scale \bjes or \jes)
  is given by the jet energy transfer function
    $W_{\rm jet} \left( E_{j}^{\,\rm rec},\ 
                        E_{j}^{\,\gen},\ 
                        \phi_{j}^{\,\gen};\ 
                        \phijes \right)$.
  In principle, different transfer functions apply to
  gluon jets and jets from different quark flavors 
  $\phi_{j}^{\,\gen}$.

  For $\phijes\neq1$, the jet transfer function is computed as
  \begin{equation}
    \label{tfjetjes.eqn}
      W_{\rm jet}(E_{j}^{\,\rm rec},
                  E_{j'}^{\,\gen},
                  \phi_{j'}^{\,\gen};\ 
                  \phijes) 
    =
      \frac{W_{\rm jet}(\frac{E_{j}^{\,\rm rec}}{\phijes},
                        E_{j'}^{\,\gen}, 
                        \phi_{j'}^{\,\gen};\ 
                        1)}
           {\phijes} 
    \ ,
  \end{equation}
  where the factor $\phijes$ in the denominator ensures the correct
  normalization in the absence of selection cuts.

  In this paper, 
  the same jet energy transfer function is used to describe light-quark
  (\uquark, \dquark, \squark, and \cquark) 
  and gluon jets\footnote{In events passing the \ttbar selection cuts,
    gluon jets arise in 
    background processes whose description is anyway only approximate;
    therefore no separate transfer function for gluon jets is introduced.}; 
  an independent transfer function is 
  used for \bquark-quark jets.
  The parametrization of the transfer function follows that of the 
  \dzero experiment given in~\cite{bib-me}, with parameters depending 
  on the jet energy and pseudorapidity.
  In a fraction of those \bquark jets that contain a semimuonic \bquark-hadron
  decay, the muon is identified, and these jets could in principle be described 
  with a separate transfer function~\cite{bib-me}
  (while the jets with unidentified semileptonic decays would still have to be
  described with one function together with all other \bquark jets).
  In this paper, only one class of \bquark jets is considered, because
  the focus is to show how an energy scale for \bquark jets can be 
  determined at all, and only one overall energy scale factor
  \bjes is determined for \bquark jets.
  Once this is achieved, it will
  be possible in principle to determine two independent
  energy scales for the different classes of reconstructed \bquark jets.

  The ability of the detector to distinguish quarks from gluons and to 
  identify the quark flavor is limited.
  Nevertheless, identification of \bquark-quarks (\bquark-tagging)
  can be useful to distinguish signal and background events, 
  or to identify the correct assignment of final-state quarks to 
  measured jets in final states like \ljets \ttbar events that contain
  both light and \bquark-quark jets.
  In this paper, we follow the approach introduced 
  in~\cite{bib-CDFljetsme} to include a
  term $W_b$ in the transfer function which describes the likelihood for
  parton $j$ with assumed
  flavor $\phi_{j}^{\,\gen}$ to be reconstructed with 
  $b$-tagging information ${\cal B}_{j}^{\,\rm rec}$.
  If $b$ tagging is used as a binary decision,
  then one simply has
  \begin{equation}
    \label{binarybtaggingweight.eqn}
      W_b \left( {\cal B}_{j}^{\,\rm rec},\ 
                 \phi_{j}^{\,\gen} \right)
    = 
      \left\{
      \begin{array}{r @{\ \ } l @{}}
            \epsilon_b \left( \phi_{j}^{\,\gen} \right)
        &
          {\rm if\ the\ jet\ }j\ {\rm is\ }b{\mhyphen}{\rm tagged\ and}
        \vspace{1.5ex}\\
          1-\epsilon_b \left( \phi_{j}^{\,\gen} \right)
        &
          {\rm otherwise,}
      \end{array}
      \right.
  \end{equation}
  where $\epsilon_b \left( \phi \right)$ is the \bquark-tagging
  efficiency for a jet from a parton of flavor $\phi$.

  Typically, \bquark-tagging algorithms yield a continuous output
  (for example, the decay length significance of a secondary vertex within
  a jet, or the output of an artificial neural network).
  Instead of a binary decision, the quantity $W_b$ can be parametrized
  as a function of this continuous value.
  Such an approach naturally makes optimal use of the information.
  As a compromise, it is possible to use several (rather than only two) bins
  in the output value.
  This is the concept used in this publication.
  Figure~\ref{btagging.fig} shows the values $W_b$ for jets in 
  \ljets \ttbar events as used in the study presented here, which
  corresponds to the \bquark-tagging performance of the \dzero 
  experiment~\cite{bib-dzerobtagging}.
  For the study in this paper, the $W_b$ functions are assumed not to 
  depend on the jet transverse momentum or pseudorapidity, but this
  will be a straightforward extension of the method for future 
  measurements.
  \begin{figure}
    \begin{center}
      \includegraphics[width=0.8\textwidth]{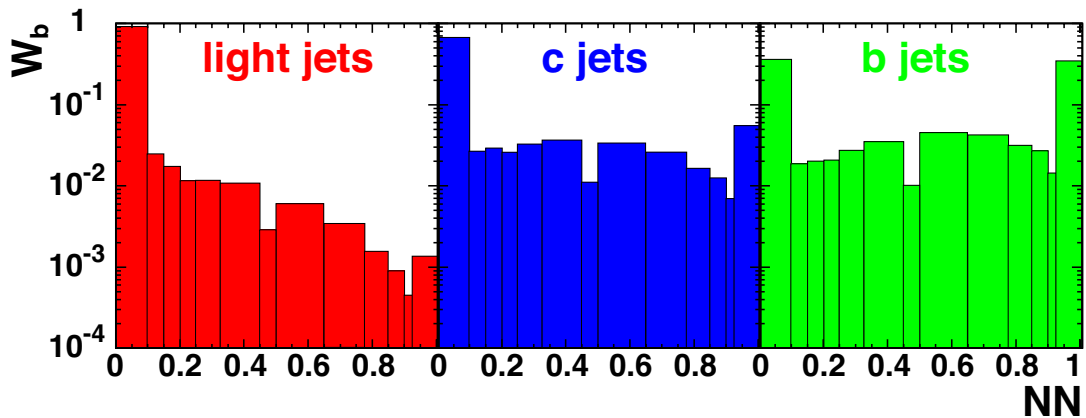}
      \caption{\captionfont\label{btagging.fig}The function $W_b$ used
        to parametrize the \bquark-tagging performance.
        The likelihood $W_b$ for a jet to be reconstructed as
        \bquark-tagged with given \bquark-tagging output ${\cal B}^{\rm rec}$ is
        shown for light, \cquark-quark, and \bquark-quark jets.
        In this paper, the output $NN$ of an artificial neural network used
        in the \dzero experiment~\cite{bib-dzerobtagging} is taken as a
        concrete example.
        The first bin contains jets that fail the \bquark-tagging preselection.
        The structure in the histogram is due to the non-equidistant binning.}
    \end{center}
  \end{figure}
\item{\bf Energetic neutrinos:}
  Neutrinos are not measured in the detector, but still an integration
  has to be performed over assumed values for all momentum components of all 
  final-state neutrinos in an event.
  Information on neutrino momenta
  can be partly inferred from mass constraints (e.g., 
  \mW or \mtop in \ttbar events).
  The additional assumption is made in this paper that events are balanced
  in the transverse plane, i.e.\ that the \ttbar system has zero transverse 
  momentum.
  This assumption is dropped in Section~\ref{systuncs.ttj.sec}, which
  means that an integration over two additional variables has 
  to be carried out.

  The presence of neutrinos in an event is typically inferred from 
  an imbalance in the transverse plane (non-zero missing transverse
  momentum \ptmissvec).
  It is not straightforward to parametrize the resolutions of the two
  \ptmissvec components since they depend on the resolutions of all other
  reconstructed objects in the event.
  Instead, the vector sum of transverse momenta of all reconstructed
  objects that are not assigned to the final state in question could be
  considered.
  In the case of \ttbar events at the Tevatron, this would be 
  calorimeter measurements outside of the jets assigned to the \ttbar
  final state.
  In this paper, however, as in~\cite{bib-me}, no corresponding 
  transfer function factor is introduced.
\end{list}

In addition to the detector resolution, 
one has to take into account the fact that the particles measured in the
detector cannot be assigned unambiguously
to specific final-state particles.
Consequently, all possibilities must be considered, and
their contributions to the transfer function summed.

The total transfer function 
can be written as
\begin{eqnarray}
  \label{tfdefinition.eqn}
    W(x,y;\,\bjes,\jes)
  & = \
    \displaystyle \sum_{i=1}^{n_{\rm comb}}
  &
    \prod_{e=1}^{n_e} \delta^{(3)} \! \left(   \vec{p}_e^{\,\rm rec}
                                             - \vec{p}_{e'}^{\,\gen} \right)
  \times
  \\
  \nonumber
  & &
    \prod_{m=1}^{n_\mu} \delta^{(2)} \! \left(   \vec{u}_\mu^{\,\rm rec}
                                               - \vec{u}_{\mu'}^{\,\gen} \right)
    W_{\mu}\left( \qoverptmurec,\qoverptmuprimegen \right)
  \times
  \\
  \nonumber
  & &
    \prod_{t=1}^{n_\tau} \delta^{(2)} \! \left(   \vec{u}_\ell^{\,\rm rec}
                                               - \vec{u}_{\tau'}^{\,\gen} \right)
    W_{\tau}\left( E_\ell^{\,\rm rec}/E_{\tau'}^{\,\gen} \right)
  \times
  \\
  \nonumber
  & &
    \prod_{j=1}^{n_j} 
    \delta^{(2)} \! \left(   \vec{u}_{j}^{\,\rm rec}
                           - \vec{u}_{j'}^{\,\gen} \right)
    W_{\rm jet} \left( E_{j}^{\,\rm rec},\ 
                       E_{j'}^{\,\gen},\
                       \phi_{j'}^{\,\gen};\ 
                       \phijes \right)
    W_b \left( {\cal B}_{j}^{\,\rm rec},\ 
               \phi_{j'}^{\,\gen} \right)
  \ ,
\end{eqnarray}
where the four lines represent the contributions from electrons,
muons, tau leptons, and jets,
respectively.
It is understood that a term only appears if the corresponding
particle is present in the final state under consideration.
The number of possible assignments of reconstructed (``rec'') particles
to final-state particles
in the process described by the matrix element (``\gen'') 
is denoted by $n_{\rm comb}$, and $i$ stands for one specific
permutation.
The symbols $n_e$, $n_\mu$, $n_\tau$, and $n_j$ stand for the 
numbers of electrons, muons, tau leptons, and quarks or gluons 
in the final state.
A reconstructed particle is denoted by
$e$, $m$, $t$, or $j$.
The symbols $e'$, $m'$, $t'$, and $j'$ stand for 
the corresponding final-state particle
assumed in the matrix element integration, which is given by the 
index $i$ of the permutation and the index of the reconstructed
particle: $e' = e'_{i,\, e}$ (and accordingly for muons, tau leptons, and 
jets).
The flavor $\phi_{j'}^{\,\gen}$ of 
final-state parton $j'$ assigned to jet $j$
is given by the permutation $i$.
The jet energy scale appropriate for jet $j$ (\bjes or \jes) is 
denoted by $S_\phi$ and selected according to the assumed flavour 
$\phi_{j'}^{\,\gen}$.
The symbol ${\cal B}_{j}^{\,\rm rec}$ stands for any output from a 
\bquark-tagging algorithm.

Because of the assumption that the transfer function factorizes into
independent contributions from the final-state particles, it may in
principle also depend on $\vec{\alpha}$.
For example, in a top quark mass measurement, smaller top quark masses
correspond to a smaller mean angular separation of jets,
which may lead to a broadening of the 
jet energy resolution.
Such effects are however typically small and are therefore neglected in the 
simplified description of the detector response with transfer 
functions; they would implicitly be taken into account in a calibration 
of the measurement with fully simulated events.

\subsection{Normalization of the Likelihood}
\label{method.normL.sec}
The normalization condition for the likelihood \pevtP for each process
is given by
\begin{equation}
    \label{normcond.eq}
    \int\limits_{x} \pevtP(x;\,\vec{\alpha},\vec{\beta})\
                    f_{\rm acc}(x)\
                    {\rm d}x
  =
    1
  \, ,
\end{equation}
where the inclusion of the factor $f_{\rm acc}(x)$ is equivalent to 
integrating only over those configurations $x$ of observed
events that pass the event selection criteria.
The normalization condition is fulfilled
according to the definition of the observable cross section 
\sigmaPobs in 
Equation~(\ref{sigmaobs.eqn}).
The calculation of \sigmaPobs is intimately
related with the normalization of the transfer function.
Both aspects are discussed in this section.

\subsubsection{Normalization of the Jet Energy Transfer Function}
\label{method.normL.tfnormjt.sec}
The normalization condition of the jet energy transfer function
used in previous implementations of the Matrix Element 
method~\cite{bib-me,bib-CDFljetsme} is given by
\begin{equation}
  \label{tfnormjt_alt.eqn}
     \int\limits_{E_{j}^{\,\rm rec}>0}
     W_{\rm jet}(E_{j}^{\,\rm rec},
                E_{j'}^{\,\gen},
                \phi_{j'}^{\,\gen};\ 
                \phijes) \,
     {\rm d}E_{j}^{\,\rm rec} 
  = 
    1
  \ .
\end{equation}
We call this a {\em process-based} normalization scheme as it
reflects the concept that a final-state quark or 
gluon gives rise to a jet of any energy (or is not reconstructed
as a jet if the energy is below the jet reconstruction 
threshold of the experiment); thus this normalization scheme does not depend on 
the event selection cuts.
A modified, {\em selection-based} 
normalization scheme which simplifies the computation of 
the normalization integral given in Equation~(\ref{sigmaobs.eqn}) but leaves
the likelihood \pevtP unchanged is introduced in this section.

We assume that the event selection requires a reconstructed object
for every charged lepton and every quark or gluon in the final state
(this means for example that the presence of four jets is required
for \ljets \ttbar events), and
that the jet selection cuts are identical for all jets.
The selection-based normalization of the transfer function is based on the 
concept that events only enter the analysis if all these reconstructed objects 
pass the corresponding selection criteria.
This means that every possible partonic final state in the integral in
Equation~(\ref{dsigmapp.eqn})
is assumed to yield an observed event that passed the event
selection.
Thus, a modified jet energy transfer function $W'_{\rm jet}$
is introduced
in the top quark mass measurement, which satisfies the condition
\begin{equation}
  \label{tfnormjt.eqn}
     \int\limits_{E_{j}^{\,\rm rec}>E_{\rm cut}(|\eta_{j}|)} 
     W'_{\rm jet}(E_{j}^{\,\rm rec},
                 E_{j'}^{\,\gen},
                 \phi_{j'}^{\,\gen};\ 
                 \phijes) \,
     {\rm d}E_{j}^{\,\rm rec} 
  = 
    1
  \ ,
\end{equation}
i.e.\ the parton under consideration is assumed to have led to a jet
that passed the selection cut $E_{j}^{\,\rm rec}>E_{\rm cut}$,
where the energy cut normally depends on the polar angle of the jet since a
transverse energy cut is used in the event selection.
Equation~(\ref{tfnormjt.eqn}) ensures that in Equation~(\ref{sigmaobs.eqn}), 
\begin{equation}
  \label{tfnorm_wprime.eqn}
  \int\limits_{x} W'(x,y;\,\vec{\beta}) f_{\rm acc}(x) {\rm d}x = 1
  \ .
\end{equation}
It is shown in Section~\ref{method.normL.sigmaobs.sec} that 
the modified denominator $\sigmaPobsprime$ which is then needed in 
Equation~(\ref{normpevtP.eqn}) to compute the likelihood \pevtP
becomes independent of the parameters $\vec{\beta}$ that describe the 
detector response.

\begin{figure}
\begin{center}
\includegraphics[width=0.48\textwidth]{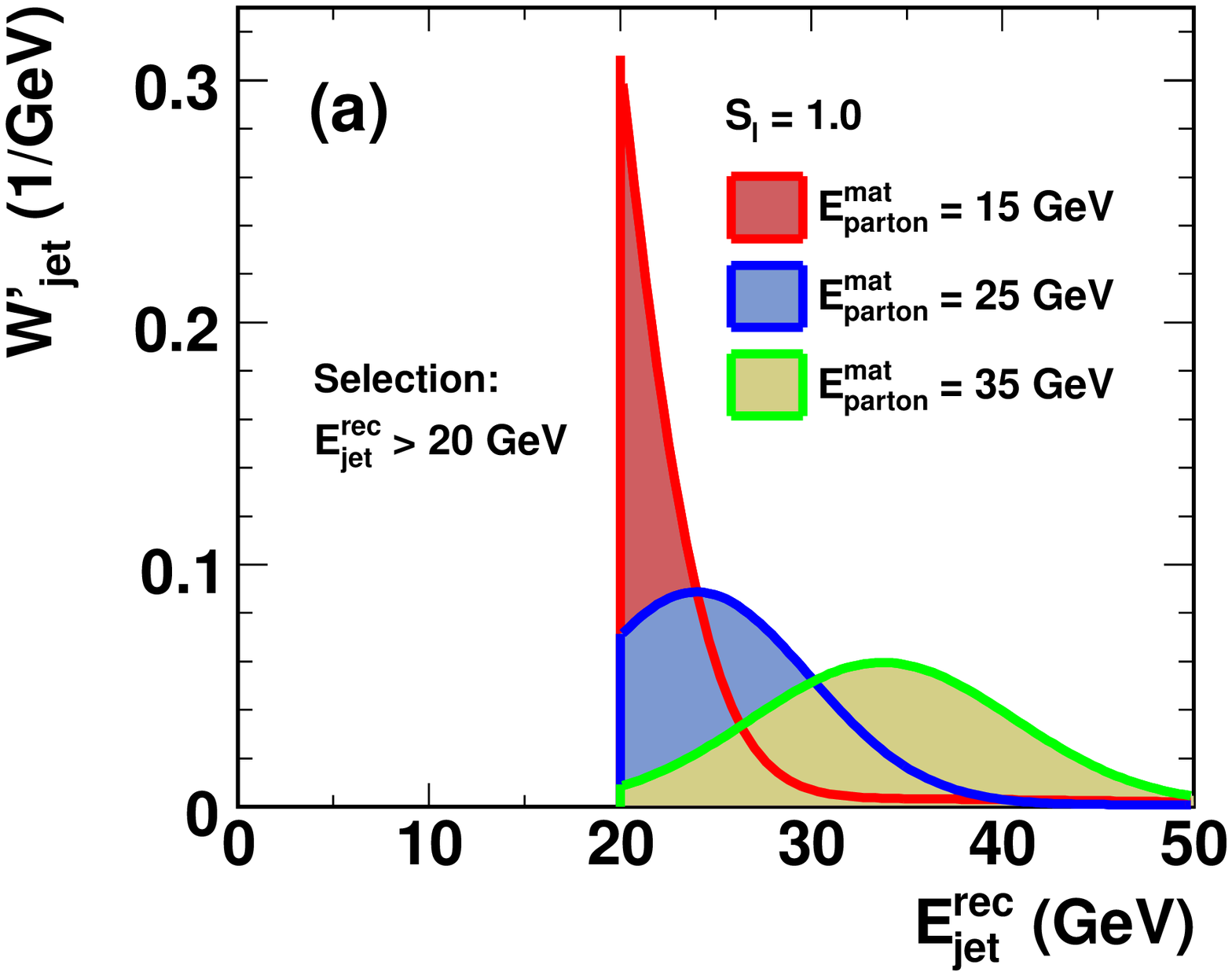}
\includegraphics[width=0.48\textwidth]{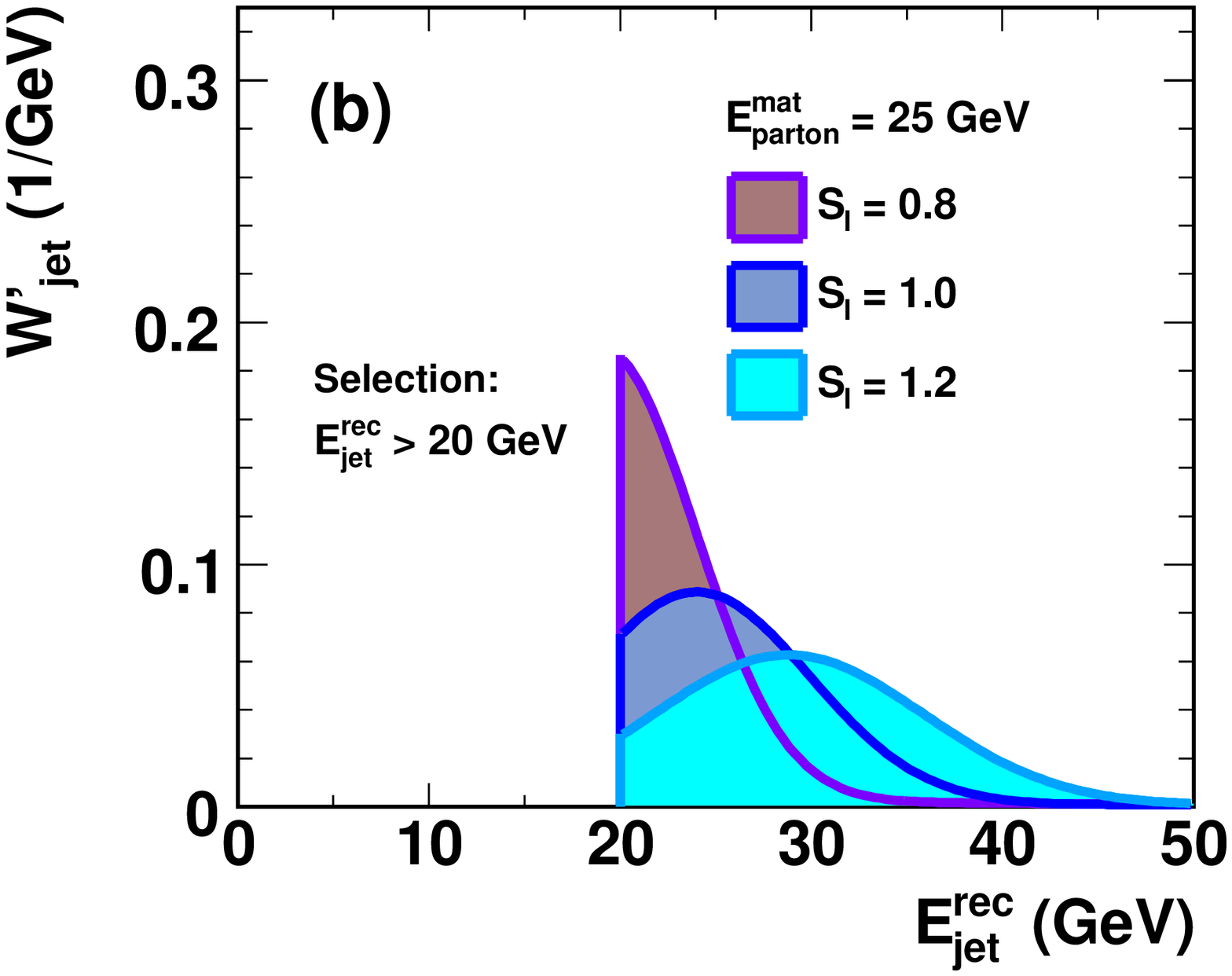}
\caption{\captionfont\label{TFnormalization.fig}
  Jet energy transfer function $W'_{\rm jet}$ in the modified
  normalization scheme.
  Plot (a) shows the transfer functions for light-quark jets at
  $\eta=0$ for an assumed value
  of $\jes=1.0$ and assumed parton
  energies of $15\,\GeV$ (red), $25\,\GeV$ (blue), and
  $35\,\GeV$ (green line).
  Plot (b) shows the same transfer function for different assumed
  \jes values (violet: $\jes=0.8$, blue: $\jes=1.0$, and
  cyan: $\jes=1.2$) and an assumed parton energy of 
  $25\,\GeV$.}
\end{center}
\end{figure}

The effect of this selection-based
normalization scheme on the jet energy transfer function
$W'_{\rm jet}$
is shown in Figure~\ref{TFnormalization.fig}(a) 
for the double-Gaussian function used
in this study:
If the parton energy is assumed to be very small, then a small reconstructed
jet energy just above the cut value is most likely.
However, the function is still normalized to unit area as it is assumed
that the parton must have given rise to a jet that passed the selection 
cut (in this example set at $E>20\,\GeV$ corresponding to $\eta=0$
for the event selection criteria of Section~\ref{samples.sec}).
The dependence of the jet energy transfer function on the parameters
$\vec{\beta}$ that describe the detector response must also be accounted
for as explained in Figure~\ref{TFnormalization.fig}(b):
For every \jes hypothesis the same event sample is considered in the 
measurement, and therefore the
event selection (in this example the minimum jet transverse energy cut)
cannot depend on the assumed \jes value.
For different assumed values of \jes, the $W'_{\rm jet}$ curve varies,
and the normalization of the curve must be adjusted to ensure that 
the normalization condition in Equation~(\ref{tfnormjt.eqn}) is satisfied.

\subsubsection{Normalization of the Muon and $\tau$ Transfer Functions}
\label{method.normL.tfnormmutau.sec}
It is assumed that every $\tau$ lepton decays to an electron
or muon that passed the event selection.
The $\tau$ energy has to be larger than the energy of the
reconstructed lepton.
Consequently, the selection-based normalization condition is
\begin{eqnarray}
\label{tfnormtau.eqn}
    \int\limits_{E^{\rm rec}_{\ell}>E_{\rm cut}(|\eta_{\ell}|)}^{E^{\rm rec}_{\ell}<E^{\gen}_{\tau'}}
    W'_\tau  \left( \frac{E^{\rm rec}_{\ell}}{E^{\gen}_{\tau'}}\right) \,
    {\rm d} \! \left( \frac{E^{\rm rec}_{\ell}}{E^{\gen}_{\tau'}}\right)  
  &
    =
  & 
    1 
  \, .
\end{eqnarray}
Because of the non-zero lower integration bound the transfer function $W_\tau$
has to be scaled with an appropriate overall factor that depends on 
the reconstructed lepton energy and pseudorapidity to arrive at the 
modified function $W'_\tau$.

In comparison with the jet energy resolution,
the muon transverse momentum resolution is good for muons close to the
minimum transverse momentum cut, and only a negligible fraction of muons
is affected by this cut.
In addition, it is assumed that final-state muons passing the selection 
cuts are always reconstructed as muons.
Thus, a calculation of the normalization corresponding to 
Equation~(\ref{tfnormtau.eqn}) can be omitted for muons.

\subsubsection{Observable Cross Section}
\label{method.normL.sigmaobs.sec}
To derive the denominator \sigmaPobsprime with which to 
normalize the likelihood \pevtP for a given process
$P$, it follows
from Equations~(\ref{normcond.eq}), 
(\ref{normpevtP.eqn}), and~(\ref{dsigmapp.eqn}) that
\begin{eqnarray}
\nonumber
    \frac{1}{\sigmaPobsprime}
    \int\limits_{x} 
    \int\limits_{y} \dsigmaP(\ppbar\to y;\,\vec{\alpha})\
                    W'(x,y;\,\vec{\beta})\
                    f_{\rm acc}(x)\
                    {\rm d}x
  & = &
    1
\\
\nonumber
  \Leftrightarrow 
    \frac{1}{\sigmaPobsprime}
    \int\limits_{y} \dsigmaP(\ppbar\to y;\,\vec{\alpha})\
    \int\limits_{x} W'(x,y;\,\vec{\beta})\
                    f_{\rm acc}(x)\
                    {\rm d}x
  & = &
    1
\\
\label{sigmaobs_independent_of_beta.eqn}
  \Leftrightarrow 
    \int\limits_{y} \dsigmaP(\ppbar\to y;\,\vec{\alpha})
  & = &
    \sigmaPobsprime
  \, ,
\end{eqnarray}
where the normalization condition for the modified transfer function $W'$
(Equation~(\ref{tfnorm_wprime.eqn}))
has been used in the last step.
The quantity \sigmaPobsprime is thus only a function of the physics
parameters $\vec{\alpha}$, but not of the detector performance
parameters $\vec{\beta}$.

In the definition of \sigmaPobsprime,
the integral over the observed events $x$ is not over the full phase space,
but instead only over that 
part of the phase space that passes the kinematic event selection.
Typically, regions of small
jet transverse energy \Et or large $|\eta|$ 
will be excluded from the integration region.

Because the normalization of the jet energy transfer function
$W'_{\rm jet}$ described in Section~\ref{method.normL.tfnormjt.sec}
accounts for the lower jet \Et cut, any 
jet energy scale dependence of \sigmaPobsprime is eliminated.
In contrast, the jet angular resolution is approximated with 
a $\delta$ function to save integration time, and this means that 
the integration over $y$ must exclude those angular regions
that do not pass the event selection.
Through the angular acceptance cuts (and through the matrix
element ${\cal M}_P$ itself, of course), \sigmaPobsprime still
depends on the physics parameters $\vec{\alpha}$.
A similar argument holds for angular acceptance cuts in the 
charged lepton selection.

The above argument is only valid if the normalization condition of
Equation~(\ref{tfnorm_wprime.eqn}) is fulfilled for the modified
transfer function $W'$.
In practice, this is difficult to implement for event selection cuts
based on quantities that depend on more than one reconstructed particle.
For example, the \ptmiss cut in Section~\ref{samples.sec} does not fulfill 
this criterion since it depends on all measured final-state particles.
Therefore, an \bjes dependence of \sigmaobsprime is taken into
account in the analysis of dilepton \ttbar events described in 
Section~\ref{enstestdilepton.sec}.
For the measurement with \ljets events, 
it is shown in Section~\ref{enstestljets.sec} that 
the \bjes and \jes dependence of \sigmaobsprime can be neglected
for the less stringent \ptmiss cut applied in the event selection.

\subsubsection{Process-Based Normalization Scheme}
\label{method.normL.altnormTF.sec}
It is possible to choose a process-based
transfer function normalization according to 
Equation~(\ref{tfnormjt_alt.eqn}).
In this case,
the dependency of the transfer function normalization 
on the parameters $\vec{\beta}$ describing the detector resolution
is not taken into account.
This means that in the last step of the derivation in 
Equation~(\ref{sigmaobs_independent_of_beta.eqn}), a dependency on 
$\vec{\beta}$ remains and \sigmaPobs
has to vary as a function of both $\vec{\alpha}$ and $\vec{\beta}$.

This process-based
scheme has the advantage that the transfer function $W$ accomodates 
the possibility of jets not passing the selection cuts.
For analyses like \mtop measurements in \ttbar events 
as described in this paper, the number of jets required in
the event selection ensures that in any event where this is the case 
an additional hard parton would have to be produced which yields a
jet passing the cuts.
In principle, such events have to be described by the signal 
process.
At the Tevatron their contribution to the event
sample is so small that they do not have to be accounted for
explicitly in the method until the final calibration step.
Consequently, the selection-based transfer function normalization described in 
Sections~\ref{method.normL.tfnormjt.sec}
and~\ref{method.normL.tfnormmutau.sec} is chosen for the studies 
described in this paper, as it
eliminates the dependency of \sigmaobsprime 
on the parameters $\vec{\beta}$ and
thus 
facilitates the simultaneous measurement of several
parameters.
This picture may change for measurements in \ttbar events 
at the LHC, where initial-state radiation becomes much more relevant.

\subsubsection{Normalization of the Background Likelihood}
\label{method.normL.bkg.sec}
The normalization of the likelihoods 
can in principle be
determined in the same way for all processes considered.
Alternatively, 
if the normalization of the likelihood \pevtPzero for one
specific process \Pzero (for example, the signal process)
has been determined as described above
and if the fraction \fPzero of events from that process in the selected
sample is left as a free parameter in the analysis,
it is possible to relate the absolute normalization 
of the likelihoods for all other processes to that of process \Pzero.
One can then make use of the fact that the fit described
in Section~\ref{method.mefit.sec} will yield a signal fraction 
\fPzero of the sample that is too small if the background
likelihood is too large and vice versa, and one can adjust
the relative normalization in the validation of the Matrix Element
method until the signal
fraction is determined correctly.
This concept can only be applied if the cross-section for the process
\Pzero is well-known (like for example for \ttbar production).
It is then helpful in particular 
if the likelihoods for background processes 
as implemented in the analysis do not depend on any of the parameters 
$\vec{\alpha}$ and $\vec{\beta}$: 
In such a case, only one normalization constant needs to be determined
for each background process.

\subsection{Fitting Procedure}
\label{method.mefit.sec}
For a given sample of selected events,
the parameters to be measured are determined as those values that maximize the
likelihood \lsample.
One wants to determine
$n_\alpha$ physics parameters, $n_\beta$ parameters describing the 
detector resolution, and \nfP fractions of events from different
processes $P$.
For every measured event, the likelihoods for each process
are calculated for an $(n_\alpha+n_\beta)$-dimensional grid of assumed parameter
values.
Given these grids of likelihood values for each process, the 
sample likelihood $\lsample(x_1,..,x_N;\,\vec{\alpha},\vec{\beta},\vec{f})$
defined in Equation~(\ref{lsample.eqn}) is available for an
$(n_\alpha+n_\beta+\nfP)$-dimensional grid of assumed parameter values.

The measurement value of a given parameter $a$ and the corresponding
uncertainty are then determined from a
one-dimensional likelihood $\lsampleoned(a)$.
The value of $\lsampleoned(a)$ is obtained by marginalization of
all other parameters; in practice, this is done by keeping the value
of $a$ constant, 
varying the assumed values of all $(n_\alpha+n_\beta+\nfP-1)$ other 
parameters, and taking the maximum \lsample value.
The one-dimensional function $-\ln\lsampleoned(a)$ is fitted with a parabola.
The parameter value that minimizes the parabola is taken to be 
the measurement value, and the measurement uncertainty is given
by the parameter values where the fitted parabola rises by $+\frac{1}{2}$
above the minimum.
By construction, this procedure takes correlations between the 
parameters into account.

\subsection{Validation With Ensemble Tests}
\label{method.enstest.sec}
To validate the measurement technique, tests are performed with
simulated events generated under the assumptions used in the 
Matrix Element method, 
i.~e.\ using the same PDF set, matrix element, and transfer function.
A pseudo-experiment emulates a measurement performed on data and
consists of events 
randomly drawn from \MC event pools for signal and background processes. 
The numbers of events taken from the different pools are chosen to 
reflect the fractions observed in the data.
An ensemble of several pseudo-experiments is performed for each of 
a number of sets of assumed input parameter values.
The range of assumed values is chosen according to previous determinations
and the expected precision of the measurement.

Taking the results from all ensembles, the
following information is obtained:
\begin{list}{$\bullet$}{\setlength{\itemsep}{0.5ex}
                        \setlength{\parsep}{0ex}
                        \setlength{\topsep}{0ex}}
\item
The relation between the expected (mean) measurement values
and the corresponding true input values.
It is expected that the method yields unbiased results if the 
Matrix Element method reflects the properties of the events.
\item
The distribution of measurement uncertainties as a function of 
input parameter values.
\item
The width $w$ of the pull distribution.
To test that the fitted uncertainties describe the actual measurement
uncertainty, the deviation of the measurement value
from the true value is divided by the fitted measurement uncertainty
in each pseudo-experiment.
The width of this distribution of deviations normalized by the measurement
uncertainties is referred to as {\it pull width}.
It is expected that $w=1$ if all features of the events are
accommodated in the method.
\end{list}
In a similar way, ensemble tests based on fully simulated events can be 
used to determine any correction of measurement values and fit uncertainties
needed when applying the method (which is based on a simplified detector 
model) to real data.

Because the computation of likelihoods is time-consuming, the size
of the pools of simulated events is usually limited and individual 
events are allowed to be redrawn, i.e.~to appear several times even
in the same pseudo-experiment.
This technique maximizes the information about the expected uncertainties
and pull widths, but it has to be taken into account when evaluating the 
uncertainties of the ensemble test results~\cite{bib-resampling}.

To summarize, the validation tests described in Sections~\ref{enstestljets.sec}
and~\ref{enstestdilepton.sec} each comprise the three following steps:
\begin{enumerate}
\item Likelihood Fit:
build one pseudo-experiment and determine
$\vec{\alpha}$, $\vec{\beta}$ and $\vec{f}$ 
(cf.~Section~\ref{method.mefit.sec});
\item Ensemble Test:
repeat Step 1 with 1000 pseudo-experiments and
obtain mean results, expected uncertainties, and pull widths; 
and
\item Validation:
repeat Step 2 for several input parameter values to
obtain calibration curves.
\end{enumerate}

\section{Simulation and Selection of \ttbar Events}
\label{samples.sec}
As an example for a concrete implementation of the Matrix Element
method described previously, the measurement of the top quark 
mass in \ljets and dilepton \ttbar events is described in this 
and the following sections.
The discussion of dilepton events is restricted to events
containing one electronic and one muonic \W decay, which yield
the most precise top quark mass measurement in dilepton events.
This section summarizes the generation of smeared events
to study the top quark mass measurement and introduces
the event selection criteria.

The characteristics of \ljets \ttbar events at a hadron collider are
the presence of one energetic isolated charged lepton,
at least four energetic jets (two of which are \bquark-quark jets), 
and missing transverse momentum due to the unreconstructed neutrino.
The main background is from events where a leptonically decaying \W is 
produced in association with four or more jets.  
Multijet background where one
jet mimicks an isolated charged lepton 
can also enter the event sample.

Dilepton \ttbar events are characterized by two
oppositely charged energetic isolated leptons (in the case considered here,
one electron and one muon), two energetic \bquark-quark
jets, and missing transverse momentum due to the two neutrinos from the 
\W decays.  
The largest physics background in the \emu channel 
is from events with $\ztautau$ decays where the \Z boson is produced in 
association with two or more jets; another background channel is the 
production of two leptonically decaying \W bosons together with two
jets.
Instrumental background arises from 
events where a leptonically decaying \W is 
produced in association with three or more jets, one of which is 
misidentified as the second isolated charged lepton.

In principle, events with leptonically decaying $\tau$ leptons 
from \W decay contribute to both the \ljets and dilepton event samples.
However, because of lower transverse momentum or transverse energy
cuts on the charged lepton(s) in the event selection (see below), these
contributions are typically small.
Thus, \ttbar events including leptonic $\tau$ decays are not simulated
for the study presented here (whereas in a real measurement, the effect
of such decays has to be accounted for).

For the study presented here, events containing a $\qqbar\to\ttbar$
reaction in a \ppbar collision at $1.96\,\TeV$ center-of-mass energy
are simulated with the \madgraph~\cite{bib-madgraph} generator.
Events are generated for each of the
different top quark masses, varied between
$160\,\GeV$ and $180\,\GeV$ in steps of $5\,\GeV$. 
The \alpgen~\cite{bib-alpgen} 
program is used to generate events containing a leptonic
\W or \Z decay in association with additional light partons;
events with \bquark quarks are simulated by smearing the light partons with
the transfer function for \bquark jets.  
To simulate the decay of a $\tau$ lepton to an electron or muon in 
\ztautaujj events, the $\tau$ transfer function shown in
\Fref{fig:tautransferfunction} is applied, while the direction of the
lepton is left unchanged.   
For the modeling of the parton
distribution functions, the leading-order PDF CTEQ5L~\cite{bib-CTEQ5L} 
is chosen.
Multijet background without leptonic \W or \Z decay is not simulated
because it was shown in~\cite{bib-me} that its effect on the measurement 
in the \ljets channel
is similar to that of additional \Wjets background\footnote{In 
  a real measurement, it is thus possible to model both
  \Wjets and multijet background by the \Wjets process to calculate the 
  likelihood \lsample, and to account for any differences between 
  \Wjets and multijet background when calibrating the measurement using
  full simulation.}.
All simulated events are passed through the parametrized detector simulation
discussed in Section~\ref{method.tf.sec}, which describes the response
of the \dzero detector~\cite{bib-me}.
No simulation of the parton shower and 
hadronization is performed since the transfer functions account for their
effects in addition to the detector resolution.
Samples with different true values of \jes are obtained by scaling the smeared
light-quark jet energies; values of \jes between $0.9$ and $1.1$ in steps of 
$0.05$ are used.
Similarly, \bquark-quark jets are scaled by \bjes, with \bjes varied between
$0.8$ and $1.2$ in steps of $0.1$, to obtain samples for different true
\bquark-jet energy scales. As the \bjes constraint is weaker than the \sjes one a wider range of generated values was studied for this observable. 
The association of final-state partons to jets is not assumed to be
known in the subsequent analysis.
The reconstructed missing transverse momentum \ptmissvec is taken to be
the negative vector sum of all other reconstructed transverse momenta
(i.e., after the smearing and scaling described above); this means that 
before smearing and scaling the \ttbar system has zero \pt.

Typical event selection criteria as used by the Tevatron experiments
are then applied to the smeared events.
Candidate \ljets events are required to contain
\begin{list}{$\bullet$}{\setlength{\itemsep}{0.5ex}
                        \setlength{\parsep}{0ex}
                        \setlength{\topsep}{0ex}}
\item
one charged lepton within a pseudorapidity range of $|\eta|<1.1$ (electrons) 
or $|\eta|<2.0$ (muons) and with a transverse energy or momentum of
at least $20\,\GeV$,
\item
four jets within $|\eta|<2.5$ and with (scaled)
transverse energies of $\et>20\,\GeV$, and
\item
missing transverse momentum with magnitude 
$\ptmiss\equiv|\ptmissvec|>20\,\GeV$.
\end{list}
The angular separation between the charged lepton and any jet is required to be
$\DeltaR\equiv\sqrt{(\Delta\eta)^2 + (\Delta\phi)^2} > 0.5$, and 
similarly, any jet-jet pair has to be separated by
$\DeltaR > 1.0$.
No \bquark-tagging requirements for the jets are included,
but \bquark-tagging information is used later in the analysis.

Similarly, dilepton events must contain
\begin{list}{$\bullet$}{\setlength{\itemsep}{0.5ex}
                        \setlength{\parsep}{0ex}
                        \setlength{\topsep}{0ex}}
\item
one electron and one muon of opposite charges
within pseudorapidity ranges of $|\eta|<1.1$ or $1.5<|\eta|<2.5$ 
(electrons)\footnote{This cut rejects electrons in the transition 
  region between the barrel 
  and endcap parts of the electromagnetic calorimeter, which has poor
  electron identification performance and is typically located at around 
  $1.1 < |\eta| < 1.5$.}
or $|\eta|<2.0$ (muons) and with a transverse energy or momentum of
at least $15\,\GeV$,
\item
two jets within $|\eta|<2.5$ and with (scaled) transverse energies of 
$\et>20\,\GeV$, and
\item
missing transverse momentum with magnitude 
$\ptmiss>30\,\GeV$.
\end{list}
The same \DeltaR separation cuts as above are applied, and in addition the 
two charged leptons are required to be separated by $\DeltaR>0.5$.

The event samples described here are
used for validating the measurement technique as discussed in 
Sections~\ref{enstestljets.sec} and~\ref{enstestdilepton.sec}.
While the exact event selection criteria are not critical to the 
method, it is mandatory to adjust the likelihood calculation 
accordingly.

\section{Likelihood Implementation for Measurements in \ttbar Events}
\label{application.sec}
This section describes the calculation of the signal and 
background likelihoods for \ljets and dilepton \ttbar events.
When the likelihood for a certain process has to be evaluated for 
many hypotheses, a dedicated implementation of the matrix element
optimized for speed is beneficial, and it is helpful to limit the 
number of evaluations of the transfer function.
Section~\ref{application.psgn.sec} discusses the evaluation of the 
signal \ttbar likelihoods for a top quark mass measurement (in the 
\ljets or dilepton channel) as an example for such a case.
In contrast, when the number of hypotheses is smaller and/or there
are many individual diagrams contributing to a process, interfacing
to routines from existing Monte Carlo generators is a powerful solution.
Such a case is the evaluation of the background likelihoods
for an \mtop measurement, which is described in 
Section~\ref{application.pbkg.sec}.
An overview of the event likelihood calculation for the different decay
channels and processes in a top quark mass measurement is given in
Table~\ref{pevt.table}.
\begin{table}[htbp]
\begin{center}
\begin{tabular}{cccc}
\hline
    channel 
  & \multicolumn{1}{c}{processes}
  & \multicolumn{1}{c}{likelihoods}
  & \multicolumn{1}{c}{parameters}
\\
\hline
    \ljets
  & \begin{tabular}{@{}c@{}}
      $\qqbar\to\ttbar$ \\
      \wjjjj
    \end{tabular}
  & \begin{tabular}{@{}c@{}}
      \pevtttbar \\
      \pevtwjjjj
    \end{tabular}
  & \begin{tabular}{@{}c@{}}
      \mtop, \bjes, \jes \\
      --
    \end{tabular}
\\
\hline
    \begin{tabular}{@{}c@{}}
      dilepton \\
      (\emu channel)
    \end{tabular}
  & \begin{tabular}{@{}c@{}}
      $\qqbar\to\ttbar$ \\
      \ztautaujj \\
    \end{tabular}
  & \begin{tabular}{@{}c@{}}
      \pevtttbar \\
      \pevtztautaujj \\
    \end{tabular}
  & \begin{tabular}{@{}c@{}}
      \mtop, \bjes \\
      -- \\
    \end{tabular}
\\
\hline
\end{tabular}
\caption{\captionfont\label{pevt.table}
Overview of the \pevt calculation
in the \ljets and dilepton channels.
The column entitled ``processes'' lists the 
signal and background processes taken into account in the calculation
of the event likelihood \pevt.
The symbol ``$j$'' refers to any
light parton, i.e.\ a \uquark, \dquark, \squark, or \cquark quark (or
antiquark) or a gluon.
The rightmost column shows the parameters on which the likelihoods \pevtP for 
each individual process depend.
In principle the background likelihoods depend on \jes, but as shown
later in the paper it is possible
to omit this dependence without introducing a significant bias
on the \mt measurement.}
\end{center}
\end{table}

\subsection{The Signal Likelihood}
\label{application.psgn.sec}
For the calculation of the signal likelihood, the procedure described
in~\cite{bib-me,bib-schiefer} has been extended and optimized.
It is based on the leading-order matrix element for the process
$\qqbar\to\ttbar$~\cite{bib-mahlonparke}.
Aspects that are unchanged from~\cite{bib-me,bib-schiefer} are only briefly
mentioned in the following.
The matrix element for the process $\glueglue\to\ttbar$ 
is not evaluated explicitly because the
top and \W propagator and decay parts of the matrix element, which contain most
of the information on the top quark mass and the separation of signal
and background events, are identical.

The correct association of reconstructed jets with the final-state
quarks is not known.
Therefore, the transfer function takes into account all possible jet-parton
assignments as described in Section~\ref{method.tf.sec}.
For a given measured event $x$, the 
convolution integral in Equation~(\ref{dsigmapp.eqn}) is calculated 
separately for each jet-parton assignment and for all different \mtop 
assumptions, while all different \bjes and \jes hypotheses are considered
simultaneously.
The integral evaluation is
performed numerically with the Monte Carlo program
\vegas~\cite{Lepage:1977sw,Lepage:1980dq}, which
has been slightly extended to achieve the simultaneous 
evaluation of several integrals with the 
same distribution of parton configurations $y$.
A single call to the routine calculating the integrand returns
an array of values for 
all assumed \bjes and \jes values under consideration. 
This diminishes the total computation time spent for a given number
of calls to evaluate the integrand, because the matrix element does not 
have to be re-evaluated when only the \bjes or \jes assumptions change.
Even more importantly, fluctuations between the
results obtained for different \bjes and \jes assumptions are reduced
because the integrand is evaluated for the same parton configurations $y$.

While the expected measurement uncertainty on \bjes and \jes is 
small relative to the resolution of jet energy measurements,
the current uncertainty on the world average \mtop value is of the same
order as the top quark width.
If the range of \mtop hypotheses to be tested in a measurement spans
a range of several times the top quark width, then the distribution of 
parton configurations $y$
at which the integrand is evaluated 
for 
one \mtop value is inappropriate for other values, and the technique
becomes inefficient.
Thus in the study presented here, the likelihoods for different \mtop
assumptions are evaluated independently.

To evaluate the signal likelihood for an event $x$ and all assumed values
of the quantities \mt, \bjes, and \jes to be measured, the following
computations are performed:
\begin{list}{$\bullet$}{\setlength{\itemsep}{0.5ex}
                        \setlength{\parsep}{0ex}
                        \setlength{\topsep}{0ex}}
\item
Loop over all top quark mass assumptions,
\item
loop over all jet-parton assignments, and
\item
use the program \vegas to compute the convolution integral 
in Equation~(\ref{dsigmapp.eqn}) for all \bjes and \jes hypotheses.
\end{list}

The integration in Equation~(\ref{dsigmapp.eqn}) is over the kinematic 
variables of the assumed parton configuration, as described in 
Section~\ref{method.pprc.sec}.
The number of dimensions is reduced by assuming perfect measurement of
some of the quantities. 
Via variable transformation the remaining integration variables
have been chosen such that where possible, they are uncorrelated,
the integrand exhibits sharp
peaks as a function of each individual variable
(this optimizes the performance of the \vegas program), and
the variable transformation involves at most a quadratic
equation (so the transformation is fast and numerically stable).
The integration variables chosen in the \ljets and dilepton channels
are summarized in Table~\ref{integrationvariables.table}.
The first two rows list variables corresponding to invariant masses
and to jet momenta, respectively.
Other variables are listed in the third row, and the final row indicates
the integration necessary because of the finite muon momentum resolution.
\begin{table}[htbp]
\begin{center}
\begin{tabular}{c@{\hspace{4ex}}c}
\hline
    \ljets channel
  & dilepton channel
\\
\hline
    \mthadsquare, \mtlepsquare, \mwhadsquare 
  & \mtonesquare, \mttwosquare  
\\
    \pqonemag                                 
  & \pbonemag, \pbtwomag
\\
    \pzblepnu                                
  & \deltapxnuonetwo, \deltapynuonetwo 
\\
    \qptmu
  & \qptmu  
\\
\hline
\end{tabular}
\caption{\captionfont\label{integrationvariables.table}Overview of 
the integration variables for the signal likelihood calculation
in the \ljets and dilepton channels.
In the \ljets channel, the integration is over the masses of the two top quarks 
and the hadronically decaying \W boson, the momentum of the 
up-type quark from the hadronically decaying \W, and the sum 
of the longitudinal momenta of the \bquark-quark
from the top quark with the leptonic \W decay and the neutrino.
In the dilepton channel, the top quark masses, \bquark-quark momenta,
and the $x$ and $y$ components of the vectorial
difference of the two neutrino momenta are taken as integration variables.
The ratio of muon charge and transverse momentum is a
further integration variable where applicable.}
\end{center}
\end{table}

The steps to evaluate the integrand for given values of the integration 
variables are:
\begin{list}{$\bullet$}{\setlength{\itemsep}{0.5ex}
                        \setlength{\parsep}{0ex}
                        \setlength{\topsep}{0ex}}
\item
Determine the momenta of all final-state particles from the values of the
integration variables.
\item
Evaluate the Jacobian determinant \detJ for the variable transformation.
\item
Calculate the value 
$|{\cal M}|^2 \fpdf(\qone) \fpdf(\qtwo)$
of the matrix element squared times PDF factors,
summing over all possible initial-state parton species.
\item
Then loop over all final-state particles,
\item
for each particle, loop over all relevant \bjes or \jes hypotheses,
if applicable, and
\item
evaluate the transfer function factor corresponding to that particle.
\item
Return the product 
\begin{eqnarray}
  \nonumber
  & &
    \dsigmaP(\ppbar\to y)\ W(x,y;\,\bjes,\jes)\ \detJ
\\
  \nonumber
  & = &
    \phantom{\times\ }
    \sum_{a_1, a_2}
    {\rm d}\qone {\rm d}\qtwo\ 
    f_{\rm PDF}^{a_1} (\qone)\ 
    \bar{f}_{\rm PDF}^{a_2} (\qtwo)\ 
    \frac{(2 \pi)^{4}\! \left|\mathscr{M}_P\left(a_1 a_2 \to y\right)\right|^{2}}
         {\qone \qtwo s}
    {\rm d}\Phi_{n_f}\
\\
  &   &
    \times\
    W(x,y;\,\bjes,\jes)\ \detJ
\end{eqnarray}
for all \bjes and \jes hypotheses.
\end{list}

The convolution integral in Equation~(\ref{dsigmapp.eqn}) has to be 
calculated for every selected event and is thus the most computing 
intensive part of the analysis.
The optimization introduced here allows the integration necessary
for the determination of three parameters in the \ljets channel
to be performed within roughly the time needed in~\cite{bib-me}
for just two parameters.

The normalization \sigmattbarobsprime
only has to be determined
once for a number of hypotheses relevant to the analysis.
This is done in a separate program based on the \vegas package 
that performs a 16-dimensional
Monte Carlo integration over the observable final-state phase space.
The phase space is generated recursively from the production of the \ttbar
pair and the subsequent two-body top and \W decays.

\subsection{The Background Likelihood}
\label{application.pbkg.sec}
There are in general many background processes that can lead to an
observed event.
It is not problematic per se to not fully account for all backgrounds 
in the event likelihood.
An incomplete background likelihood will lead to a shift of the 
measured top quark mass value (apart from an increased statistical 
uncertainty); the shift will in general depend 
on the top quark mass itself and on the fraction of events in the 
sample that are not accounted for in the overall likelihood.
The shift is determined in the calibration procedure.
When a background term is omitted in the event likelihood, the 
situation will thus be quantitatively, but not qualitatively
different from that in an analysis that includes this term in the 
likelihood.

If several different background
processes have similar kinematic characteristics, it is also
possible to approximately describe the total background by
the likelihood for only one of the background processes, multiplied
by the total background fraction, cf.\ Equation~(\ref{pevt.eqn}).
This technique has been applied by both CDF and
\dzero in the Matrix Element analyses in the \ljets channel, where a
likelihood for QCD multijet production is not explicitly calculated.

Only leading-order background processes to \ttbar events
and only the most important ones among them are considered explicitly
in this paper.
To take into account all individual diagrams, 
routines from existing Monte Carlo generators are used to 
compute the likelihood for generic processes.
They take into account 
the relative importance of the various subprocesses that contribute
and perform a statistical sampling of all possible spin, flavor, and color
configurations.
Because the background likelihood does not depend on the top quark
mass, it does not have to be computed for as many different
assumptions as the signal likelihood and it is possible to evaluate
the matrix element without a dedicated routine optimized for speed.

The generic background process taken into account in the \ljets channel is
the production of a leptonically decaying \W boson in association with
four additional light partons, \wjjjj.
Events with a leptonically decaying \W boson
and four partons that include heavy-flavor
quarks are not considered separately because their kinematic
characteristics are very similar to those of \wjjjj events.

The \vecbos~\cite{bib-vecbos}
generator is used to calculate the background likelihood \pevtwjjjj.
The jet
directions and the charged lepton are taken as well-measured.
The integral in Equation~(\ref{dsigmapp.eqn}) is
performed by generating Monte Carlo events with quark energies
distributed according to the jet transfer function.
In these Monte Carlo events, the neutrino transverse momentum is 
given by the condition that the transverse momentum of the \wjets
system be zero, while the invariant mass of the charged lepton and
neutrino is assumed to be equal to the \W mass to obtain the 
neutrino $z$ momentum (both solutions are considered).
All 24 possible assignments of jets to quarks in the 
matrix element are considered and their contributions to the likelihood
summed.
Monte Carlo events are drawn according to the appropriate
jet resolution functions
for the four reconstructed jets, the likelihood \pevtwjjjj is computed for
each of these events, and their mean value is used in the subsequent analysis.
The study described in
Section~\ref{enstestljets.sec} supports that it is
not necessary to compute the likelihood
\pevtwjjjj for different \jes values; only the
value $\pevtwjjjj\left(\jes=1\right)$ is used.

The measurement in the dilepton (\emu) channel considers
background from events containing a \Z boson decaying via $\tau$ leptons to an
electron and a muon (plus neutrinos) and two additional light partons
explicitly in the event likelihood.
The likelihood is calculated using the \vecbos generator as above, including
the transfer function for leptonic $\tau$ decays described in 
Section~\ref{method.tf.sec}.
Again, the jet
directions and the charged lepton are taken as well-measured,
and the integral in Equation~(\ref{dsigmapp.eqn}) is
performed by generating Monte Carlo events with quark energies
distributed according to the jet transfer function.
The energies of the two $\tau$ leptons are then given by the condition 
that the transverse momentum of the \zjets
system be zero.
Both possible assignments of jets to quarks are considered, and as
above, only the value $\pevtztautaujj\left(\jes=1\right)$ is used.

\section{Application of the Technique to \ttbar Events in the Lepton+Jets Channel}
\label{enstestljets.sec}
The method is validated using smeared parton-level simulated \ttbar
and \wjjjj events, generated as described in Section~\ref{samples.sec}.
In this study, pool sizes of 1500 events for the \ljets \ttbar signal  
process and 850 \wjjjj background events are available. 
In order to model processes not covered by the method, two additional  
samples are generated. 
Out of the \wjjjj background sample 450  
events are modified into \wbbjj events by randomly assigning  
two light partons as \bpartons and smearing them according to the  
\bquark quark transfer functions. 
With this sample, effects of heavy  
flavor content in the background can be studied. 
The other test sample  
is composed of 800 \ljets \ttbar events that contain an  
additional parton from initial- or final-state radiation.

As the \ejets and \mujets decay channels only differ in the momentum
resolution of the lepton and whether a transfer function is used to
parametrize it, no qualitative difference between measurements in
the two channels is expected. 
This was verified in~\cite{bib-PHdiss}. 
Thus, in the following only the \ejets decay  
channel will be considered.
The different angular acceptance cuts for electrons and 
muons lead to different signal fractions for the two channels, but
the conclusions from the studies described here are still valid since
they have been performed for a wide range of signal fractions.

In Section~\ref{enstestljets.sig.sec} the method is tested on ensembles  
containing signal events only. 
Section~\ref{enstestljets.bkg.sec}  
describes studies performed on samples including background events.

\subsection{Signal-Only Studies}
\label{enstestljets.sig.sec}
The method is first tested with ensembles only  
containing signal events. 
For these studies, 1000 pseudo-experiments  
are composed of 100 \ejets events each, and background likelihoods are  
not included. 
The reconstructed fit  
observables should resemble the generated input values within  
statistical uncertainties and the pull widths are expected to be equal  
to unity (within uncertainties).

\begin{figure}[tbp]
\begin{center}
\includegraphics[width=0.5\textwidth]{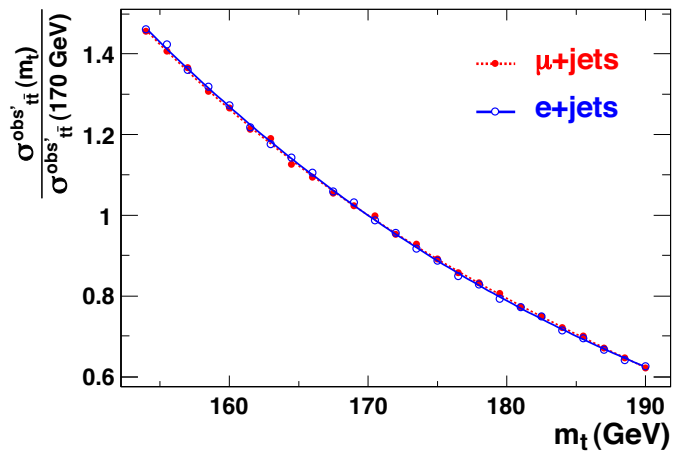}
\caption{\Ljets channel: Normalization function \sigmattbarobsprime
  for the \ttbar likelihood in the \ljets channel
  as a function of the top quark mass for 
  the \ejets (blue) and \mujets channels (red circles).
  The normalization functions are plotted relative to the fitted values
  for $\mt=170\,\GeV$.  An overall scale
  factor is irrelevant for the subsequent analysis because it is
  absorbed by the normalization procedure for the background likelihood.}
\label{fig:ljets_nrm}
\end{center}
\end{figure}

The likelihood normalization \sigmattbarobsprime as a function of
the top quark mass as defined in  
Section~\ref{method.normL.sec} is given in Figure~\ref{fig:ljets_nrm} for  
the \ejets and \mujets channels
(only the \ejets function is further used).
These functions are fitted with \thirdorder
polynomials as a function of the top quark mass.
The two channels yield different  
normalization functions as the detector acceptance differs for the two  
lepton types.

\begin{figure}[tbp]
\begin{center}
\begin{tabular}{@{\hspace{-10pt}}c@{\hspace{-17pt}}c@{\hspace{-17pt}}c@{}}
          \includegraphics[width=0.37\textwidth]{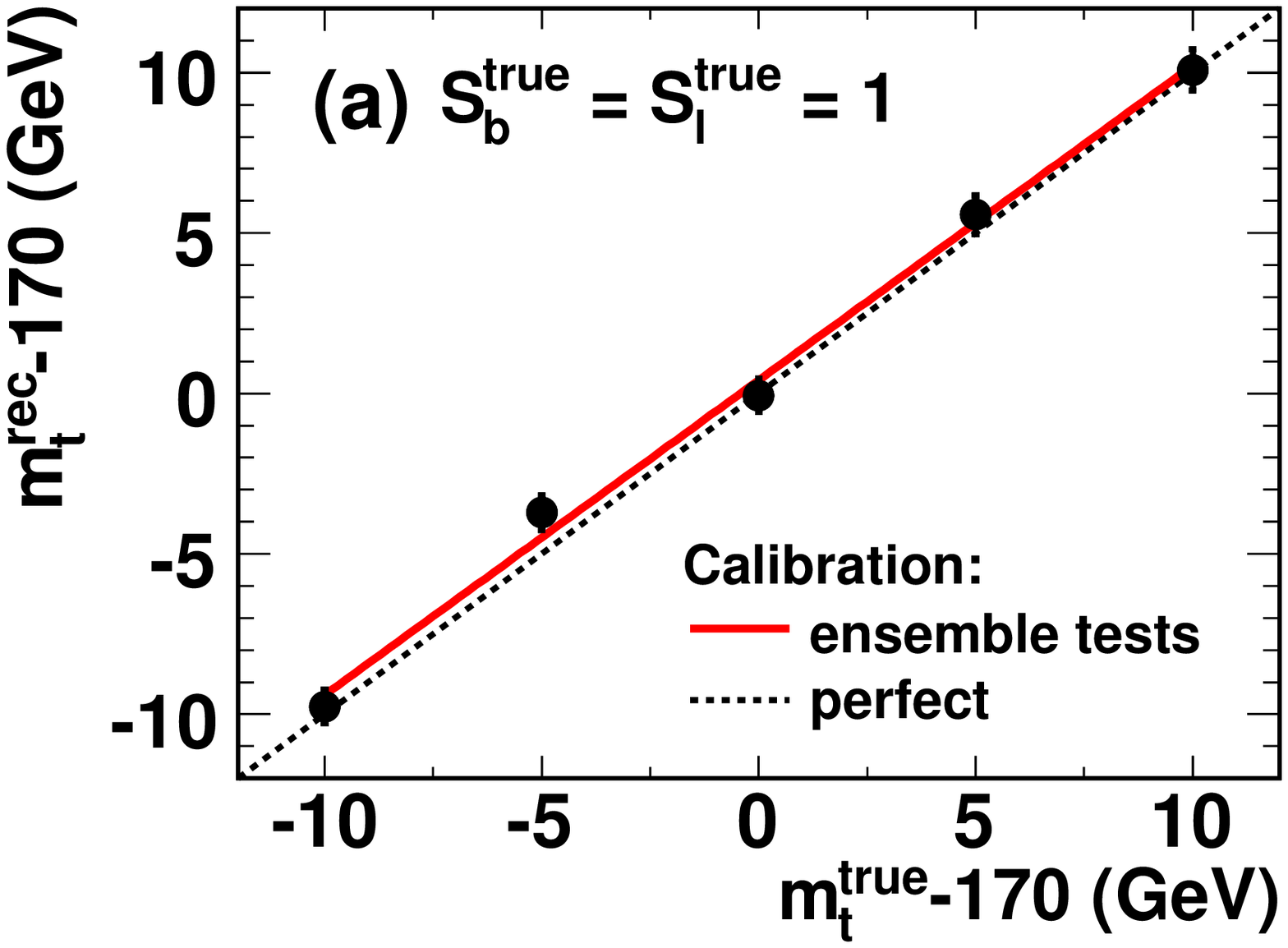}
&
          \includegraphics[width=0.37\textwidth]{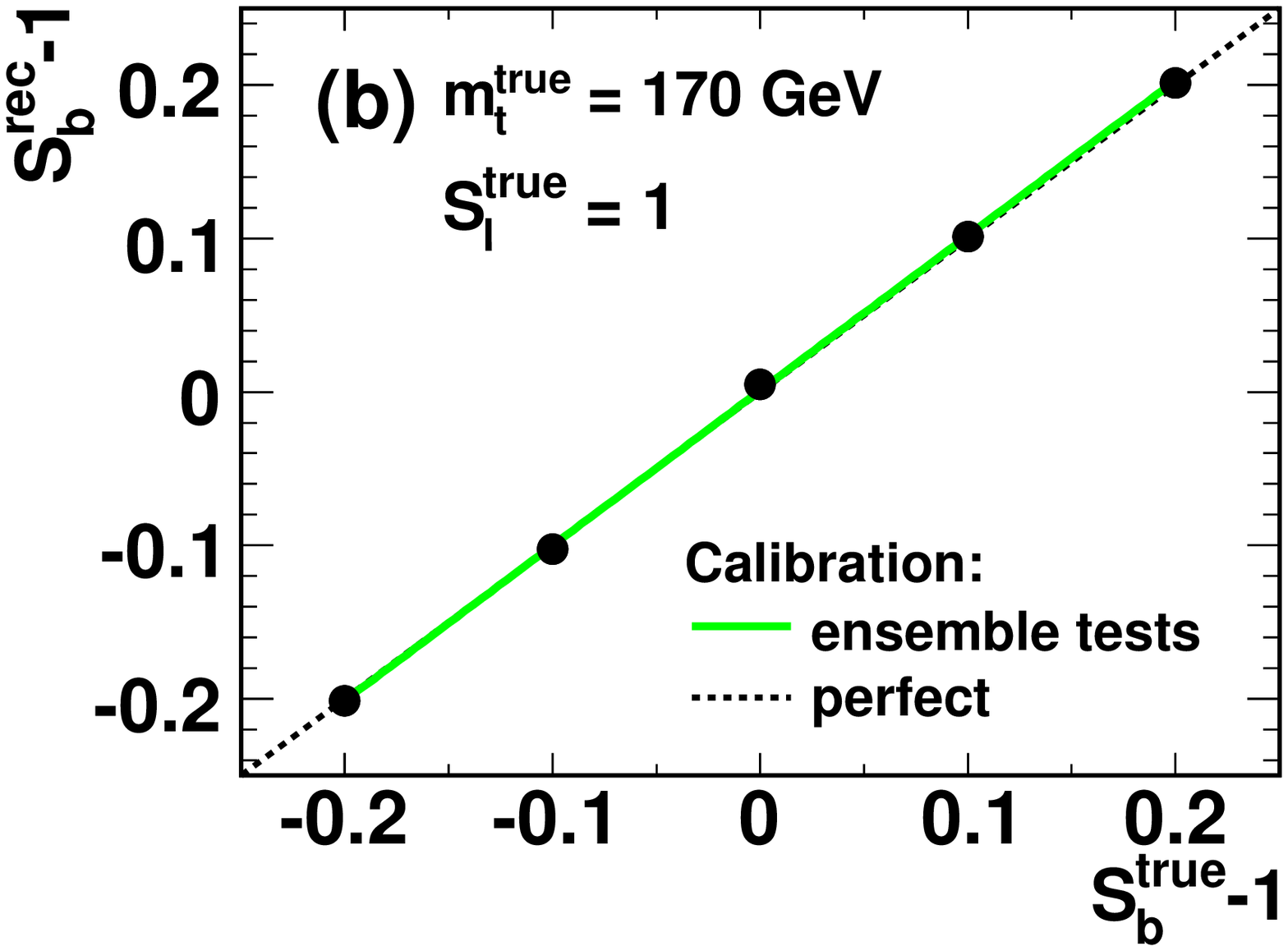}
&
          \includegraphics[width=0.37\textwidth]{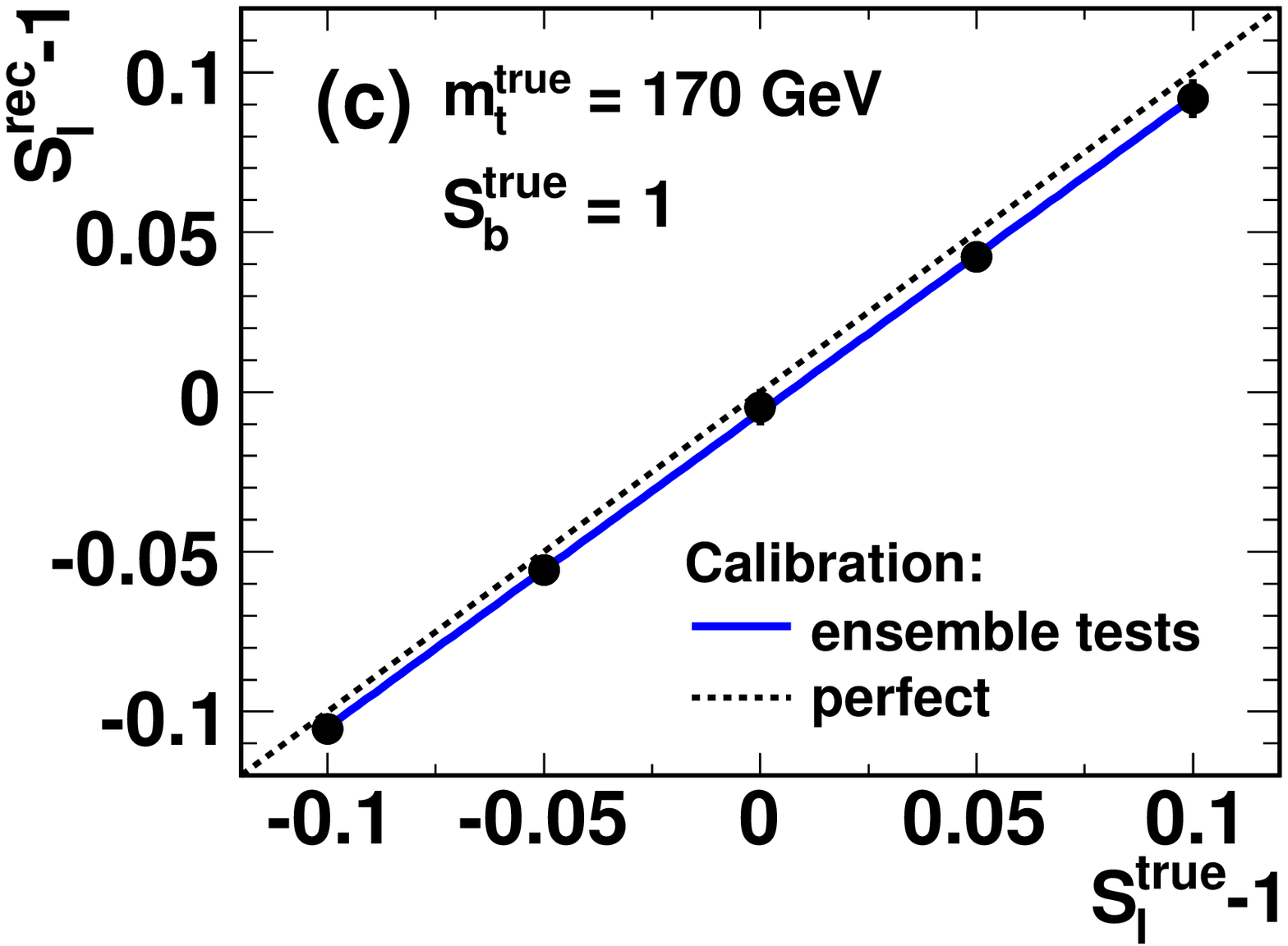}
\\
          \includegraphics[width=0.37\textwidth]{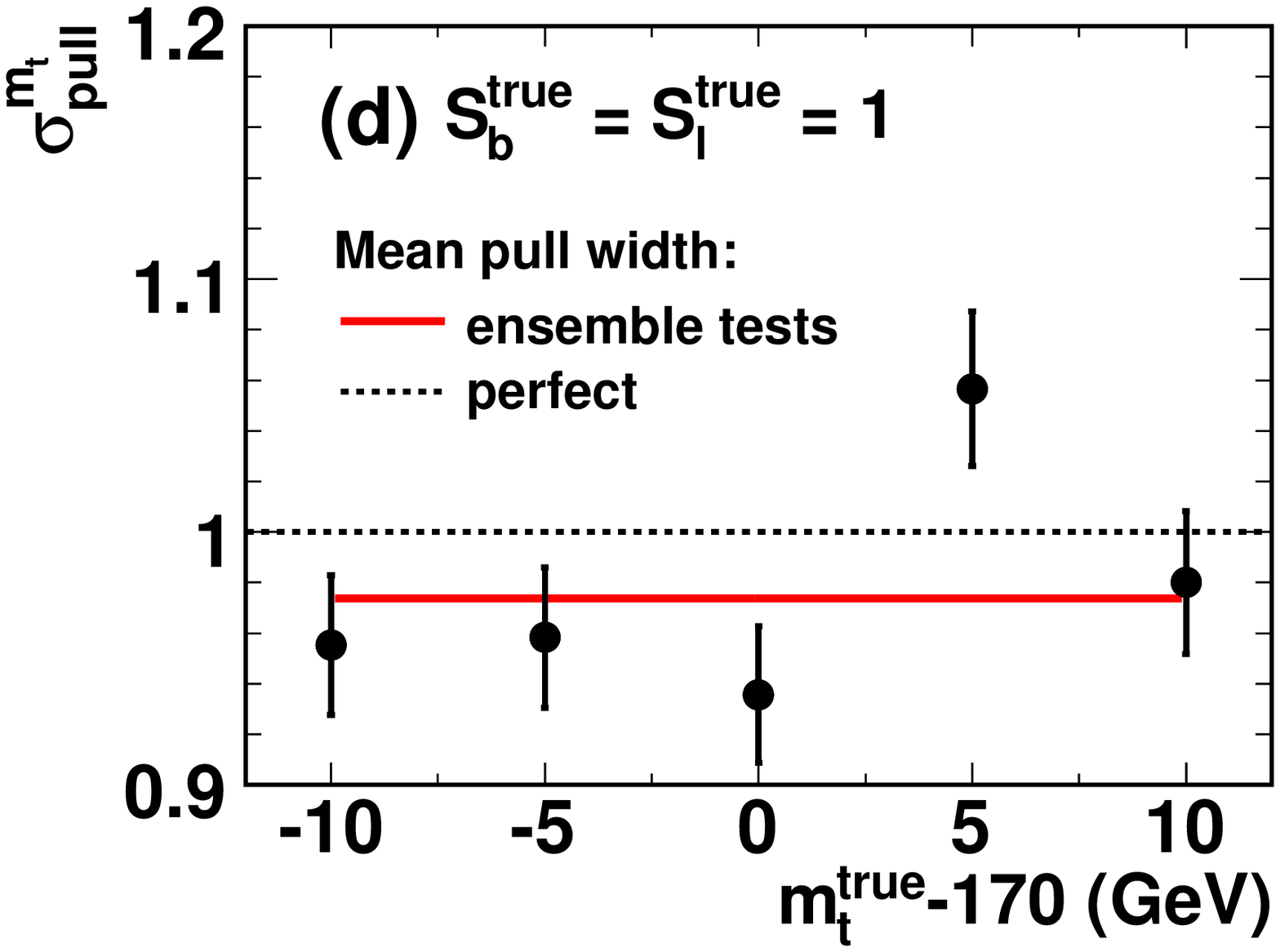}
&
          \includegraphics[width=0.37\textwidth]{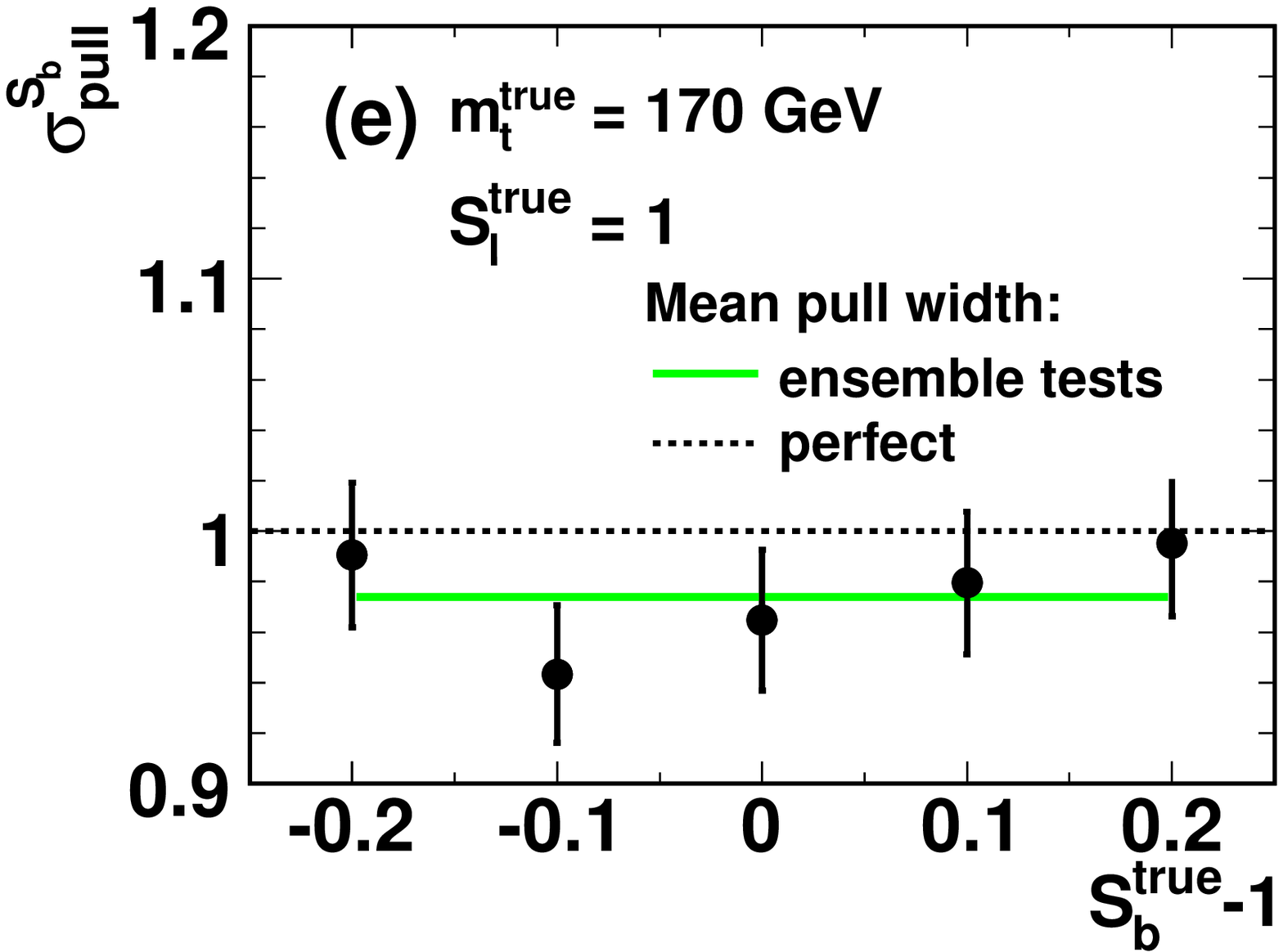}
&
          \includegraphics[width=0.37\textwidth]{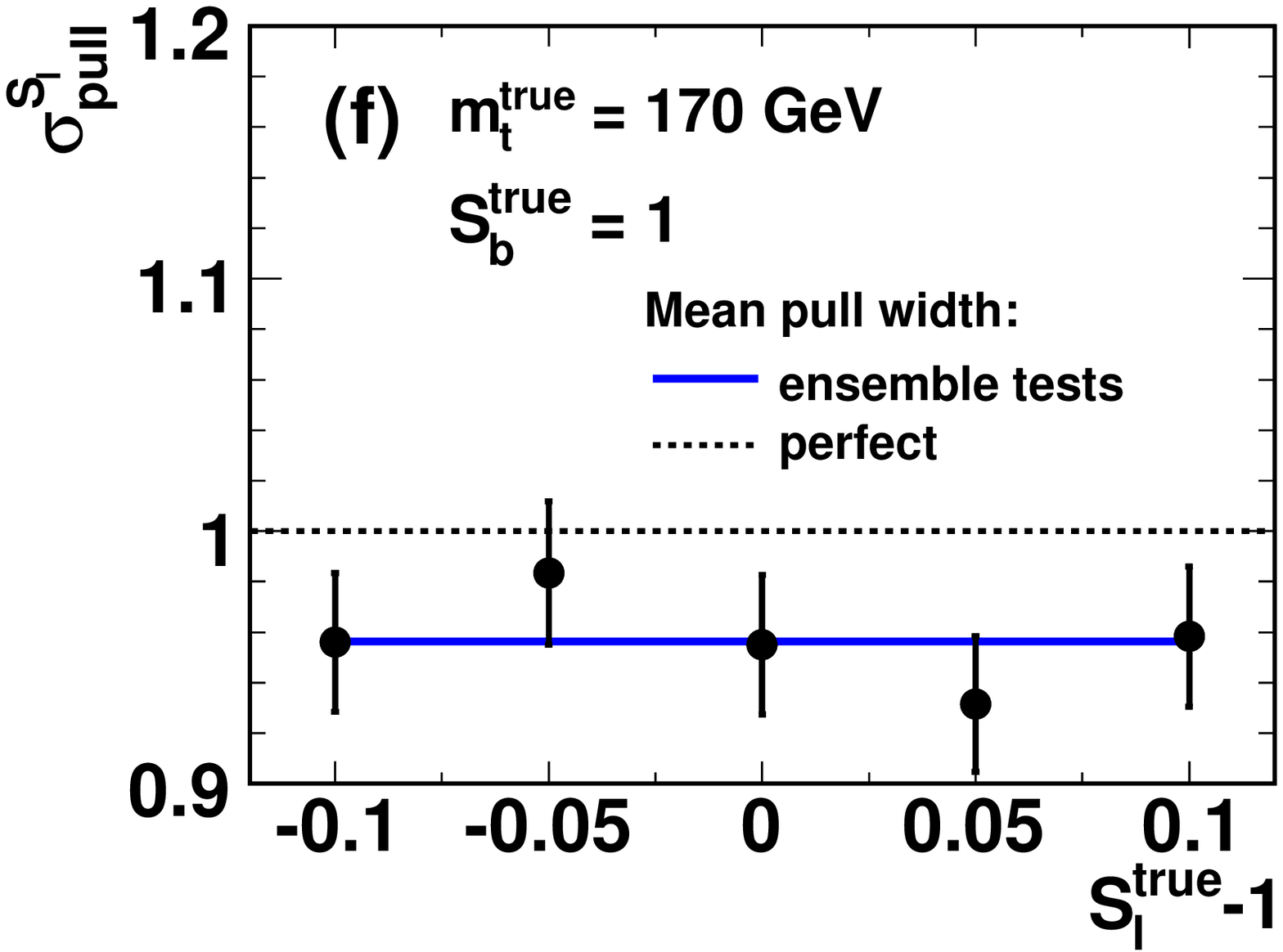}
\end{tabular}
\caption{\Ljets channel: Measurement of \mt, \sbjes, and 
\sjes in pure signal ensembles.  
Reconstructed (``rec'') vs.\ true values are shown in
plots (a)-(c), the widths of the pull distributions 
vs.\ true values in plots (d)-(f).
In plots (a)-(c) the solid lines show the results of straight-line fits,
while in plots (d)-(f) they indicate the mean values.
Event samples with the same \mt but different
\sbjes or \sjes values are correlated since they are obtained by 
scaling the final-state quark energies as described in 
Section~\ref{samples.sec}.}
\label{fig:ljets_fit_psgn}
\end{center}
\end{figure}

In Figure~\ref{fig:ljets_fit_psgn}, results for the measurements 
of the three fit  
observables \mt, \sbjes, and \sjes can be found.
In this and the similar figures that follow, the error bars
represent the uncertainties arising from 
limited statistics in the ensemble tests.
The reconstructed values reproduce the generated ones well, and 
the deviations between fitted and true values are adequately described by the
fitted statistical uncertainties.
The pull widths in Figure~\ref{fig:ljets_fit_psgn}(d) are on average
two standard deviations below the expectation.
A similar conclusion cannot be drawn from
Figures~\ref{fig:ljets_fit_psgn}(e) and (f) since the same events are used
for all \bjes and \jes values except for a rescaling of jet energies.
The results show that the method works in this test case and that  
the \bquark-jet energy scale can be measured together with the top
quark mass and light-jet energy scale.

\begin{figure}[tbp]
\begin{center}
\includegraphics[width=0.8\textwidth]{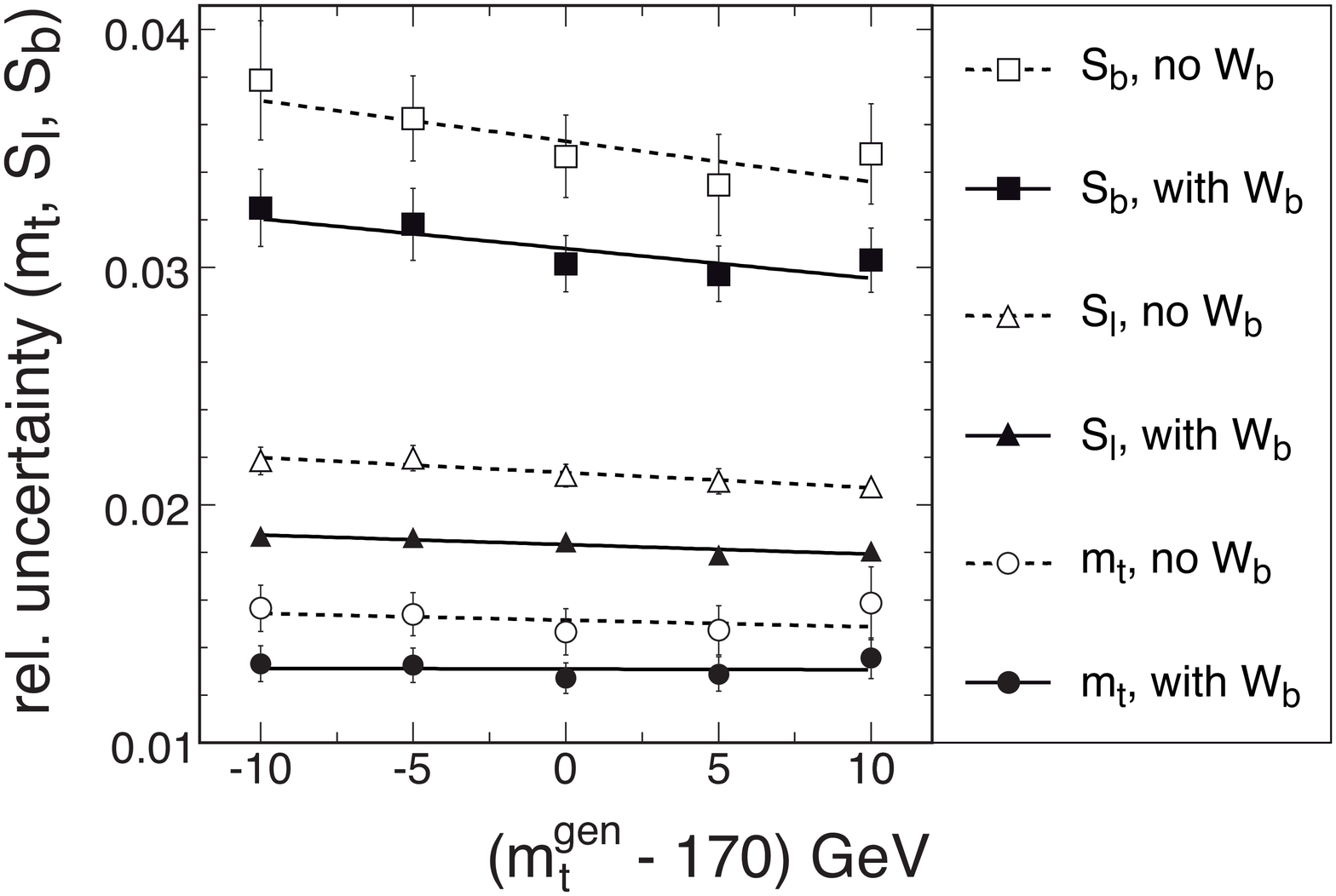}
\caption{\Ljets channel: Expected statistical
uncertainties of the \mt (circles), \bjes (squares), and 
\jes (triangles) determination in pseudo-experiments with
signal events only. The filled
markers are the results with the full transfer function included,
whereas the open markers lack the $W_b$ factor ($b$
identification). Solid and dashed lines indicate 
straight-line fits to the points.}
\label{fig:ljets_error_psgn}
\end{center}
\end{figure}

In order to quantify the gain from the inclusion of \bID likelihoods
as the factor $W_b$ in Equation~(\ref{tfdefinition.eqn}),
the expected statistical uncertainties on the measurement quantities \mt, 
\bjes, and \jes are depicted in Figure~\ref{fig:ljets_error_psgn}. 
These statistical uncertainties
correspond to the hypothetical case of
an integrated luminosity at the Tevatron of about 
0.8~fb$^{-1}$ (for one experiment and one decay channel) without any background.
The solid lines show the expected uncertainties obtained when using the full
transfer function, while the uncertainties given by the dashed lines
are obtained when the factor $W_b$ is omitted from the transfer
function.
For all three quantities an  
improvement of about 15\% on the expected relative statistical uncertainty can  
be achieved because of the additional \bID information. 
No systematic deviation between the measurement values 
obtained with and without \bID likelihoods is observed.
The results without $W_b$ are shown for illustration only; in the
final studies with \ljets events the factor $W_b$ is included.

\subsection{Studies Including \wjjjj and \wbbjj Background}
\label{enstestljets.bkg.sec}
For the background studies, ensembles are composed of 1000 
pseudo-experiments with  
$200$ events each, using different signal fractions.
The case of a signal fraction $f_{\ttbar}=50\,\%$ 
corresponds to an integrated luminosity at one Tevatron experiment of about 
0.8~fb$^{-1}$ (for one decay channel). 
Two 
sources of background are studied, \wjjjj and \wbbjj events.

Background from \wjjjj events, containing a  
leptonically decaying \W and four light partons, is described  
by the background likelihood (see Section~\ref{application.pbkg.sec})
and thus accounted for.
To study the dependence of the method on the background fraction, it is  
varied between 0\% and 90\% in 10\% steps. 
\begin{figure}[tbp]
\begin{center}
\begin{tabular}{@{\hspace{-10pt}}c@{\hspace{-16pt}}c@{\hspace{-16pt}}c@{}}
          \includegraphics[width=0.37\textwidth]{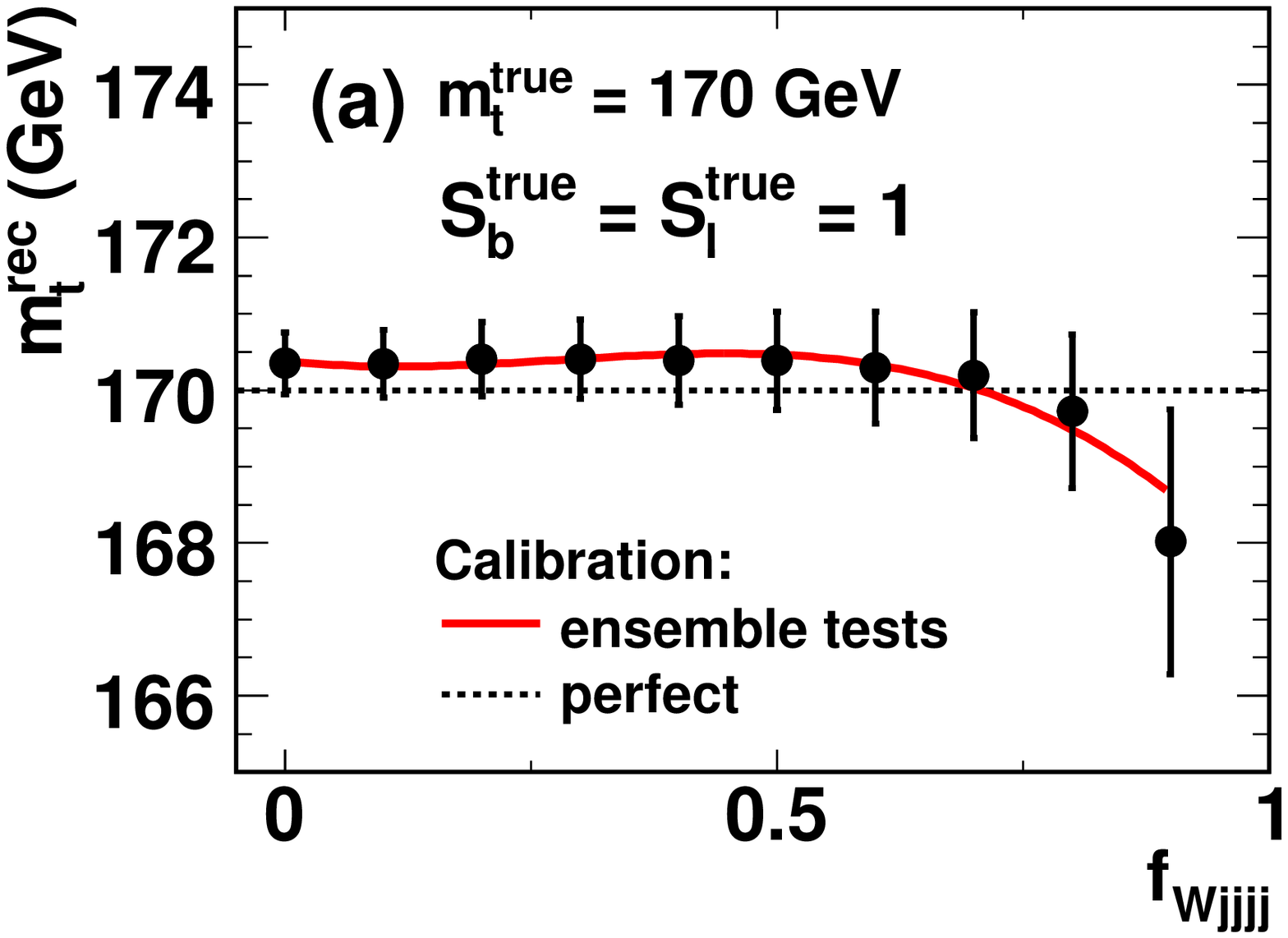}
&
          \includegraphics[width=0.37\textwidth]{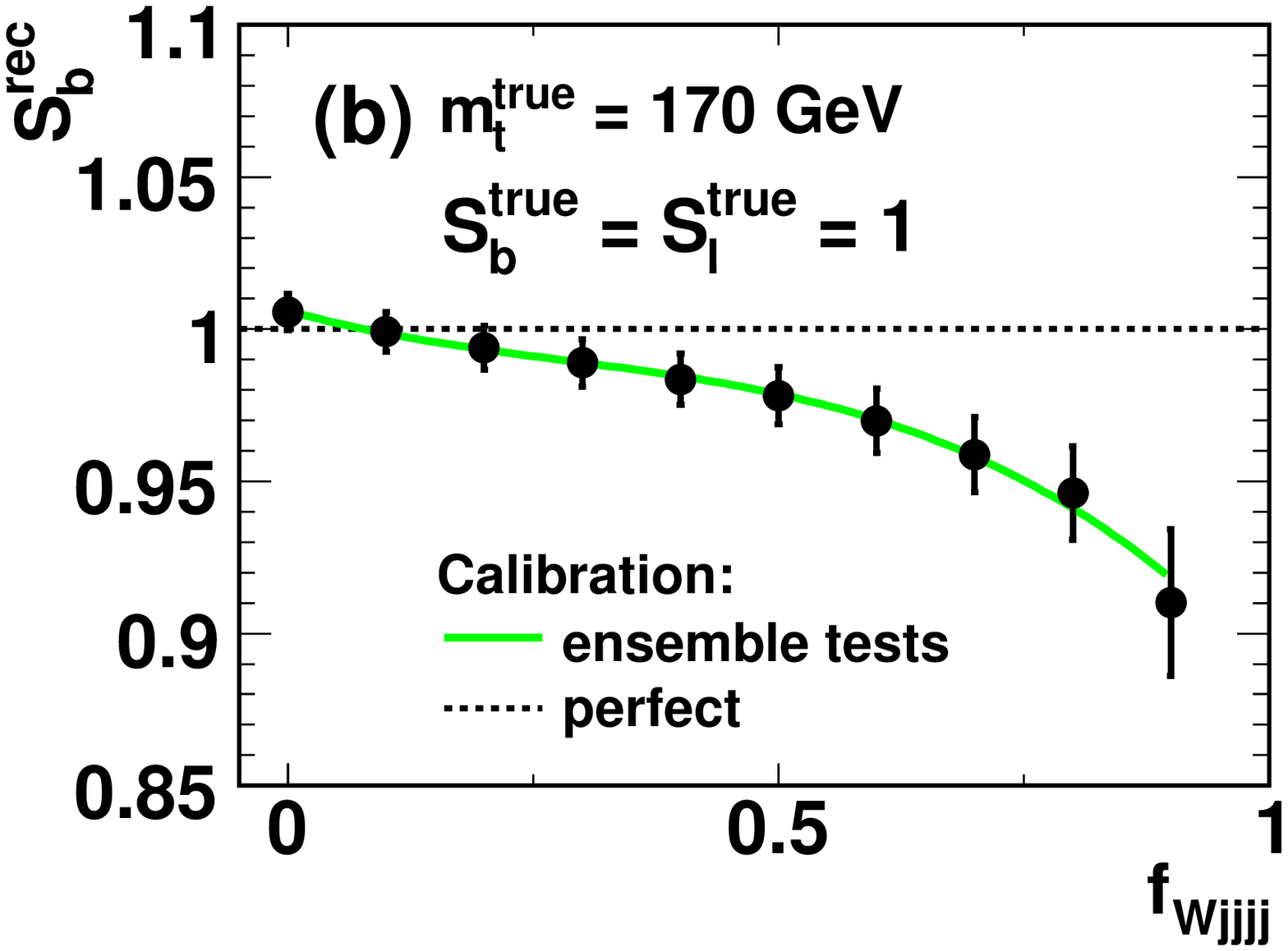}
&
          \includegraphics[width=0.37\textwidth]{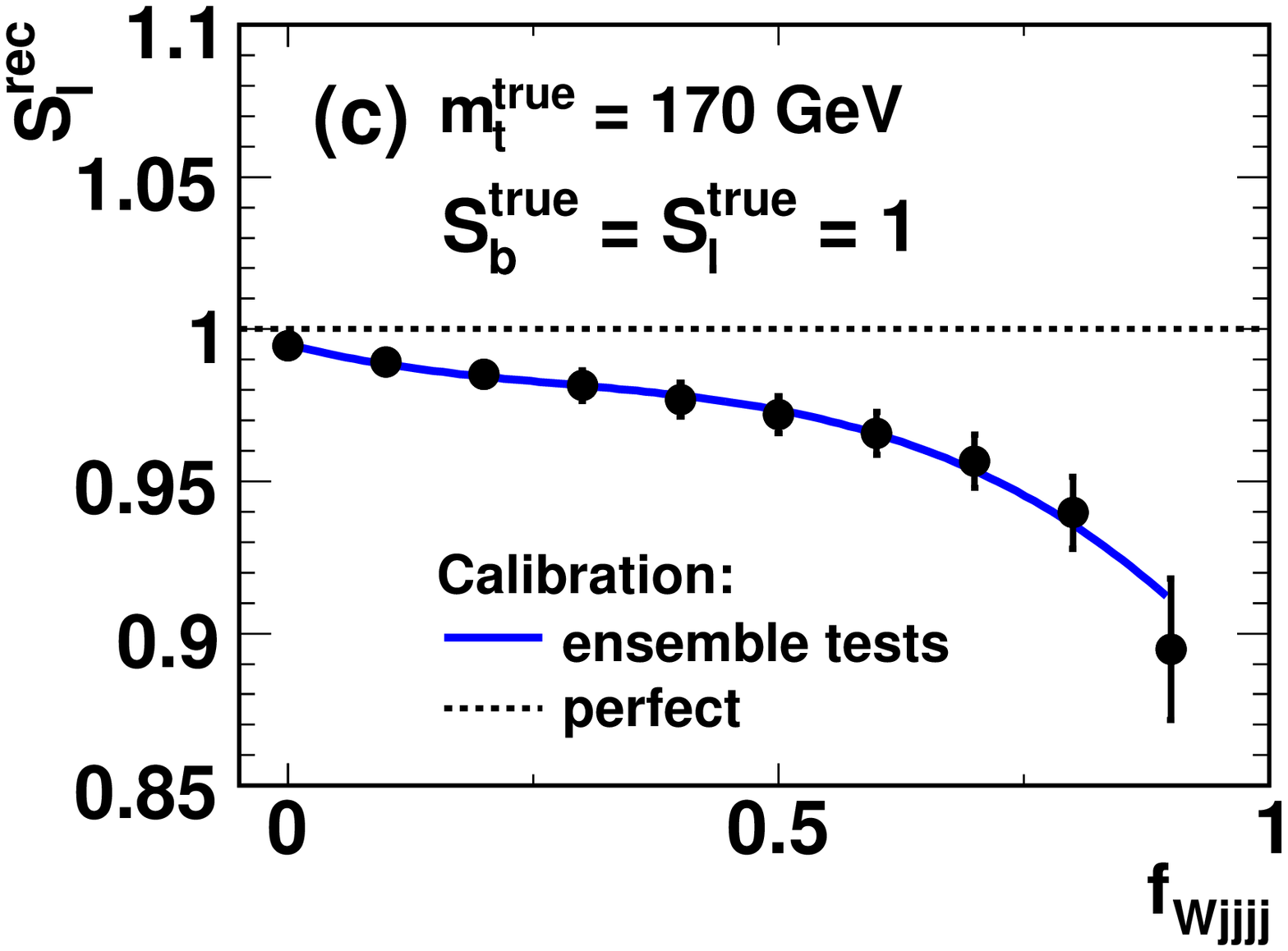}
\end{tabular}
\caption{\Ljets channel: Measurement of (a) \mt, (b) \sbjes, and (c)
\sjes in ensembles including \wjjjj background.  Reconstructed values
are shown as a function of the background fraction.  
The individual points in each plot are correlated because the 
ensembles are drawn from the same event pools.
The lines
indicate the results of \thirdorder polynomial fits to the points.} 
\label{fig:ljets_f_wjjjj}
\end{center}
\end{figure}
Figure~\ref{fig:ljets_f_wjjjj} shows the results for ensembles
with true values of $\mt=170\,\GeV$ and $\bjes=\jes=1$. 
The top quark mass fit yields the expected results even for 
background fractions significantly larger than those observed in the data.
The two jet energy scales show deviations from the expected values
when background is included;
these deviations increase with the background fraction.
This effect is not unexpected since the background likelihood is  
calculated only for the $\bjes=\jes=1$ hypothesis.
However, this simplification is appropriate when
the goal of the analysis is a measurement of the top quark mass, 
while the determination of the jet energy scales is only performed
to reduce the systematic uncertainties on this measurement.
The lack of an exact modelling of the jet energy scale in background events
does not limit the precision of the top quark mass determination.

\begin{figure}[tbp]
\begin{center}
\includegraphics[width=\textwidth]{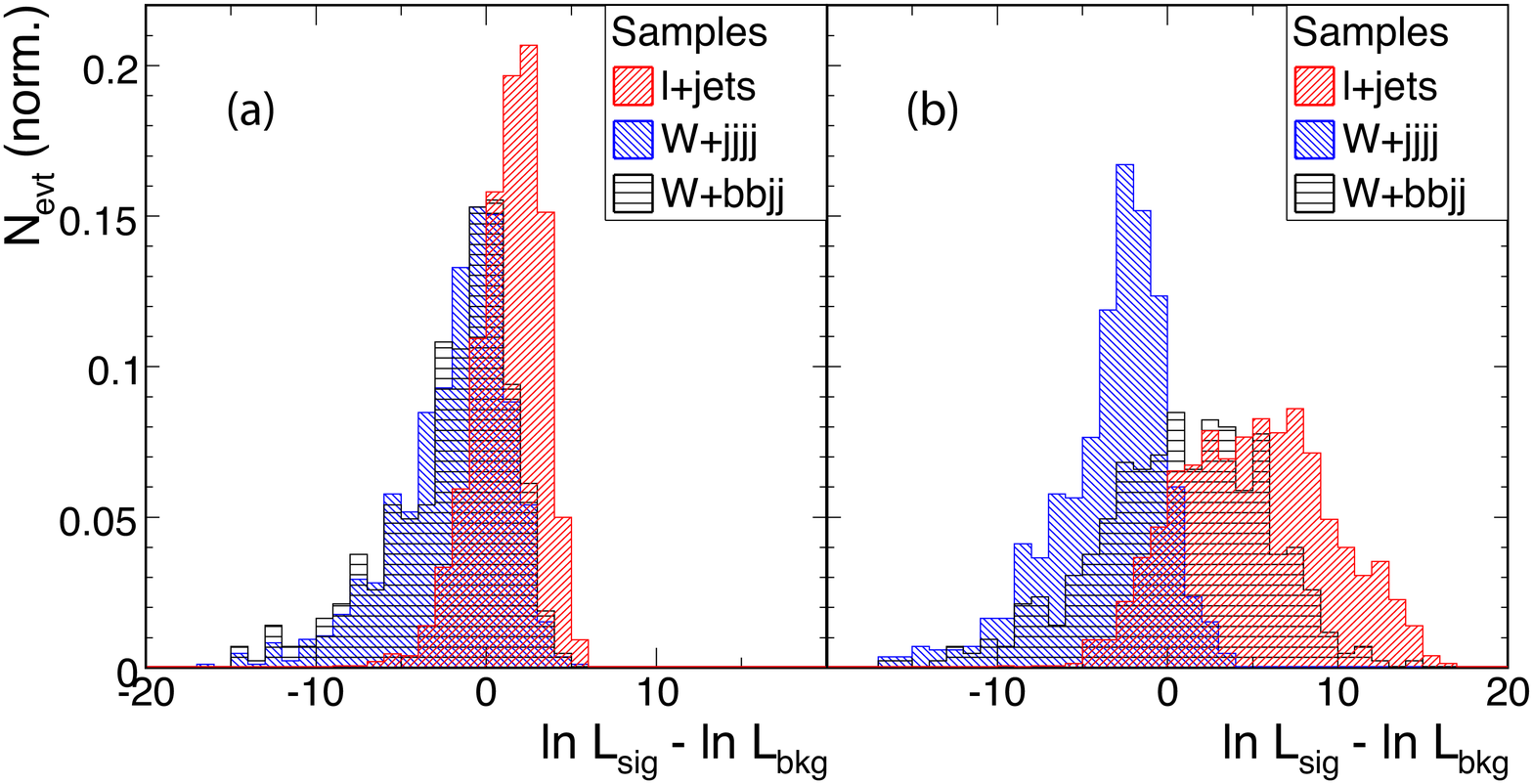}
\caption{\Ljets channel: Difference between the 
  log-likelihood values $\ln\pevtttbar$ and $\ln\pevtwjjjj$
  for \ljets \ttbar signal as well as \wjjjj and \wbbjj 
  background events.
  Each individual distribution is normalized.  
  In plot (a), the $W_b$ factor has been omitted in the 
  transfer function, while it is included in plot (b).
}
\label{fig:ljets_separation}
\end{center}
\end{figure}

Background from 
\wjets events containing \bquark quarks is topologically very similar to  
\wjjjj background and is thus not treated as a separate process
in the likelihood calculation.
Nonetheless, if one includes \bID information in the analysis, \wbbjj
events have to be considered carefully.

Figure~\ref{fig:ljets_separation} shows the difference between the 
log-likelihood values $\ln\pevtttbar$ and $\ln\pevtwjjjj$
for \ttbar signal as well as \wjjjj and \wbbjj 
background events.
The topological information alone already allows for a discrimination
between signal and background.
But as expected, there is no separation
between background without (\wjjjj) and with (\wbbjj) \bquark~jets;
such a separation only arises when \bID information is included.
Figure~\ref{fig:ljets_separation}(a) is shown for illustration only;
in the final studies with \ljets events the factor $W_b$ is included
in the transfer function as given in Equation~(\ref{tfdefinition.eqn}).

Ensembles with $\mt=170\,\GeV$ and $\bjes=\jes=1$ 
are created that have a fixed total  
fraction of background (50\%), but the fraction of \wbbjj
events within this background is varied. 
This means that an absolute fraction of  
$f_{\wbbjj} = 0.5$ corresponds to pseudo-experiments in which the  
background consists solely of \wbbjj events.
\begin{figure}[tbp]
\begin{center}
\begin{tabular}{@{\hspace{-10pt}}c@{\hspace{-17pt}}c@{\hspace{-17pt}}c@{}}
          \includegraphics[width=0.37\textwidth]{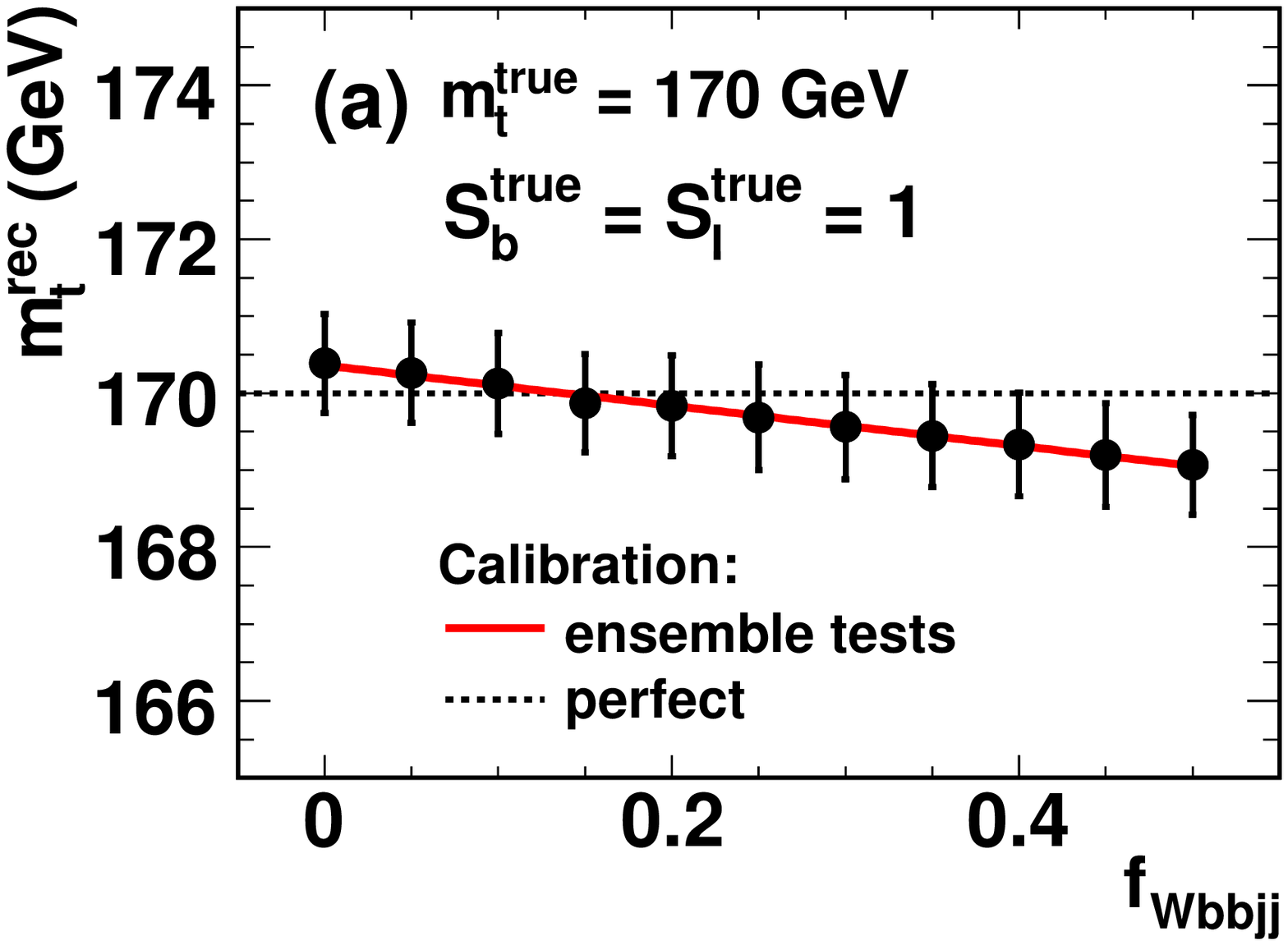}
&
          \includegraphics[width=0.37\textwidth]{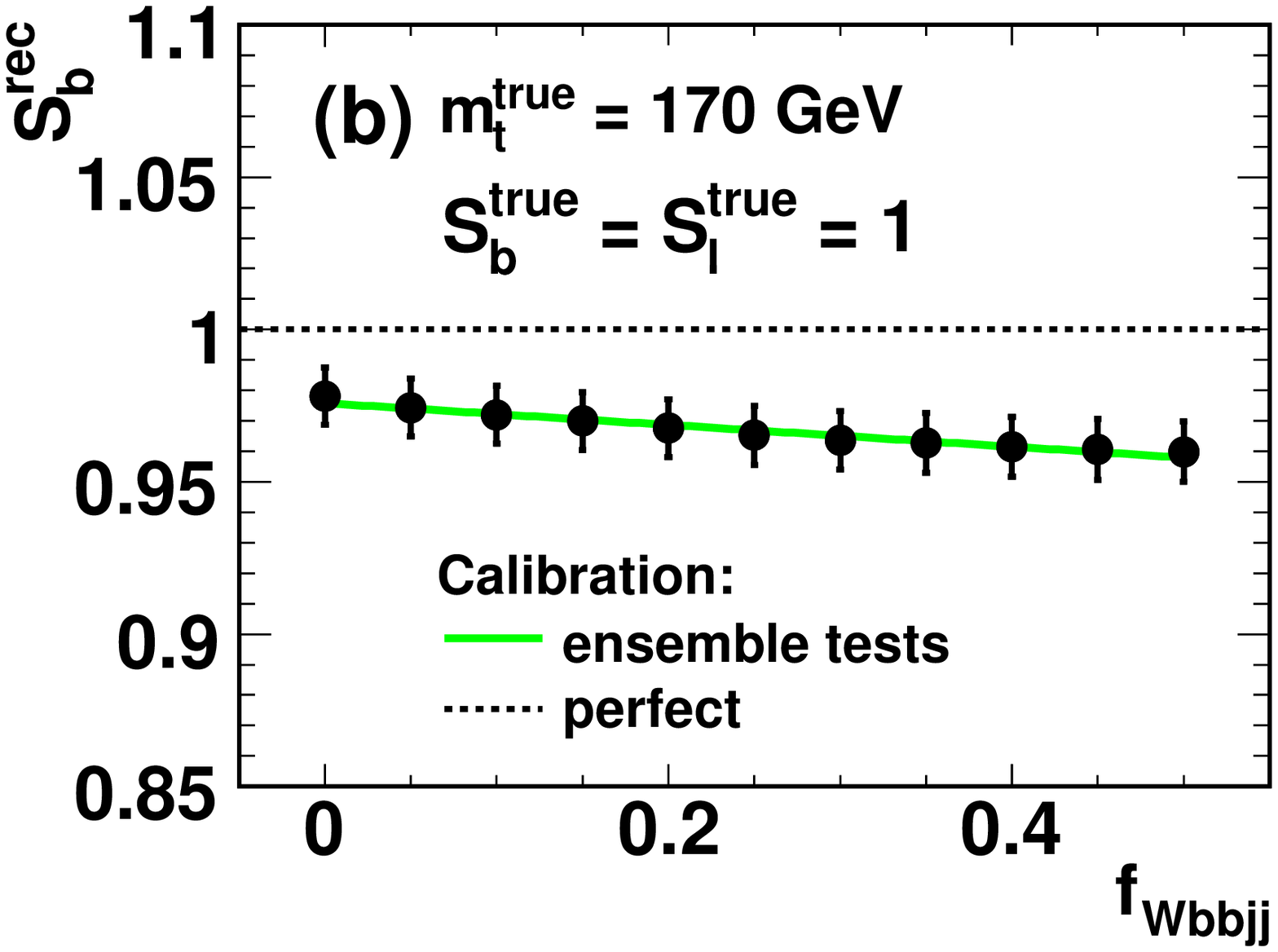}
&
          \includegraphics[width=0.37\textwidth]{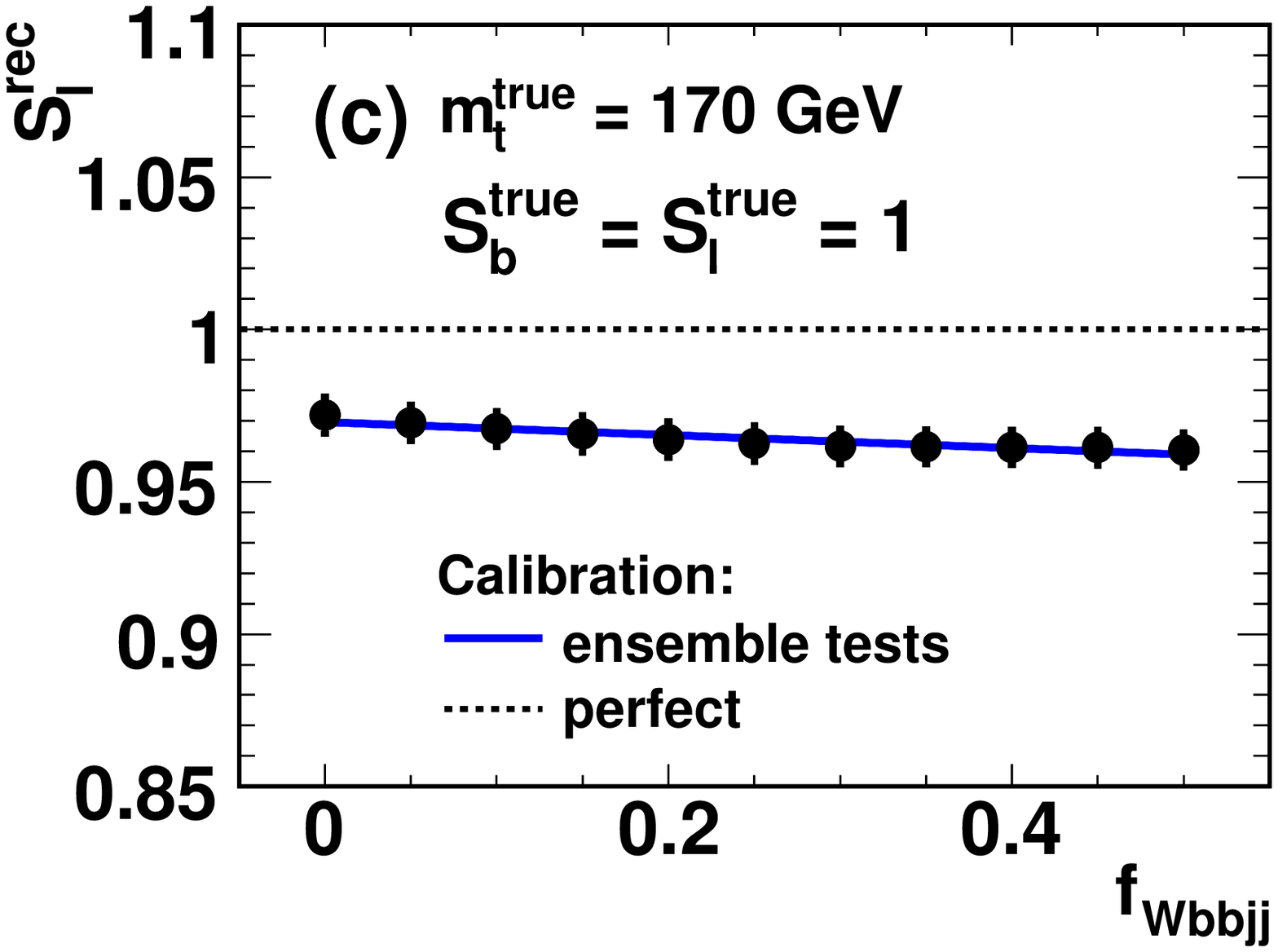}
\end{tabular}
\caption{\Ljets channel: Measurement of (a) \mt, (b) \sbjes, and (c)
\sjes in ensembles including \wjjjj and \wbbjj background.  
Reconstructed values are shown vs.\ the absolute \wbbjj background
fraction. The total fraction of background is fixed to 50\%. 
The uncertainties on the individual points are correlated.
The lines
indicate the results of straight-line fits to the points.}
\label{fig:ljets_f_wbbjj}
\end{center}
\end{figure}
Figure~\ref{fig:ljets_f_wbbjj} shows the  
results for the three measurement quantities versus the absolute fraction  
of \wbbjj events. 
The results indicate that 
there is only a weak dependence of the fit results on the  
fraction of \wbbjj events for all three fit observables.
Consequently, only small systematic uncertainties arise in the 
calibration of the measurement, and it is
justified that \wbbjj events are not accounted for 
explicitly in the event likelihood.
Note that the results for $f_{\wbbjj} = 0$ correspond to the values
for $f_{\wjjjj} = 0.5$ in Figure~\ref{fig:ljets_f_wjjjj} and that 
for \bjes and \jes, deviations between fitted and true values are not 
unexpected for $f_{\wjjjj} > 0$ as explained above.

For an integrated luminosity of $12\,{\rm fb}^{-1}$, a signal fraction
$\ftop=50\,\%$, and absolute background fractions of
$\fwjjjj=40\,\%$ and $\fwbbjj=10\,\%$, the expected
statistical uncertainties in the \ljets channel (combining
the \ejets and \mujets channels)
obtained by a single Tevatron experiment
are found to be
\begin{eqnarray}
\nonumber
  \sigma_{\mt}(\textrm{lepton+jets})   & = & 0.45\ \GeV \, , \\
  \sigma_{\bjes}(\textrm{lepton+jets}) & = & 0.0064     \, ,\ {\rm and}\\
\nonumber
  \sigma_{\jes}(\textrm{lepton+jets})  & = & 0.0039     \, .
\end{eqnarray}
The slopes of the calibration curves are between $0.91$ and $0.97$ and 
have been accounted for,
and the uncertainties have been multiplied with the pull widths 
between $0.94$ and $1.04$.

\section{Application of the Technique to \ttbar Events in the Dilepton Channel}
\label{enstestdilepton.sec}
This section describes the application of the Matrix Element 
method for a simultaneous measurement of the top quark mass 
and the \bquark-jet energy scale in dilepton \ttbar events in 
the \emu channel.
As in the \ljets channel,
to minimize computing time, the parton-level studies described here
have been
performed assuming perfectly measured lepton momenta, i.e.\ the 
generated leptons are
not smeared, and the additional integration over the inverse muon
transverse momentum is not carried out.
This approach is valid here because the aim is to study the behavior
of the measurement method when applied to pseudo-experiments with 
different background compositions, and because it has been verified 
that the conclusion from the parton-level 
tests stays the same and is not affected by this choice. 
To validate the integration over the inverse muon momentum, an
additional test has been performed using smeared 
leptons~\cite{bib-DrarbAlexander}.

In the dilepton channel, \bquark-tagging information cannot help to 
select the correct assignment of jets to partons as in the \ljets case
(except in events with significant gluon radiation).
Since the background to \ttbar dilepton events is small, no
\bquark-tagging information has been used in the studies 
shown in this section, i.e. the factor $W_b$ has been omitted
from the transfer function in Equation~(\ref{tfdefinition.eqn}).

\begin{figure}[ht]
\centering
\includegraphics[width=0.53\textwidth]{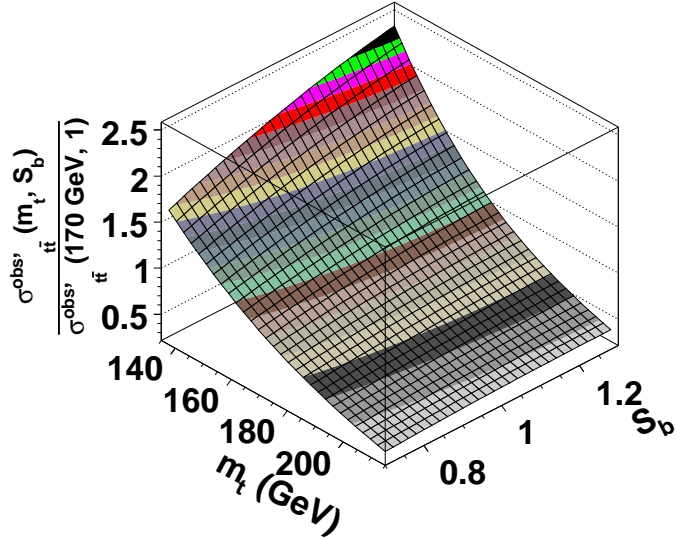}
\caption{\label{fig:nrmmtopbjes} Dilepton channel:
  Normalization function \sigmattbarobsprime
  for the \ttbar likelihood in the dilepton channel
  as a function of the top quark mass and \bquark-jet energy scale,
  normalized to the fitted value for $\mt=170\,\GeV$ and
  $\bjes=1$.}
\end{figure}

As the missing transverse momentum \ptmiss depends on the
reconstructed jet energies, the 
normalization of the signal likelihood depends not only on \mtop, but also
on \sbjes. 
For a given value of \sbjes, the normalization is calculated as a
function of the top quark mass and fitted with a \thirdorder
polynomial, similar to what is shown in \Fref{fig:ljets_nrm}.
Each of the four parameters of
the polynomials as a function of \sbjes are then fitted in turn
with a quadratic function. 
The resulting two-dimensional normalization
is shown in \Fref{fig:nrmmtopbjes}.
The relative normalization of the background and signal likelihoods
is derived as
described in Section~\ref{method.normL.bkg.sec}.

Similar to Section~\ref{enstestljets.sec}, 
ensemble tests under different hypotheses are described
in the following.

\subsection{Signal-Only Studies}
\label{sec:plmtopbjessgnonly}
In a first step, ensemble tests with pure signal events are
performed, and the signal likelihood is taken as the event likelihood.
For each of nine calibration points in the (\mtop,~\sbjes) plane,
1000 pseudo-experiments are performed.
Each of the pseudo-experiments is built of 50 \ttbar events in the 
\emu channel, corresponding
to an integrated luminosity of about $1.4\ {\rm fb^{-1}}$ 
at the Tevatron~\cite{bib-d0dileptonxs}.
At all calibration points, the 
generated and the measured values of the top quark mass and \bquark-jet 
energy scale are in excellent agreement. 
The uncertainty on \mtop (\sbjes) does not depend on \sbjes (\mtop),
and increases with \mtop (\sbjes) as expected. 
The pull width is
always consistent with unity within uncertainties.

\subsection{Studies Including \ztautaujj and \ztautaubb Background}
\label{sec:plmtopbjessgnzjj}
In the next step, the dominant source of background is added, 
i.e.\ \ztautaujj events where the \Z boson decays into an
electron and a muon via two $\tau$ leptons. Accordingly, the 
\ztautaujj likelihood is included in the event likelihood.
 
\begin{figure}[ht]
\centering
\begin{tabular}{cc}
  \includegraphics[width=0.45\textwidth]{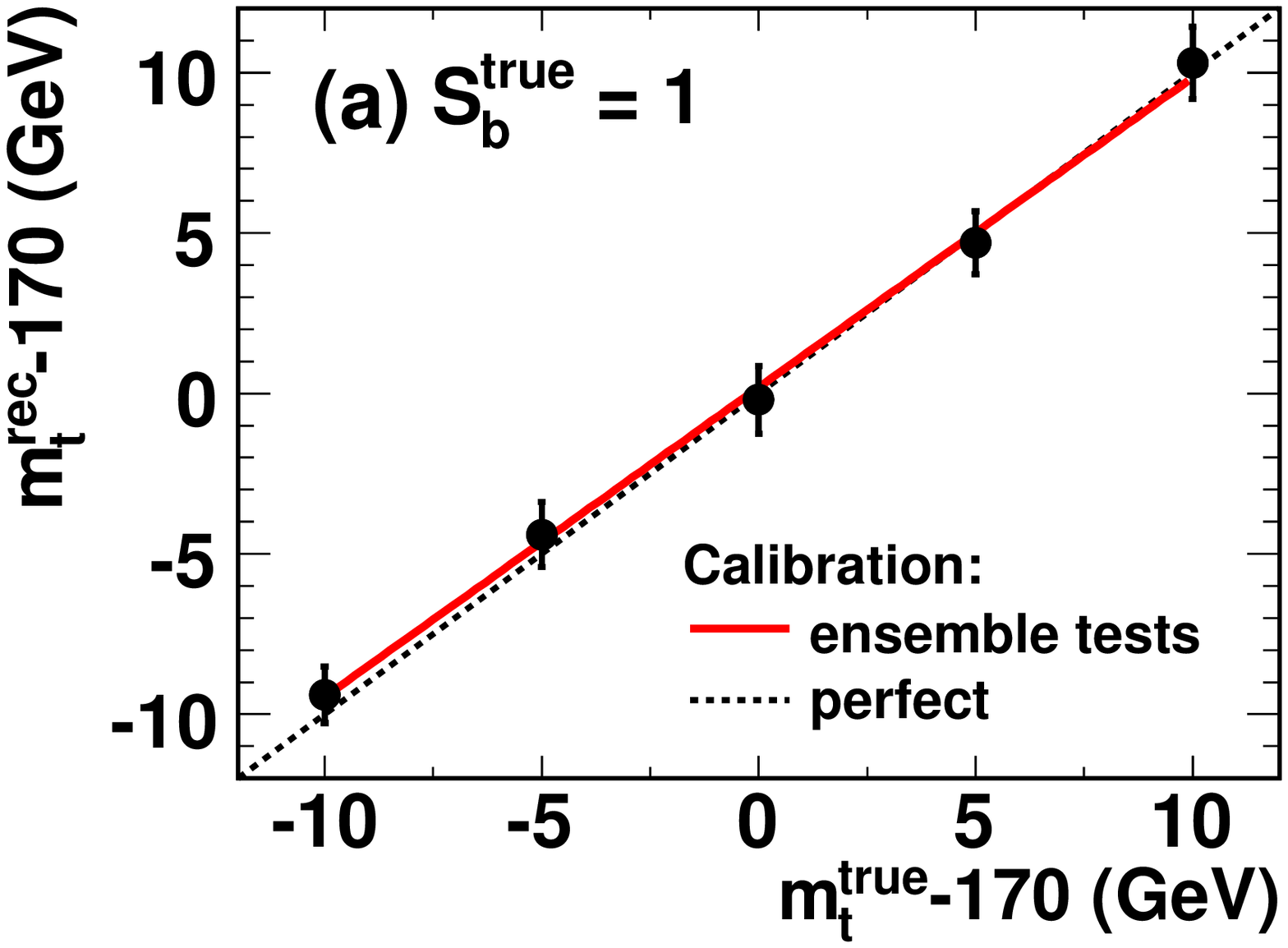}
 &
  \includegraphics[width=0.45\textwidth]{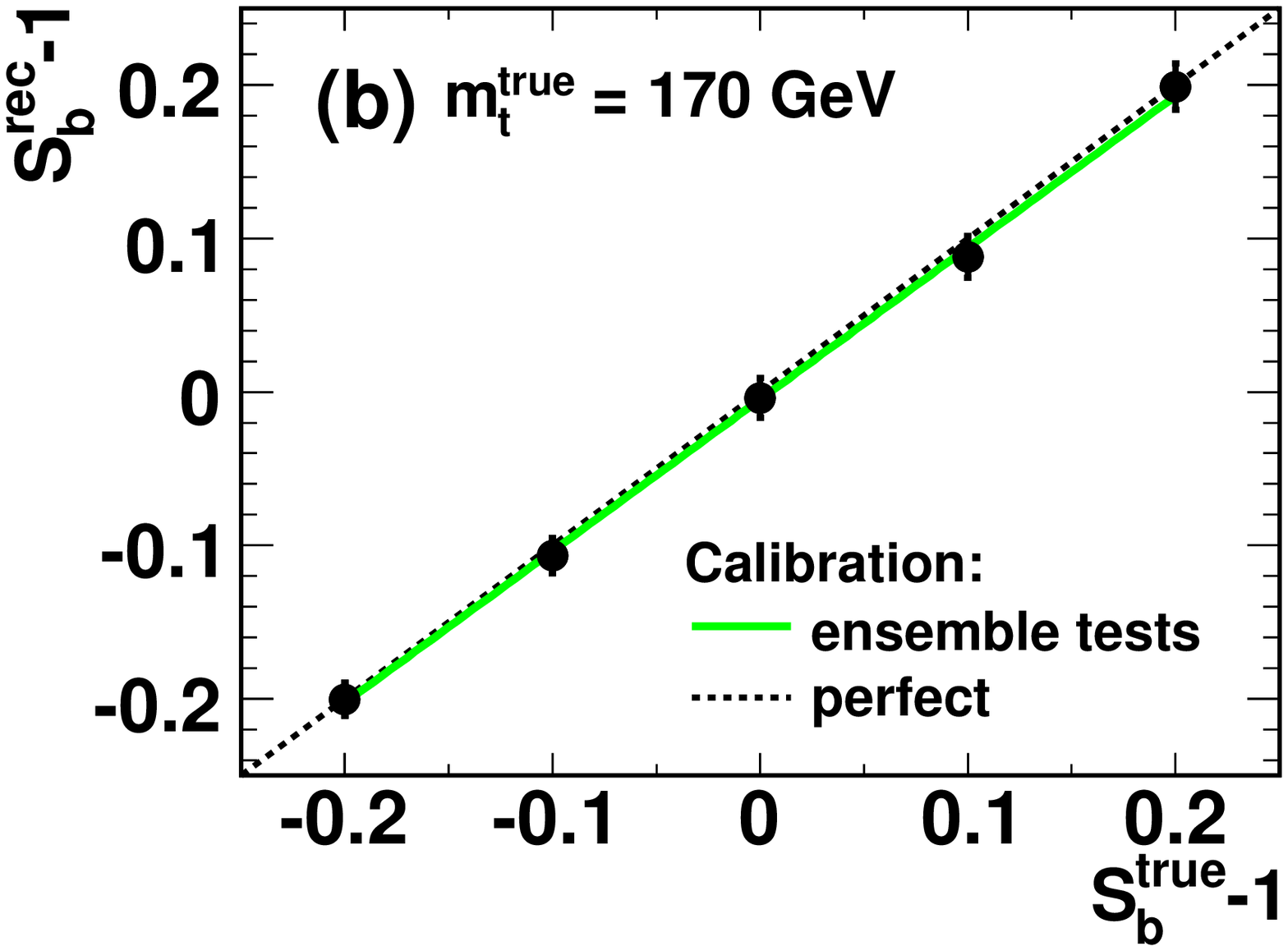}
 \\
  \includegraphics[width=0.45\textwidth]{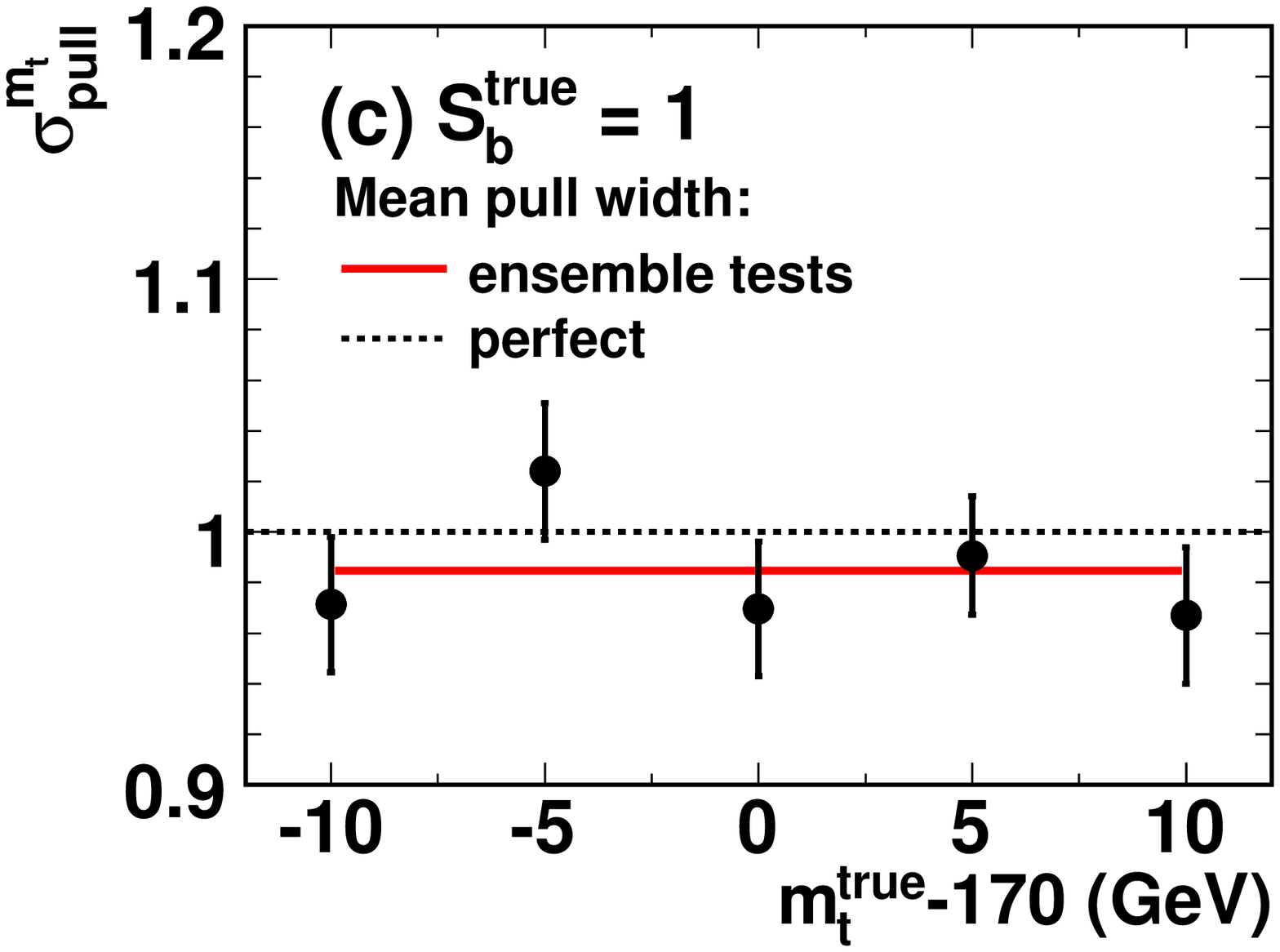}
 &
  \includegraphics[width=0.45\textwidth]{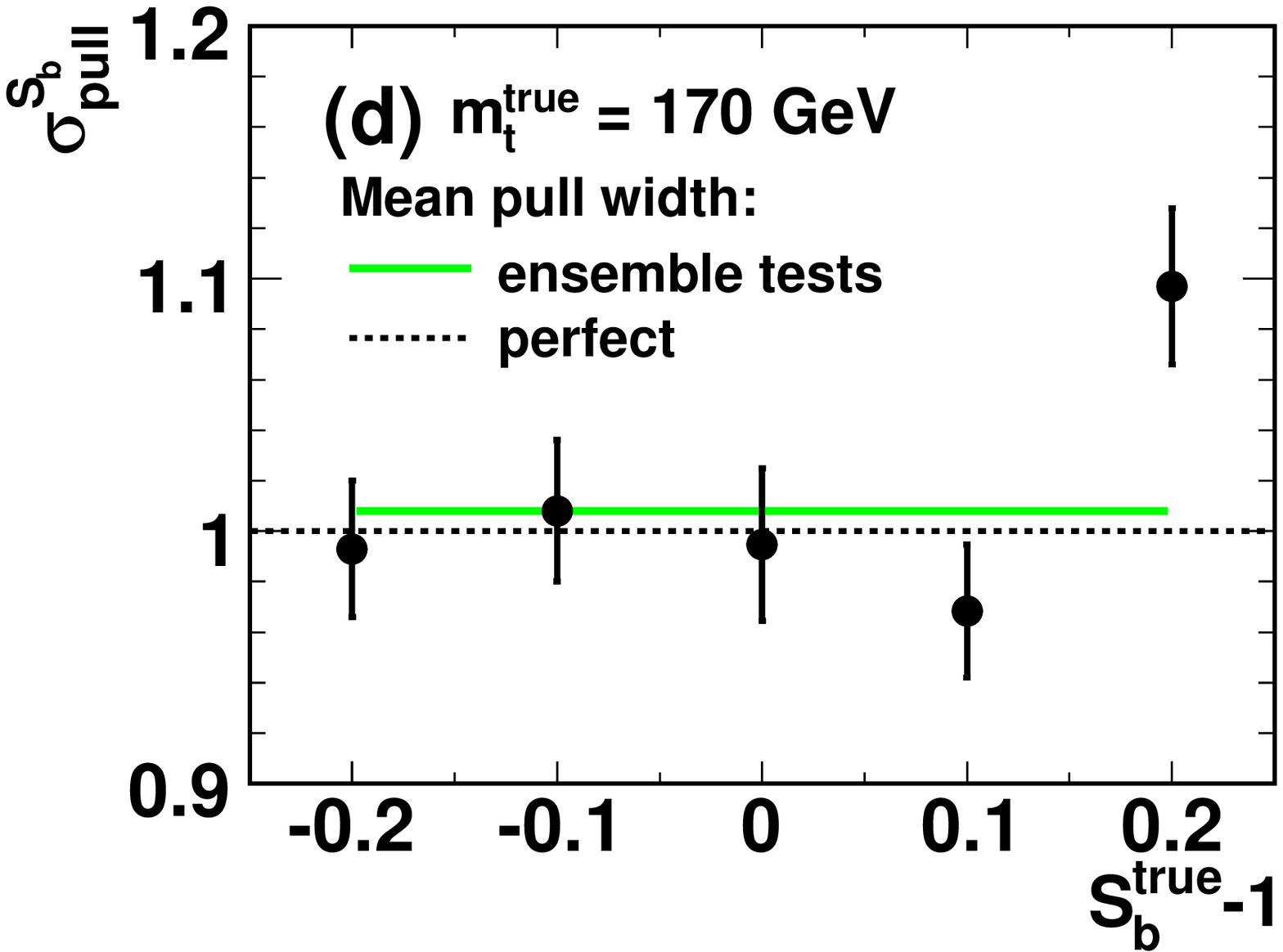}
\end{tabular}
\caption{\label{fig:plmtbjesememzjjcalibmtbjes} 
  Dilepton channel: 
  Measurement of \mt and \sbjes
  in ensembles including 30\% of \ztautaujj background.  
  Reconstructed (``rec'') vs.\ true values are shown in
  plots (a) and (b), pull widths vs.\ true values in plots (c)-(d).
  In plots (a) and (b) the lines show the results of straight-line fits,
  while in plots (c) and (d) they indicate the mean values.}
\end{figure}

In each of the 1000 pseudo-experiments, 50 events are used, 
and the fraction of \ztautaujj events is varied
from 10\% to 50\% in steps of 10\%. 
Within statistical uncertainties,
the measured values of \mtop and \sbjes do not depend on the
fraction of \ztautaujj events. 
\Fref{fig:plmtbjesememzjjcalibmtbjes} shows the calibration curves for
\mtop and \sbjes for pseudo-experiments containing 30\% of \ztautaujj events.
This fraction
corresponds roughly to the total fraction of background events 
selected by the \dzero experiment in
the \emu channel~\cite{bib-d0dileptonxs}.
Both calibration
curves are in excellent agreement with the expectation. 
The pull widths of the \mtop and \sbjes measurements are consistent with 
unity for all ensembles.

To study the effect of jets from \bquark quarks, \ztautaubb
events are also included in the pseudo-experiments. 
These are described by the \ztautaujj likelihood; no dedicated likelihood
for \ztautaubb events is included.
The sum of the fractions of \ztautaujj and \ztautaubb events is kept
at 30\%, and the absolute contribution from \ztautaubb is varied between
3\% and 15\% in steps of 3\%.
Within statistical uncertainties, no effect
from the jet flavor can be observed on the mean expected measurement 
values or the widths of the pull distributions.

\subsection{Studies Including \ztautaujj and \wwjj Events}
\label{sec:plmtopbjessgnzjjwwjj}
Additional contamination of the selected dilepton data sample 
comes from \wwjj events. 
The expected fraction in the \emu channel compared to
\ztautaujj events is about one fourth~\cite{bib-d0dileptonxs}.
A study has been performed where 
each of the 1000 pseudo-experiments is composed on average of 35
signal, 12 \ztautaujj, and 3 \wwjj events.  
The \wwjj events are not described by a dedicated likelihood because
their contribution to the event sample is small.

\begin{figure}[ht]
\centering
\begin{tabular}{cc}
  \includegraphics[width=0.45\textwidth]{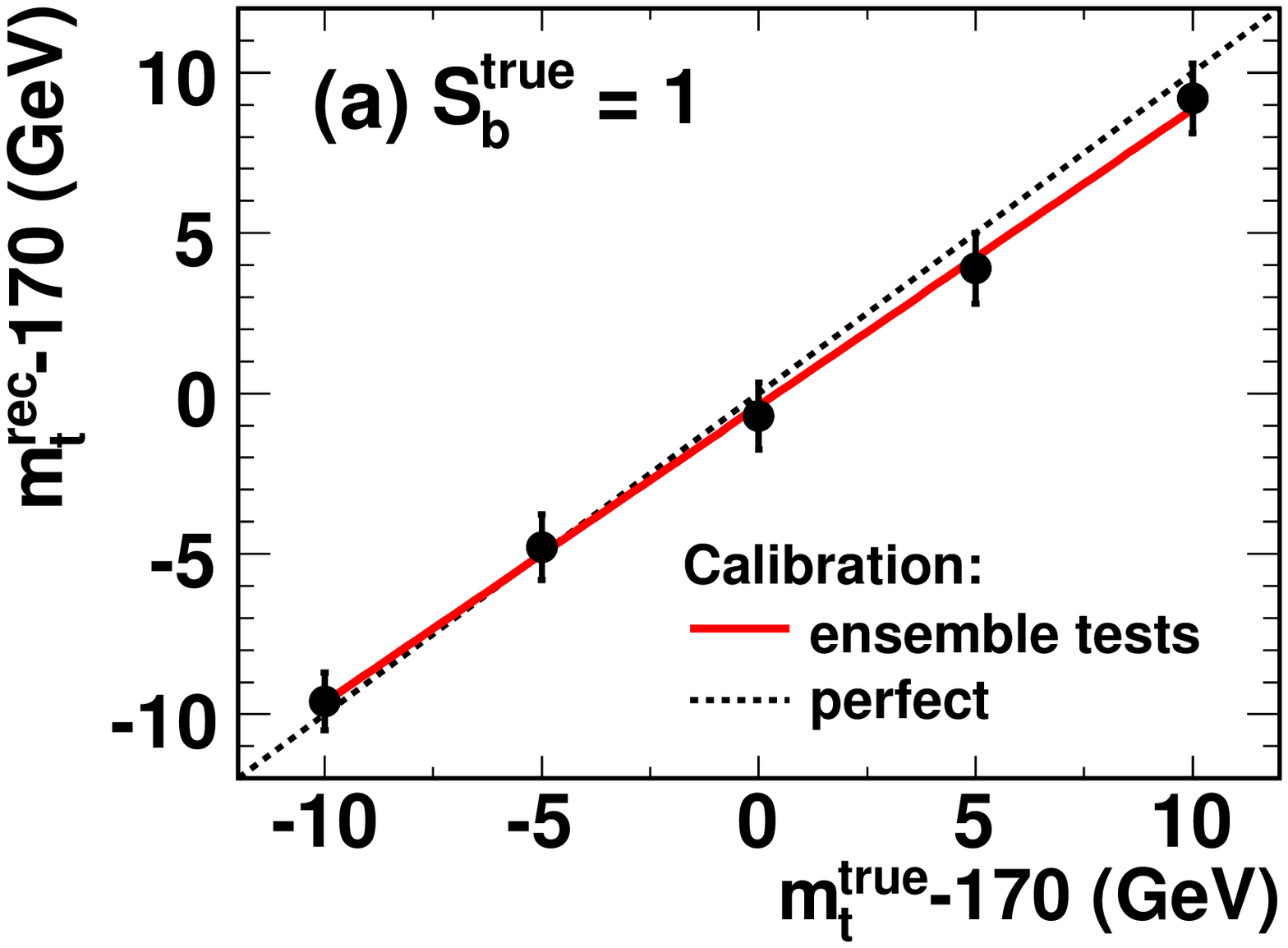}
 &
  \includegraphics[width=0.45\textwidth]{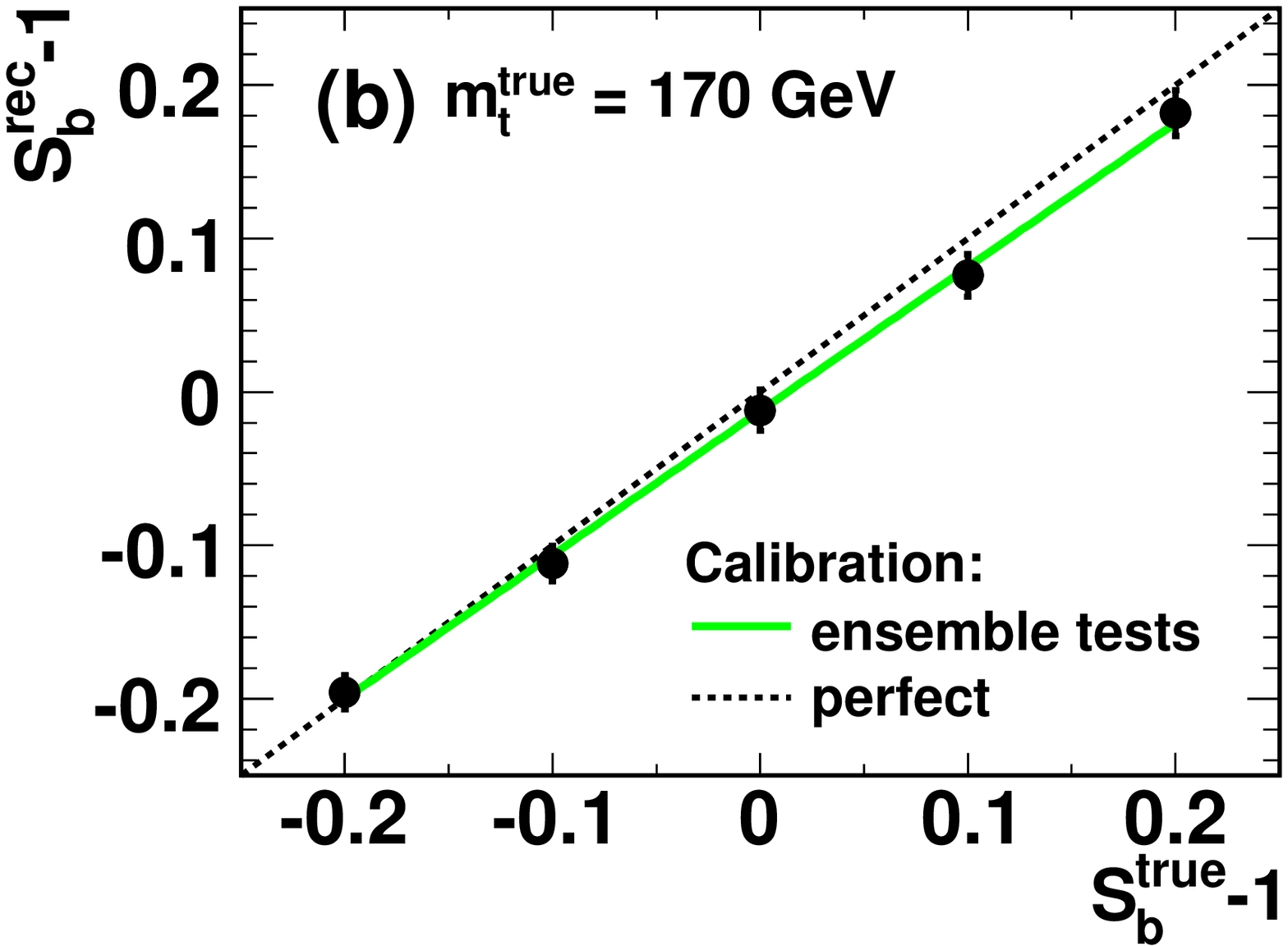}
 \\
  \includegraphics[width=0.45\textwidth]{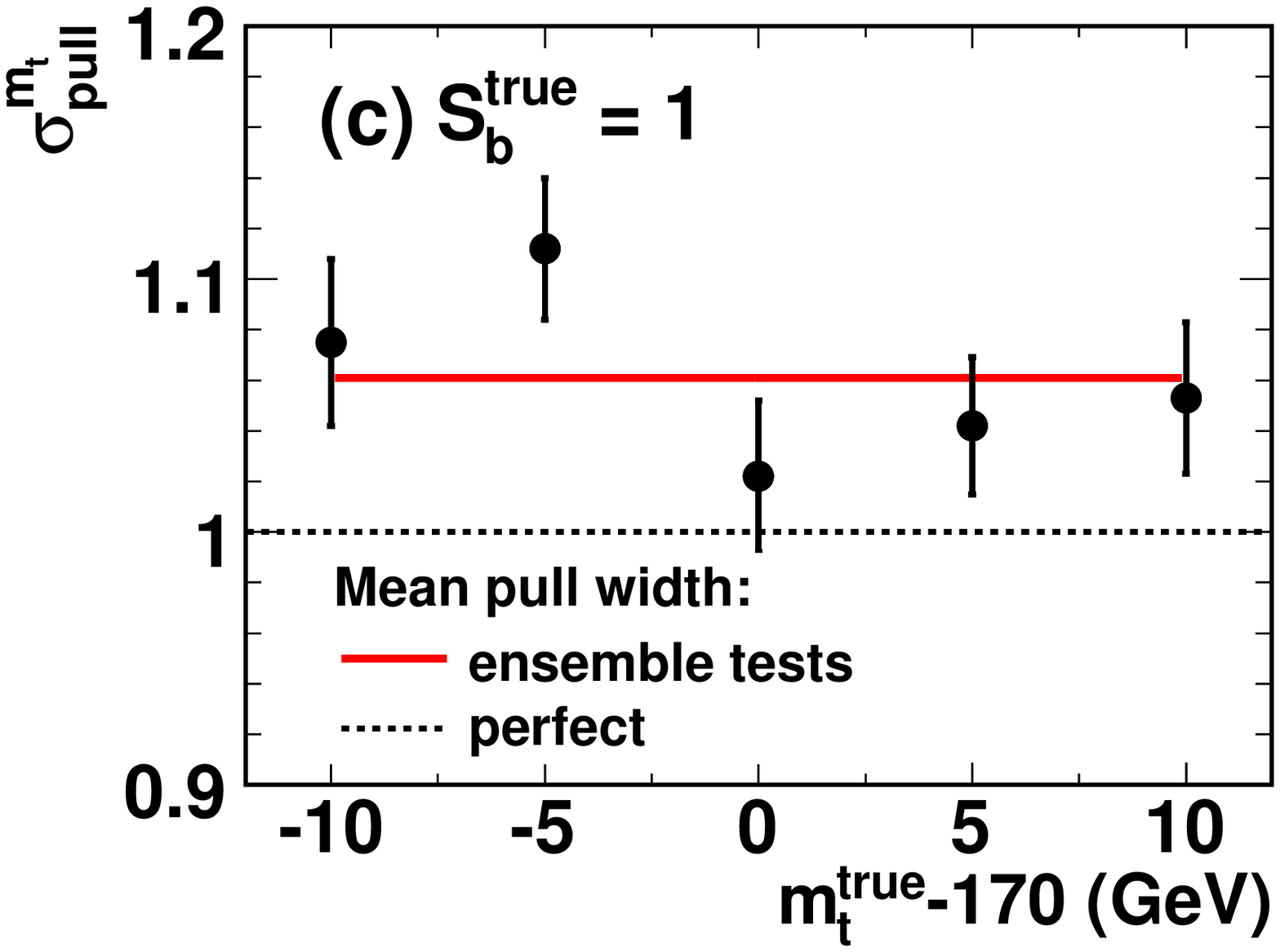}
 &
  \includegraphics[width=0.45\textwidth]{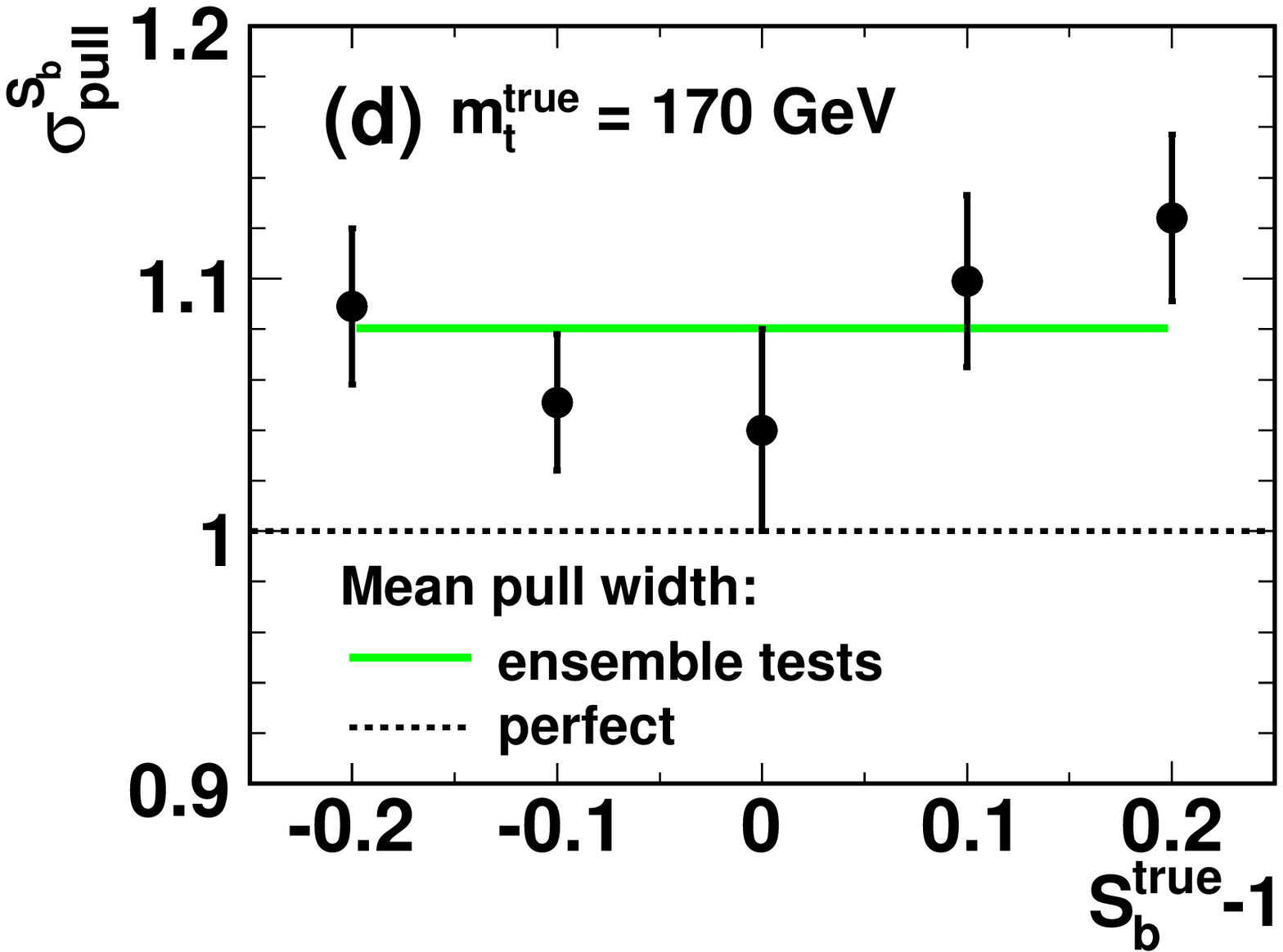}
\end{tabular}
\caption{\label{fig:plmtbjesememzjjwwjjcalibmtbjes}
  Dilepton channel: 
  Measurement of \mt and \sbjes
  in ensembles including 24\% of \ztautaujj and 6\% of \wwjj background.  
  Reconstructed (``rec'') vs.\ true values are shown in
  plots (a) and (b), pull widths vs.\ true values in plots (c)-(d).
  In plots (a) and (b) the lines show the results of straight-line fits,
  while in plots (c) and (d) they indicate the mean values.}
\end{figure}

\Fref{fig:plmtbjesememzjjwwjjcalibmtbjes} shows the calibration curves
of the top quark mass and the \bquark-jet energy scale. 
Their slopes degrade slightly to
$92\pm6\%$ and $94\pm4\%$, respectively.
The pull widths increase to $1.06\pm0.01$ and $1.08\pm0.01$.
In a measurement, the fitted values have to be adjusted according to the 
calibration curve, and the \sbjes 
uncertainty has to be scaled by the pull width.

Note that in this case it is not expected to obtain 
perfect calibration curves, because the \wwjj background is not described with 
a separate likelihood.
This study shows that it is possible to perform the measurement even when
a background source is not accounted for in the likelihood.

For an integrated luminosity of $12\,{\rm fb}^{-1}$, a signal fraction
$\ftop=70\,\%$, and absolute background fractions of
$\fztautaujj=24\,\%$ and $\fwwjj=6\,\%$, the expected
statistical uncertainties in the \emu channel 
obtained by a single Tevatron experiment
are found to be
\begin{eqnarray}
  \sigma_{\mt}(\emu)   & = & 2.3\ \GeV \ {\rm and}\\
\nonumber
  \sigma_{\bjes}(\emu) & = & 0.028     \, .
\end{eqnarray}
The slopes of the calibration curves in
Figures~\ref{fig:plmtbjesememzjjwwjjcalibmtbjes}(a) and (b) 
have been accounted for, 
and the uncertainties have been multiplied with the pull widths shown
in Figures~\ref{fig:plmtbjesememzjjwwjjcalibmtbjes}(c) and (d).
The statistical correlation between the \mtop and \bjes measurements
is $-55\,\%$.

\section{Systematic Uncertainties}
\label{systuncs.sec}
For a measurement, uncertainties in the properties of 
the full simulation (a \geant-based detector simulation for all relevant
processes) used for the calibration 
have to be accounted for by systematic uncertainties on the measurement result.
These systematic uncertainties are not necessarily equal in magnitude 
to the corrections derived in the calibration.
Systematic uncertainties arise from three sources: modeling of the 
detector performance, uncertainties in the method itself, and
modeling of the physics processes for \ttbar production and background.

In the first top quark mass measurements, 
the largest systematic uncertainties related to
the detector performance originated from the absolute jet energy scales
\bjes and \jes.
With the technique described in this paper, these uncertainties can be
absorbed into the statistical uncertainty.
Uncertainties on the top quark mass 
due to the $|\eta|$ or energy dependencies of the jet energy
scales or due to other detector effects like energy dependent efficiencies
are typically much smaller.

An uncertainty arises from the finite event samples used to 
calibrate the method, which is reflected in uncertainties on the 
calibration curves shown e.g.\ in Figures~\ref{fig:ljets_fit_psgn}
and~\ref{fig:plmtbjesememzjjwwjjcalibmtbjes}.
These uncertainties can be reduced when larger simulated 
event samples are used.
Since all other effects are accounted for by the uncertainties on the 
properties of the full simulation used in the calibration, no
additional uncertainties are assigned to the measurement method
itself.

A significant systematic uncertainty in previous top quark mass measurements
was due to the uncertainty in modeling of initial- and final-state gluon 
radiation.
The most basic uncertainty is related to the overall fraction of events with 
significant radiation.
Since the jet energy scales are measured from the data, it can be 
expected that the method is insensitive to the amount of (soft)
gluon radiation off the final-state quarks, while knowledge of 
the amount of initial-state radiation is important.
Dedicated ensemble tests to study events
with significant radiation have been performed and are
described in the following section.

\subsection{Studies of \ttbar Events with Initial- and Final-State Radiation}
\label{systuncs.ttj.sec}
The model described so far does not account for
\ttbar events with an additional hard
parton from initial- or final-state radiation (\ttj).
When the \ttbar events are replaced by such \ttj events, ensemble
tests yield deviations of about $4\,\GeV$ from the nominal top quark
mass in both the \ljets and dilepton channels.
Thus, the method presented so far relies on the 
knowledge of \fttj,
and an uncertainty on \fttj directly translates into an uncertainty
on the top quark mass.

\begin{figure}[ht]
\centering
\begin{tabular}{cc}
  \includegraphics[width=0.45\textwidth]{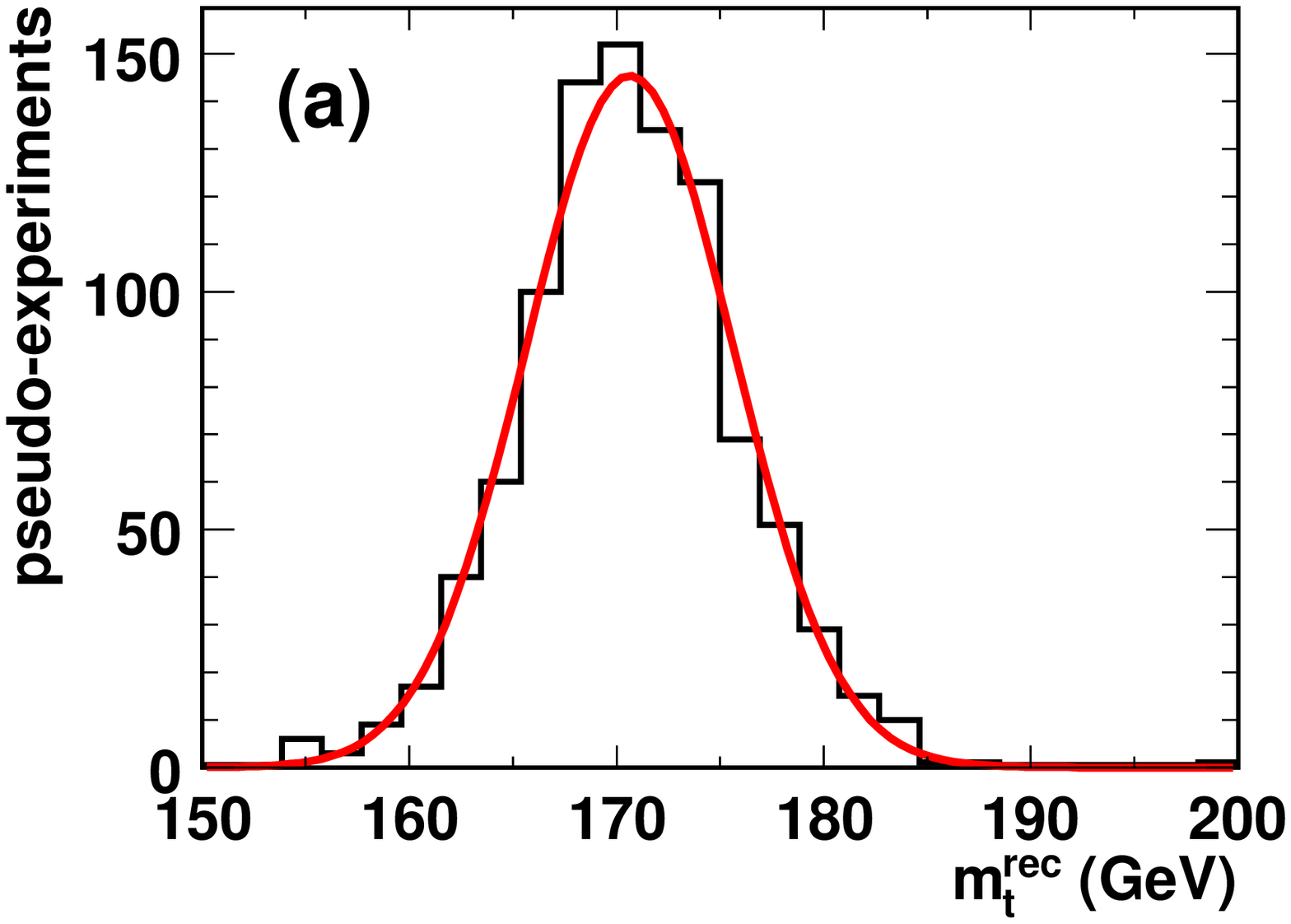}
 &
  \includegraphics[width=0.45\textwidth]{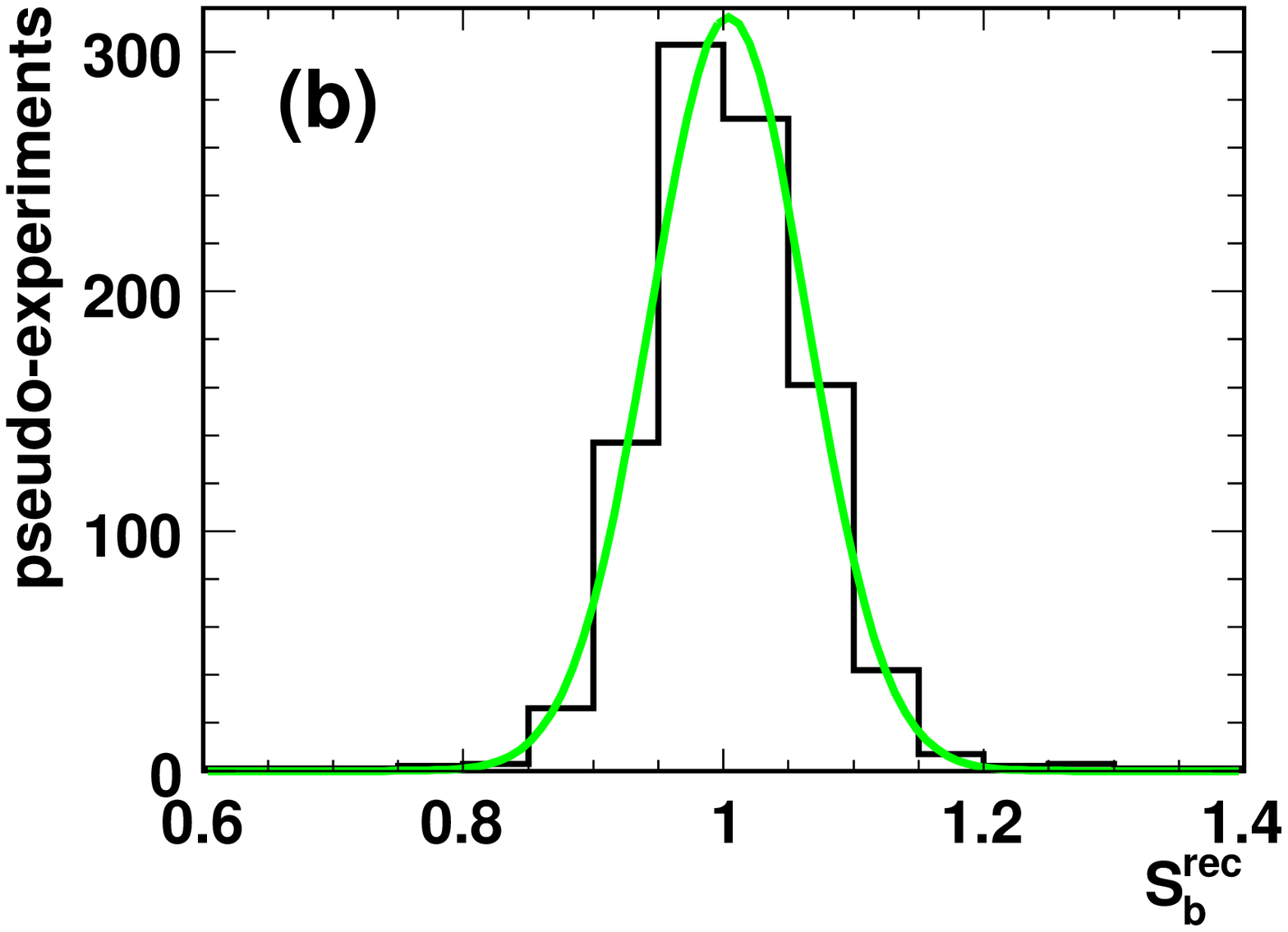}
\end{tabular}
\caption{\label{enstestdilepton_withttbarpt.fig}
  Dilepton channel: 
  Measurement of \mt and \sbjes
  in pseudo-experiments of 50 dilepton \ttbar signal events
  with non-zero \ttbar transverse momentum.
  True values of $\mt=170\,{\rm GeV}$ and $\bjes=1.0$ were used.
  Shown are the distributions of (a) the reconstructed top quark mass
  and (b) the reconstructed \bquark-jet energy scale.}
\end{figure}

An ensemble test has been performed with \ttj events using
an extended model, which has first been described in~\cite{bib-DrarbAlexander}.
The assumption of zero transverse momentum of the \ttbar system is
dropped, an additional integration over the two transverse momentum 
components of the \ttbar system is performed in the likelihood
calculation, and an additional factor $W_{\ptttbar}$ 
is introduced which describes the likelihood
to obtain a \ttbar system with a given transverse momentum.
The ensemble test is performed with pseudo-experiments containing
50 dilepton \ttj signal events.
The true top quark mass is $170\,{\rm GeV}$ and the true \bquark-jet
energy scale is $1.0$.
Figure~\ref{enstestdilepton_withttbarpt.fig} 
shows the results of this ensemble test which 
yields expected central measurement values 
for the top quark mass of $170.6\pm1.0\,{\rm GeV}$
and for the \bquark-jet energy scale of $1.004\pm0.013$,
consistent within uncertainties with the input values.
The pull widths are consistent with $1.0$ in both cases.

This test shows that it is possible to adequatly describe \ttj events
in the method.
To reduce the systematic error on the top quark mass that arises from 
the uncertainty on the fraction of \ttj events in the data sample,
the method should be extended in the future by
introducing the fraction of such events as an additional unknown parameter 
to be measured from the data, similar to the parameters \bjes and \jes.

\section{Conclusions}
\label{conclusions.sec}
The Matrix Element method is a powerful analysis tool that has been
applied with great success in measurements of the top quark mass, 
the discovery of electroweak single top quark production, and searches
for the Higgs boson.
In this paper, a detailed introduction into the method is given
with the aim of facilitating its application to further measurements.
The principle of the method is introduced, and details concerning the 
description of the detector response are given.

It has been proposed previously to overcome the current limitation
in top quark mass measurements arising from experimental systematic
uncertainties by a simultaneous determination of the top quark mass
as well as the absolute energy scales for both \bquark-quark and light-quark
jets.
The paper discusses how this strategy can be implemented naturally
in the Matrix Element method for measurements in both \ljets and 
dilepton events at hadron colliders.
It is shown that the limiting systematic
uncertainty in current measurements (arising from the absolute
energy scale for \bquark-quark jets) can be overcome.
In the future, it should be possible to render the method stable
also against systematic uncertainties related to the fraction of events with 
significant initial- or final-state radiation.

In conclusion, we have given a general introduction to the Matrix
Element method, and we have shown how
future measurements of the top quark
mass can be performed with the Matrix Element method in order to reduce
the experimental systematic error.

\section*{Acknowledgements}
The authors would like to thank Gaston Gutierrez and Juan Estrada for their 
fundamental contributions to the development of the Matrix Element
method, many of which are part of the foundation for the work presented here.
Also, the authors would like to thank Raimund Str\"ohmer
for his careful reading of the manuscript and his very valuable comments,
and all their colleagues at the Tevatron experiments
\dzero and CDF for many helpful discussions.
All authors have previously been employed at Munich University (LMU),
where a substantial part of the work towards this paper has been
performed, and would like to thank Dorothee Schaile, Otmar Biebel, and all 
members of the LMU experimental particle physics group.


\begin{thebibliography}{99}
\bibitem{bib-originalmem}
  V.~M.~Abazov {\it et al.},
  Nature {\bf 429} (2004) 638;\\
  V.~M.~Abazov {\it et al.},
  Phys.\ Lett.\  B {\bf 617} (2005) 1;\\
  K.~Kondo,
  J.\ Phys.\ Soc.\ Jpn.\  {\bf 60} (1991) 836;\\
  R.~H.~Dalitz and G.~R.~Goldstein,
  Phys.\ Rev.\  D {\bf 45} (1992) 1531.

\bibitem{bib-mtopmem}
  A.~Abulencia {\it et al.},
  Phys.\ Rev.\ Lett.\  {\bf 99} (2007) 182002;\\
  T.~Aaltonen {\it et al.},
  Phys.\ Rev.\ Lett.\  {\bf 102} (2009) 152001;\\
  V.~M.~Abazov {\it et al.},
  Phys.\ Rev.\ Lett.\  {\bf 101} (2008) 182001.

\bibitem{bib-singletop}
  V.~M.~Abazov {\it et al.},
  Phys.\ Rev.\ Lett.\  {\bf 103} (2009) 092001;\\
  T.~Aaltonen {\it et al.},
  arXiv:1004.1181 [hep-ex] (2010).

\bibitem{bib-habil}
  F.~Fiedler,
  habilitation thesis at Munich University (2007),
  arXiv:1003.0521.

\bibitem{bib-mtopbjes}
  F.~Fiedler,
  Eur.\ Phys.\ J.\  C {\bf 53} (2008) 41.

\bibitem{bib-pdg}
  C.~Amsler {\it et al.}, Phys.\ Lett.\ B {\bf 667} (2008) 1,
  and 2009 partial update for the 2010 edition.

\bibitem{bib-geant}
  R.~Brun and F.~Carminati, CERN Programming Library Long Writeup 
  {\bf W5013} (1993).

\bibitem{bib-me}  
  V.~M.~Abazov {\it et al.},
  Phys.\ Rev.\ D {\bf 74} (2006) 092005.

\bibitem{bib-CDFljetsme}
  T.~Aaltonen {\it et al.},
  Phys.\ Rev.\  D {\bf 79} (2009) 072001.

\bibitem{bib-dzerobtagging}
  V.~M.~Abazov {\it et al.},
  Phys.\ Rev.\  D {\bf 75} (2007) 092007.

\bibitem{bib-resampling}
  R.~Barlow,
  ``Application of the bootstrap resampling technique to particle
  physics experiments,'' MAN/HEP/99/4 (2000),
  {\tt
  http://www.hep.man.ac.uk/preprints/manhep99-4.ps}

\bibitem{bib-madgraph}  
  F.~Maltoni and T.~Stelzer,
  JHEP {\bf 0302} (2003) 027.

\bibitem{bib-alpgen}  
  M.~L.~Mangano, M.~Moretti, F.~Piccinini, R.~Pittau and A.~D.~Polosa,
  JHEP {\bf 0307} (2003) 001.

\bibitem{bib-CTEQ5L}  
  H.~L.~Lai {\it et al.},
  Eur.\ Phys.\ J.\ C {\bf 12} (2000) 375.

\bibitem{bib-schiefer}
  P.~Schieferdecker,
  PhD thesis at Munich University (2005),
  FERMILAB-THESIS-2005-46.

\bibitem{bib-mahlonparke}  
  G.~Mahlon and S.~J.~Parke,
  Phys.\ Lett.\ B {\bf 411} (1997) 173.

\bibitem{Lepage:1977sw}
  G.~P.~Lepage,
  J.\ Comput.\ Phys.\  {\bf 27} (1978) 192.
 
\bibitem{Lepage:1980dq}
  G.~P.~Lepage,
  Cornell preprint CLNS:80-447 (1980).

\bibitem{bib-vecbos}  
  F.~A.~Berends, H.~Kuijf, B.~Tausk and W.~T.~Giele,
  Nucl.\ Phys.\ B {\bf 357} (1991) 32.

\bibitem{bib-PHdiss}
  P.~Haefner,
  PhD thesis at Munich University (2008),
  FERMILAB-THESIS-2008-51.

\bibitem{bib-DrarbAlexander}
  A.~Grohsjean,
  PhD thesis at Munich University (2008),
  FERMILAB-THESIS-2008-92.

\bibitem{bib-d0dileptonxs}
  V.~M.~Abazov {\it et al.},
  Phys.\ Lett.\  B {\bf 679} (2009) 177.

\end{thebibliography}
\end{document}